\documentclass[12pt]{iopart}

\usepackage{iopams}  
\usepackage{graphicx}
  \expandafter\let\csname equation*\endcsname\relax
  \expandafter\let\csname endequation*\endcsname\relax
\usepackage{amsmath,amsfonts,amssymb}
\usepackage[margin=1in]{geometry}
\usepackage{tikz}
\usepackage{bm}   
\usepackage{cite} 
\usepackage{dsfont}  
\usepackage{url}

\begin{document}

\title[Phase diagram of the triangular-lattice Potts 
antiferromagnet]{Phase diagram of the triangular-lattice 
Potts antiferromagnet} 

\author{Jesper Lykke Jacobsen$^{1,2,3}$,
Jes\'us Salas$^{4,5}$, and 
Christian R. Scullard$^{6}$}

\address{${}^1$Laboratoire de Physique Th\'eorique, D\'epartement de Physique 
de l'ENS, \'Ecole Normale Sup\'erieure, UPMC Univ. Paris 06, CNRS, PSL Research 
University, 75005 Paris, France}
\address{${}^2$Sorbonne Universit\'es, UPMC Univ. Paris 06, \'Ecole Normale 
Sup\'erieure, CNRS, Laboratoire de Physique Th\'eorique (LPT ENS), 75005 Paris,
France}
\address{${}^3$Institut de Physique Th\'eorique, CEA Saclay, 91191 
Gif Sur Yvette, France}
\address{${}^4$Grupo de Modelizaci\'on, Simulaci\'on Num\'erica y Matem\'atica
Industrial, Universidad Carlos III de Madrid, Avda.\/ de la Universidad, 30, 
28911 Legan\'es, Spain} 
\address{${}^5$Grupo de Teor\'{\i}as de Campos y F\'{\i}sica 
Estad\'{\i}stica, Instituto Gregorio Mill\'an, Universidad Carlos III de 
Madrid, Unidad Asociada al Instituto de Estructura de la Materia, CSIC, 
Madrid, Spain}  
\address{${}^6$Lawrence Livermore National Laboratory, Livermore, CA 94550, 
USA}

\eads{\mailto{jesper.jacobsen@ens.fr}, 
      \mailto{jsalas@math.uc3m.es}, 
      \mailto{scullard1@llnl.gov}}

\begin{abstract}
We study the phase diagram of the triangular-lattice $Q$-state Potts model in
the real $(Q,v)$-plane, where $v={\rm e}^J-1$ is the temperature variable.
Our first goal is to provide an obviously missing feature of this 
diagram: the position of the antiferromagnetic critical curve. 
This curve turns out to possess a bifurcation point with two branches 
emerging from it,
entailing important consequences for the global phase diagram.
We have obtained accurate numerical estimates for the position of this curve by 
combining the transfer-matrix approach for strip graphs with toroidal 
boundary conditions and the recent method of critical polynomials.
The second goal of this work is to study the corresponding $A_{p-1}$ RSOS model
on the torus, for integer $p=4,5,\ldots,8$. We clarify its relation 
to the corresponding Potts model, in particular concerning
the role of boundary conditions. For certain values of $p$, we identify
several new critical points and regimes for the RSOS model and
we initiate the study of the flows between the corresponding field theories.
\end{abstract}

%
%
\section{Introduction}
\label{sec:intro}

The concept of universality in statistical physics states that the field 
theory describing the continuum limit of a given lattice model is often 
insensible to the microscopic details of the latter, as long as a suitable 
set of interaction constants is adjusted so as to place the model
at its critical point%
\cite{Essam_63,Widom_65a,Widom_65b,Kadanoff_66,Fisher_67,Wilson_75,Kadanoff_76} 
(see also the more recent reviews 
\cite[and references therein]{Fisher_98,Pelissetto_Vicari_02}). 
This issue has been particularly well tested in two dimensions, where
many models of interest turn out to be exactly solvable \cite{Baxter_book}. 
In particular, the Ising model can be solved on essentially any planar 
lattice \cite{Kasteleyn61}, and its ferromagnetic critical point leads 
in all cases to the same field theory of free Majorana fermions. 
However, under very particular circumstances, different, lattice-dependent 
scaling behaviour can be produced as well. This is the case for 
fully-frustrated models such as the zero-temperature Ising model on the 
triangular lattice \cite{Stephenson64}, which gives rise instead to a 
free bosonic continuum limit \cite{BloteHilhorst82,NienhuisHilhorstBlote84}. 
Similar remarks can be made about hard lattice gases, which
can produce both a universal scaling behaviour of the Lee-Yang type, 
and other critical points whose continuum limit depends on the underlying 
lattice structure \cite{Assis14}.

It is clearly of interest to investigate the issue of universality in richer 
and more complicated models, in which adjustable parameters give rise to 
critical lines along which the critical exponents vary continuously. One case 
that has been studied in details is that of fully-packed loop models,
where the adjustable parameter is the fugacity of a loop. These models have 
been solved both on the hexagonal \cite{KondevGierNienhuis96} and square
\cite{JacobsenKondev98,KondevJacobsen98} lattices, and the continuum 
limits turn out to be different. Whilst certainly surprising at first sight, 
this difference can be traced back to the number of bosonic degrees of freedom 
of the field theory, which is found to depend on the lattice structure 
\cite{KondevIJMPB97}, again via a mechanism of full frustration.

The $Q$-state Potts model \cite{Potts52,Wu82} provides a next natural test bed, 
containing two adjustable parameters---the number of states $Q$ and the 
temperature variable $v = {\rm e}^J-1$, where $J$ denotes the dimensionless 
coupling constant---, for which one could hope to see a mixture of universal
and non-universal behaviour within one single model. For instance, the Potts 
models on the square \cite{Baxter73} and triangular lattice 
\cite{BaxterTemperleyAshley78} are both exactly solvable (integrable) 
at their ferromagnetic phase transition, giving rise
to a continuum limit of a twisted free boson, known as the Coulomb gas. 
On the other hand, both models possess another integrable case---a pair of 
mutually dual antiferromagnetic (AF) transition curves for the square-lattice 
model \cite{Baxter82}, and the zero-temperature AF transition 
\cite{Baxter86,Baxter87} (also known as the chromatic polynomial) 
for the triangular-lattice model---which are both related to continuum 
theories involving two coupled bosons. These field theories
are however profoundly different, the first one involving one compact and one 
non-compact boson \cite{JacobsenSaleur06}, while the second one has been 
found recently \cite{VJS16} to involve one or two compact bosons depending 
on the regime (that is, roughly speaking, the range of $Q$-values).

In order to complete this intricate picture, it seems highly desirable to 
acquire complete knowledge about the phase diagram and the flow between the 
various critical points present in each theory. For the square-lattice
Potts model we have essentially already a full understanding 
\cite{Saleur91,JacobsenSaleur06}. The goal 
of the present paper is thus to investigate
as far as possible the phase diagram of the triangular-lattice model, 
and to complete our understanding of
its possible types of critical behaviour, outside the particular curves 
which have already been examined using integrability techniques. 
To this end, we shall employ a variety of numerical approaches, including the
study of limiting curves of partition function zeroes, a recent method 
involving the roots of a certain graph polynomial,
and direct diagonalisation of the transfer matrix (TM). The resulting 
universality classes will be discussed and
classified, when possible, in terms of conformal field theories (CFT)
\cite{DiFrancesco_97}.

One feature that is worth special attention is the determination of the 
AF critical curve. We shall show here that it bifurcates at a T-point, which 
in turn accounts for the two distinct regimes that have already been 
identified in the zero-temperature case \cite{Baxter86,Baxter87}. Moreover, 
we exhibit an unusual phenomenon for $Q > 4$, where pairs of complex-conjugate 
dominant eigenvalues coexist throughout
regions of the $(Q,v)$ phase diagram, rather than just along curves.

The second half of the paper is devoted to a detailed study of the 
$A_{p-1}$ type RSOS 
restriction \cite{Pasquier87,SaleurBauer89,PasquierSaleur90}
of the triangular-lattice Potts model. It is obtained when $Q$ is 
equal to a Beraha number, $B_p = 4 \cos^2 (\pi/p)$ with $p \ge 3$
an integer, and corresponds to the reformulation of the generic Potts model 
as a restricted height model with local interactions. 
A previous study of the corresponding square-lattice model for $p=5,6,7$ 
was reported in \cite[figures~23--25]{JacobsenSaleur06}. 
The plots of the central 
charge $c$ against the variable $x=v/\sqrt{Q}$ showed interesting new features,
including critical points and phases, including some with 
completely unexpected values of $c$.
We shall see here that the triangular-lattice RSOS model has a very rich 
phase diagram as well,
involving both critical points and critical phases, several of which are 
absent from the corresponding model on the square lattice. As a first step in 
their understanding, we provide accurate numerical values of 
the central charge for these theories.

We have laid out the paper as follows. 
In section~\ref{sec:setup} we have included all 
needed background material to make this paper as self-contained 
as possible. Section~\ref{sec:tools} is devoted to explain the main 
technical tools used in our computations: i.e., the TM approach and 
the method of critical polynomials (CP). 
The discussion of our numerical results is contained in 
section~\ref{sec:results}, in particular, the position in the 
$(Q,v)$-plane of the AF critical curve of the triangular-lattice 
$Q$-state Potts model. The next two sections are devoted to the
study of the triangular-lattice RSOS model on the torus: in 
section~\ref{sec:RSOS}, we focus on the description of this model 
and its structural properties, while in section~\ref{sec:RSOS.results}, 
we discuss the numerical results we obtain for $4 \le p \le 8$. Finally, 
section~\ref{sec:concl} deals with the physical implications of our 
findings and prospects for future work. The amplitudes that appear in the 
TM method of section~\ref{sec:tools_TM} are collected in \ref{app.coef}. 

%
%
\section{Basic setup}
\label{sec:setup}

%
%
\subsection{The $Q$-state Potts model} 
\label{sec:potts}

Let $G=(V,E)$ be an unoriented graph (not necessarily simple) with vertex set 
$V$ and edge set $E$, and attach to each vertex $i \in V$ 
a spin variable $\sigma_i \in \{1,2,\ldots,Q\}$, where $Q \in \mathbb{N}$. 
The partition function of the $Q$-state Potts model (in the 
\emph{spin representation}) then reads
\begin{equation}
 Z^{\rm Potts}_G(Q,J) \;=\; \sum_{\{\sigma_i\}} \, 
    \prod_{\langle i,j\rangle} \,
    \mathrm{e}^{J \, \delta_{\sigma_i, \sigma_j}} \,,
 \label{eq:ZPotts}
\end{equation}
where $J = K/T$, the interaction energy between adjacent spins is 
$-K \delta_{\sigma_i,\sigma_j}$, and $T$ denotes the temperature.  

A useful way of rewriting \eqref{eq:ZPotts} is the so-called 
{\em Fortuin--Kasteleyn (FK) representation} \cite{FK1972}: 
\begin{equation}
 Z^{\rm Potts}_G(Q,K) \;=\; \sum_{A \subseteq E} v^{|A|} \, Q^{k(A)} 
                      \;\equiv\; Z^{\rm FK}_G(Q,v) \,,
\label{def.Z_FK}
\end{equation}
where $|A|$ is the number of edges in the subset $A \subseteq E$, 
$k(A)$ denotes the number of connected components in the spanning subgraph 
$(V,A)$, and the temperature-like variable $v$ is given by
\begin{equation}
v \;=\; e^{J} - 1 \,.
\label{def:v}
\end{equation}
Therefore, the Potts-model partition function is a polynomial in the 
variables $Q,v$. This FK representation of the Potts model is useful to
make sense of the model when the original variables $Q,J$ are beyond 
their original physical values: i.e., when $Q$ takes non-integer values and/or 
the coupling constant $J$ takes imaginary values (i.e., $v<-1$). The usual 
AF regime corresponds to $v\in[-1,0)$, while the ferromagnetic one is given 
by $v\in (0,\infty)$. Furthermore, in the most general setting, one can 
promote the variables $(Q,v)$ to complex numbers.

There are many studies in the literature that have aimed at elucidating 
the phase diagram of the $Q$-state Potts model, mostly in two dimensions. 
We find mainly two different approaches. 
The first one is based on the investigation of the zeros of the partition 
function \eqref{def.Z_FK} for recursive strip graphs of the square and 
triangular lattices with various boundary conditions (see 
\cite{ChangSalasShrock02,ChangJacobsenSalasShrock04} and references therein). 
The partition function was usually obtained by using TM  
techniques (see section~\ref{sec:tools_TM}).  
In those studies, the authors considered the partition-function zeros for 
integer values of $Q\ge 2$ (resp.\/ real values of $v\ge -1$) in the complex 
$v$ (resp.\/ $Q$) plane. However, recent studies \cite{JacobsenSalas13} 
considered the real zeros of $Z_G(Q,v)$ in the \emph{real} plane $(Q,v)$. 
This alternative approach provided useful insights about the phase diagram 
of the corresponding Potts models.

A more modern approach is based on the use of a CP introduced by two of 
the present authors  
\cite{JacobsenScullard12,JacobsenScullard13,Jacobsen14,Jacobsen15,JS16}. This 
method has provided more precise results on more general families of graphs  
(see section~\ref{sec:tools_CP}).

In Statistical Mechanics one is mainly interested in the behavior of the system
when the number of degrees of freedom diverges. Therefore, the main object is
not the partition function $Z_G(Q,v)$ (which does not converge to a function 
in the thermodynamic limit $|V|\to \infty$), but the \emph{free energy 
density} (per site): 
\begin{equation}
f_G(Q,v) \;=\; \frac{1}{|V|} \, \log Z_G(Q,v) \,.  
\label{eq.free_energy}
\end{equation}
In particular, the thermodynamic limit is taken by choosing a suitable  
sequence of graphs $G_n=(V_n,E_n)$ converging in some sense to an infinite 
lattice $G_\infty$ (e.g., one satisfying the van Hove criteria). 
Then the \emph{infinite-volume free energy} is defined as the limit
\begin{equation}
f_{G_\infty}(Q,v) \;=\; \lim_{n\to\infty} f_{G_n}(Q,v) \;=\; 
             \lim_{n\to\infty} \frac{1}{|V_n|} \, \log Z_{G_n}(Q,v)\,.  
\label{eq.infiniteV_free_energy}
\end{equation}
Phase transitions then correspond to singularities in $v$ for fixed $Q$ 
of this limiting function $f_{G_\infty}(Q,v)$.

\medskip

\noindent
{\bf Remarks.} 1. In graph theory the two-variable polynomial 
$Z^{\rm FK}_G(Q,v)$ [cf.~\eqref{def.Z_FK}] is known---after an innocuous 
change of variables---as the Tutte polynomial \cite{Tutte54}. 

2. In the FK representation one has to be very careful about the thermodynamic
limit, as it does not commute with other limits. In particular
\begin{equation}
\lim_{n\to\infty} \lim_{Q\to Q_*}    f_{G_n}(Q,v)  \;\neq \; 
\lim_{Q\to Q_*}   \lim_{n\to\infty}  f_{G_n}(Q,v) 
\end{equation}
for certain values of $Q_*$. As we will discuss later, these limits do not
commute when $Q_*$ is a Beraha number $B_p$:
\begin{equation}
B_p \;=\; 4 \cos^2 \left( \frac{\pi}{p} \right) \,, \qquad 
     \text{integer $p \ge 2$.}
\label{eq.Bp}
\end{equation}

3. The infinite-volume free energy \eqref{eq.infiniteV_free_energy} 
does exist and it is a continuous function of $v,Q$ when both 
parameters belong to their non-degenerate physical ranges (i.e., 
positive integer $Q$ and $v\in (-1,\infty)$ using the spin
representation, or when $Q> 0$ and $v\ge 0$ using 
the FK representation). 
In those cases a probabilistic interpretation of the model exists. 
In other cases, the very existence or uniqueness of the limit 
\eqref{eq.infiniteV_free_energy} is not even clear.

4. It is worth mentioning that studying the real zeros in the real 
$(Q,v)$-plane for $v\ge 0$ does not provide any useful information, as there
are no \emph{real} partition-function zeros for finite graphs in this range. 
In this case one has to consider the more `classical' approach of 
\cite[and references therein]{ChangSalasShrock02,ChangJacobsenSalasShrock04},
or the method of CP to be discussed in section~\ref{sec:tools_CP} below.

%
%
\subsection{The square lattice}
\label{sec:setup_sq}

\begin{figure}
\begin{center}
\begin{tabular}{cc}
\includegraphics[width=200pt]{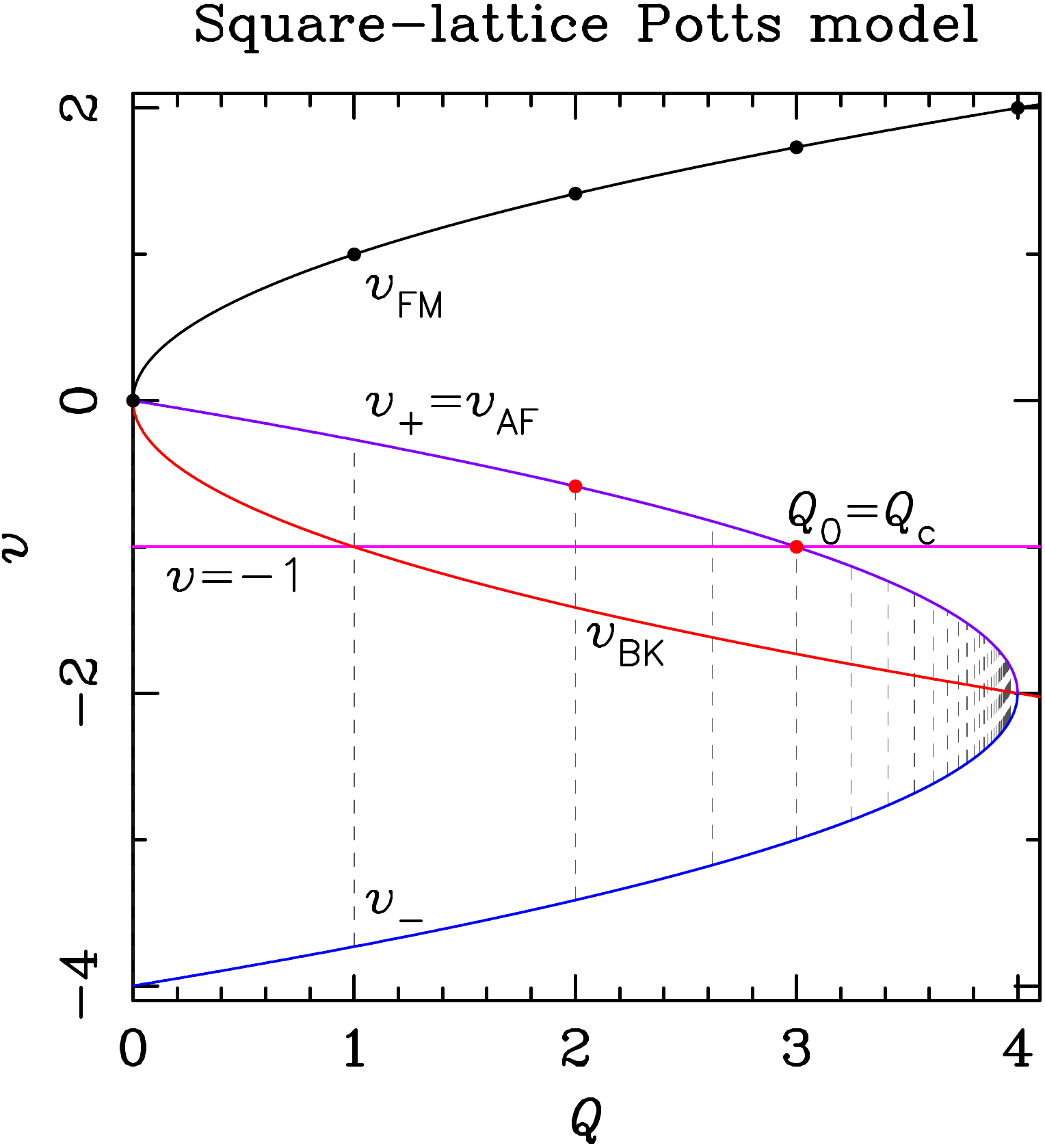} & 
\includegraphics[width=200pt]{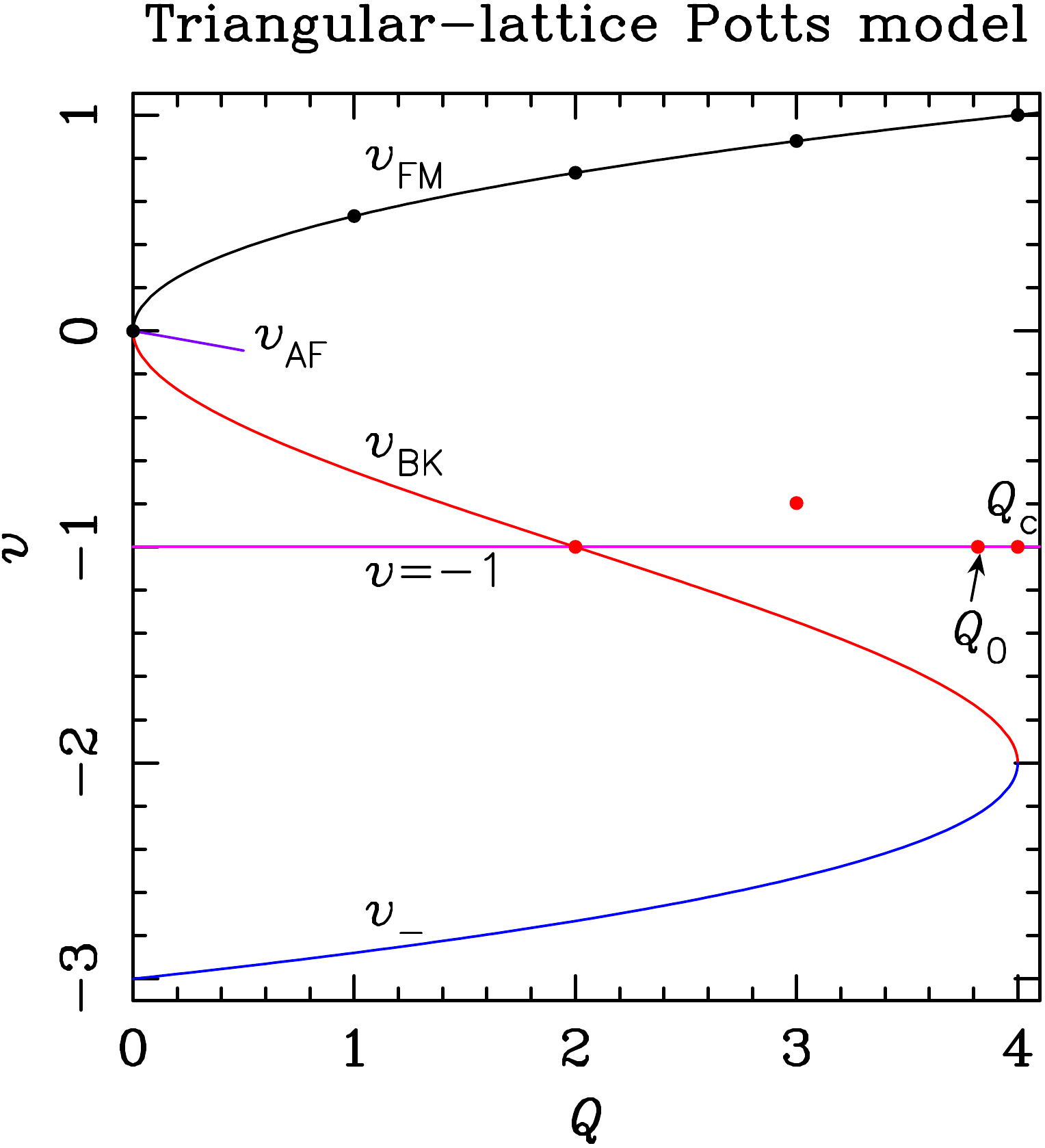} \\[2mm]
\qquad(a) & \qquad(b) \\
\end{tabular}
\end{center}
\caption{(a) Phase diagram of the square-lattice $Q$-state Potts model 
in the real $(Q,v)$ plane. The curve $v_{\rm F}$
\eqref{FM_square}$_+$ is the ferromagnetic critical one, and its analytic 
continuation $v_{\rm BK}$ \eqref{FM_square}$_-$, stays inside 
the BK phase bounded by the
curves \eqref{AF_square}. The dashed vertical lines within this phase show the
location of the first 33 Beraha numbers $B_p$ \eqref{eq.Bp}; along these 
lines the BK phase does not exist. The horizontal (pink) line corresponds 
to $v=-1$, and the solid dots depict the known critical points for integer 
values of $Q$.
(b) \emph{Provisional} phase diagram (as based on previous work) 
of the triangular-lattice $Q$-state 
Potts model in the real $(Q,v)$ plane. The three branches of the cubic 
\eqref{cubic_tri} are expected to play the same role as the curves 
$v_\text{FM}$, $v_\text{BK}$, and $v_{-}$ for the square-lattice model, 
respectively. The solid thin line emerging from the origin represents the 
starting values of the expected antiferromagnetic critical curve $v_{AF}$ 
for this model (see \cite{JSS05}). 
}
\label{fig:sq-tripd}
\end{figure}

Even though our goal is to complete the phase diagram of the Potts model
on the triangular lattice, we need another model where all the important
ingredients are exactly known. We will choose the closely related 
square-lattice Potts model [see figure~\ref{fig:sq-tripd}(a)].  

%
%

Baxter \cite{Baxter73,Baxter82} found the exact free energy in the 
infinite-volume limit \eqref{eq.infiniteV_free_energy} along certain
curves in the $(Q,v)$-plane. On these solvable curves, this model becomes
integrable. He found four integrable curves (see figure~\ref{fig:sq-tripd}(a)):
the first two are self-dual 
\cite{Baxter73} 
\begin{equation}
 v \;=\; \pm \sqrt{Q} \,, 
 \label{FM_square}
\end{equation}
and the other two are mutually dual AF curves \cite{Baxter82}
\begin{equation}
 v_\pm \;=\; - 2 \pm \sqrt{4-Q} \,.
 \label{AF_square}
\end{equation}

{}From the expressions for the bulk free energy $f_\infty(Q,v)$ 
\eqref{eq.infiniteV_free_energy} along these curves, one can obtain 
information about the nature of the phase transitions (when these curves
are critical). Baxter found that when moving across the self-dual curves 
\eqref{FM_square} in the $v$-direction, the system undergoes a first-order 
(resp.\/ second-order) phase transition when $Q > 4$ (resp.\/  
when $0 \le Q \le 4$). However, when crossing \eqref{AF_square} 
the transition is simultaneously first \emph{and}
second order \cite{JacobsenSaleur06}. Moreover, when moving across $Q=4$ in 
the direction along the integrable curves,
the free energy exhibits an essential singularity \cite{Baxter_book}. There 
are also singularities in the surface and corner free
energies in this case \cite{Jacobsen10,VernierJacobsen12}.

Within the critical regime $0 \le Q \le 4$ much further information can be 
obtained. Along the integrable curves one can compute
various critical exponents, which will typically vary continuously with $Q$. 
These exponents provide crucial information that can
help identifying the CFT describing the continuum limit. This program has 
been carried out for the selfdual curves \eqref{FM_square}, and the 
corresponding CFT is found to be that of a compactified boson, that can be 
effectively described by the Coulomb gas (CG) technique \cite{JacobsenReview}. 
This is arguably the simplest possible CFT with continuously varying exponents.

The coupling constant $g$ of the CG that describes the continuum limit 
along the selfdual curves \eqref{FM_square} is related to the temperature-like
parameter $v$ as: 
\begin{equation}
 v \;=\; -2 \cos(\pi g) \,, \qquad Q \;=\; v^2 \,.
\end{equation}
This CG coupling constant satisfies $\frac{1}{2} \le g \le 1$ for the 
ferromagnetic branch $v = +\sqrt{Q}$, 
while $0 < g \le \frac{1}{2}$ for the other (`unphysical selfdual' 
\cite{Saleur91}) branch $v = -\sqrt{Q}$. The 
thermal operator has critical exponent \cite{Nienhuis84,JacobsenReview}
\begin{equation}
 x_T \;=\;  \frac{3}{2g} - 1
 \label{thermexp}
\end{equation}
and is conjugate to a perturbation in the temperature variable $v$ around the 
critical curve \eqref{FM_square}. In particular, the temperature perturbation
is relevant ($x_T \le 2$) along the upper branch of \eqref{FM_square}, 
which is identified with the ferromagnetic critical curve 
$v_\text{FM}(Q)=+\sqrt{Q}$. 
On the other hand, this temperature perturbation is irrelevant ($x_T > 2$) 
along the lower branch of \eqref{FM_square}, which will hence act as a basin 
of attraction for a finite range of $v$-values. 

This basin is delimited by the AF curves \eqref{AF_square} 
and is called the Berker--Kadanoff (BK) phase \cite{Saleur91}.
The BK phase is however unphysical, in the sense that all the scaling levels 
corresponding to the CG description disappear from the spectrum
whenever $Q$ is equal to a Beraha number $B_p$ \eqref{eq.Bp} (these numbers are
shown in figure~\ref{fig:sq-tripd} as vertical dashed lines). 
Two distinct representation theoretical mechanisms are responsible for this 
phenomenon. First, the multiplicity (also known as amplitude, or quantum 
dimension) of certain eigenvalues of the 
corresponding TM vanishes at $Q = B_p$, as can be seen from a 
combinatorial decomposition of the Markov trace \cite{Richard06,Richard07}. 
Second, the representation theory of the quantum group $U_\mathfrak{q}(sl_2)$ 
for $\mathfrak{q}$ a root of unity (with 
$\sqrt{Q} = \mathfrak{q} + \mathfrak{q}^{-1}$) guarantees that other 
eigenvalues are equal in norm at $Q = B_p$ \cite{PasquierSaleur90,Saleur91}, 
and as their combined multiplicity is zero, they vanish from the spectrum as 
well. As a result, for $Q = B_p$ only local observables remain, and these 
can be realised in an equivalent RSOS height model \cite{Pasquier}.

Finally, the continuum limit along the curves $v_\pm$ \eqref{AF_square} 
that bound the BK phase, gives rise to a more exotic CFT 
\cite{JacobsenSaleur06,IkhlefJacobsenSaleur08} with one compact and one 
non-compact boson that couple to form the ${\rm SL}(2,\mathbb{R})/{\rm U}(1)$ 
black hole sigma model \cite{IkhlefJacobsenSaleur12}, 
familiar in the string-theory context \cite{Witten91,DijkgrafVerlinde92}. 
This same continuum limit was recently found to arise also from the
seemingly unrelated spin-one Izergin-Korepin loop model on the square lattice 
\cite{VJS:a22}, which contains as a special case a theta-point 
collapse transition of lattice polymers \cite{VJS_16}.

Because the chromatic line $v=-1$ intersects the BK phase in the 
interval $Q\in [0,3]$ for the square lattice, and the energy is an irrelevant 
operator in the BK phase, this scenario can be checked by studying the 
chromatic zeros in the complex $Q$-plane along the $v=-1$ line. Detailed 
studies of these zeros with different boundary conditions (free, cylindrical, 
cyclic, and toroidal) have been carried out 
\cite{SalasSokal01,JacobsenSalas01,JacobsenSalas06,JacobsenSalas07,transfer5}, 
and they confirm this scenario (with some minor details that depend on 
the boundary conditions). In particular, in the thermodynamic limit, chromatic 
zeros accumulate around certain points (due to the Beraha-Kahane-Weiss (BKW)
theorem \cite{BKW1,BKW2,BKW3}) that are confined to the region $0 \le Q \le 3$.
 
For $Q > 3$ the chromatic line renormalises to infinite temperature 
($v=0$) and is hence non-critical.

\medskip

\noindent
{\bf Remarks.} 1. The lower branch of \eqref{FM_square} (i.e.,  
$v_\text{BK}=-\sqrt{Q}$) is the analytic continuation of the upper branch 
of \eqref{FM_square}, viz., of the critical ferromagnetic curve $v_\text{FM}$.

2. The 3-state Potts antiferromagnet on the square lattice is critical on the
chromatic line $v=-1$, but when the temperature is non-zero (i.e., $-1 < v <0$)
it is disordered; so it renormalises to $v=0$. 

%
%
\subsection{The Berker-Kadanoff phase}

The BK phase actually exists on any lattice and governs a finite part of the 
AF ($v<0$ and $Q > 0$) part of the phase diagram for the following reasons  
(see \cite{JacobsenSaleur08bdrychrom}).
First, the ferromagnetic transition curve $v_{\rm F}$---generalising the 
upper branch of \eqref{FM_square}---must exist on any lattice, by the 
universality of the CG description with $\frac{1}{2} \le g \le 1$. In the 
$(Q,v)$-plane this curve must have a vertical tangent at the origin, since 
the corresponding symplectic fermion CFT (with central charge $c=-2$) is 
known to describe spanning trees \cite{DuplantierDavid88}.
Indeed, if that tangent was not vertical, we would have instead a 
spanning-forest model \cite{JSS05}, which corresponds to perturbing the 
free fermion CFT by a four-fermion term \cite{CJSSS04}, which would render 
it non-critical \cite{CJSSS04,JacobsenSaleur05}.
This means that the critical curve $Q(v)$ has a vanishing derivative 
$Q'(0) = 0$ on any lattice.%
\footnote{This can indeed be checked explicitly for the square and triangular 
lattices, where the exact critical curve is known; see \eqref{FM_square} 
and \eqref{cubic_tri}. Other explicit confirmations for a larger family of 
lattices, including all the Archimedean ones, have been obtained from the 
method of CP \cite{Jacobsen14}.
}
Barring the (unlikely) accident that also $Q''(0) = 0$, we infer
that the critical curve continues analytically from the first 
quadrant $Q,v>0$ into the fourth quadrant $Q>0$, $v<0$. 
Invoking again the universal CG description, now with $g < \frac{1}{2}$, 
we have $x_T > 2$ by \eqref{thermexp}, and a finite range of $v$-values will 
indeed be `controlled' by the BK critical curve $v_\text{BK}$.
This BK phase is bounded in the $(Q,v)$ plane by two smooth curves $v_\pm$ 
such that for any $Q\in[0,Q_\text{c})$, $0\ge v_{+}(Q) > v_{-}(Q)$. The two
curves merge at some value $Q_\text{c} > 0$, signaling the termination of the 
BK phase. The inequality $Q_\text{c} \le 4$ for two-dimensional models 
follows from quantum group results \cite{Saleur91}. 

The upper curve is usually identified with the critical AF 
curve, $v_{+} = v_\text{AF}$, and hits the point $(Q,v)=(0,0)$ at a finite 
negative slope. This value is related to the critical coupling of the 
corresponding spanning-forest model \cite{JSS05}.

Some unusual features of the BK phase have already been pointed out
in the previous section about the square-lattice model, but they hold 
true for any two-dimensional lattice:
At $Q=B_p$, there are vanishing amplitudes and eigenvalue cancellations, so 
that the actual ground state is deeply buried in the spectrum of the 
corresponding TM. We now present another related argument, this time directly 
in the continuum limit, leading to the conclusion that the BK phase
should be dismissed as unphysical.
The critical behaviour of the BK phase can be obtained by analytic 
continuation of the CG results for the ferromagnetic transition 
($g \ge \frac{1}{2}$). It turns out \cite{Saleur90} that some of the critical 
exponents (namely, magnetic and watermelon exponents) become negative when 
$g< \frac{1}{2}$. This means that the analytic continuation of the 
ferromagnetic ground state is not anymore the lowest-energy state in
the BK phase, or alternatively, that correlation functions will not decay 
but rather grow with distance. Such features are clearly unphysical and 
non-probabilistic.

The existence of the BK phase has been verified for all Archimedean 
lattices in \cite{JacobsenScullard13,Jacobsen14}, and its extent in 
the $(Q,v)$ plane has been accurately estimated. It was found that in most, 
but not all \cite{JacobsenScullard12}, cases $Q_\text{c} = 4$.

The critical behaviour on the chromatic line $v=-1$ (for general 
$Q \in \mathbb{R}$) then depends in a crucial way on its 
intersection with the BK phase. We have seen above that this intersection
is the interval $0 \le Q \le 3$ for the square lattice. The remainder of the 
interval $[0,4]$, namely $3 < Q \le 4$ in this case, will then have a 
different behaviour. For the square lattice this behaviour is just 
non-critical. The triangular lattice however offers a much more interesting 
alternative, as we shall see presently.

\medskip

\noindent
{\bf Remark.} For non-planar recursive lattices \cite{JacobsenSalas13}, the 
BK phase is also conjectured to exist, and the value for $Q_\text{c}$ might
be much greater than 4. 

%
%
\subsection{The triangular lattice}
\label{sec:setup_tri}

The $(Q,v)$ phase diagram on the triangular lattice appears to be 
considerably more complicated than was the case on the square lattice.
(See figure~\ref{fig:sq-tripd}(b) for a graphical summary of the
main results known \emph{before} this work.)
In particular the part close to $Q=4$ possesses several intricate features, 
as witnessed by numerical studies of different types
\cite{JacobsenSalasSokal03,JSS05,JacobsenSalas06,JacobsenSalas07}.
This model is integrable along the three branches of the 
cubic \cite{BaxterTemperleyAshley78}
\begin{equation}
 v^3 + 3 v^2 \;=\; Q \,.
 \label{cubic_tri}
\end{equation}
Fortunately, the chromatic line 
\begin{equation}
 v \;=\; -1
 \label{chrom_tri}
\end{equation}
is also integrable \cite{Baxter86,Baxter87} and understanding it in detail 
should provide valuable information that will also be useful for
disentangling the full $(Q,v)$ phase diagram. This programme, initiated in 
\cite{Nienhuis82}, has been completed 
in \cite{VJS16}: by \emph{exactly} mapping the triangular-lattice Potts 
AF at zero temperature to a spin-1 integrable vertex model with a suitably 
chosen twist, whose critical behavior for $Q\in[0,4]$ has been obtained in 
terms of CFTs.

Let us parameterise the cubic curve \eqref{cubic_tri} by
\begin{equation}
 v \;=\; -1 + 2 \cos \left( \frac{2\pi(1-g)}{3} \right) \,, \qquad
 Q \;=\; 4 \cos^2 \left(\pi g \right)
\end{equation}
where $-\frac12 \le g \le 1$. The range $0 < g \le 1$ then has then same CG 
interpretation as for the square lattice: 
$\frac{1}{2}\le g\le 1$ corresponds to $v_\text{FM}$, and 
$0<g\le \frac{1}{2}$ to $v_\text{BK}$. 
The interval $-\frac{1}{2} \le g < 0$ (corresponding to $v_{-}$)
cannot be interpreted within this CG, but has been shown numerically 
\cite{JacobsenSaleur06} to belong to the same universality class 
\cite{IkhlefJacobsenSaleur08,IkhlefJacobsenSaleur12}
as the {\em antiferromagnetic} curve \eqref{AF_square} for the square-lattice 
Potts model (with the same value of $Q$).

This implies that the middle branch $v_\text{BK}$ of \eqref{cubic_tri} 
will control the BK phase, which should extend from the lower branch 
of \eqref{cubic_tri}, that we hence identify with $v_-$, up to some  
AF transition curve $v_\text{AF}=v_{+}$, that unfortunately has not yet been 
determined analytically. In this work we aim at a precise numerical computation
of the antiferromagnetic critical curve, $v_{\rm AF}$. 

This curve cannot be the chromatic line \eqref{chrom_tri}, since the 
latter is not always above the middle branch of \eqref{cubic_tri}. At the 
origin, $(Q,v) = (0,0)$, the AF curve describes a spanning-forest problem 
and its slope is known numerically \cite{JSS05}:
\begin{equation}
\left. \frac{{\rm d} v_\text{AF}}{{\rm d}Q} \right|_{Q=0} \;=\; 
-0.1753 \pm 0.0002\,.
\label{def_der_Q=0}
\end{equation}
(See the small curve emerging from the origin in figure~\ref{fig:sq-tripd}(b).) 

Along the chromatic line $v=-1$, Baxter has found \cite{Baxter86,Baxter87} 
three distinct analytic expressions for the TM eigenvalue $g_i$, 
whose logarithm is the bulk free energy (per site), and determined the regions
of the complex $Q$-plane where each $g_i$ is dominant. In particular,
he found that $g_1$ corresponds to a non-critical phase which includes the
interval $Q \in (-\infty,0) \cup (4,\infty)$. 
In this region, 
exact expressions for the surface and corner free energies are also known
\cite{Jacobsen10}, and the contributions from corners of angle 
$\frac{\pi}{3}$ and $\frac{2\pi}{3}$ can even be determined
separately \cite{VernierJacobsen12}. In the limit $Q \to 4^+$, these corner 
free energies develop essential singularities that have been determined from 
the asymptotic analysis of the exact expressions \cite{VernierJacobsen12}.

The BK phase is characterised by $g_3$ and corresponds to the interval 
$Q\in (0,Q_0)$ where \cite{Baxter87,JacobsenSalasSokal03}
\begin{equation}
 Q_0 = 3.819\,671\,731\,239\,719 \cdots \,.
 \label{eq.Q0}
\end{equation}
This BK phase is called Regime~I in \cite{VJS16}. Finally, $g_2$ is 
dominant in the interval $Q\in (Q_0,4)$, and defines a new critical phase 
(Regime~IV) whose thermodynamic limit is governed by the coset CFT \cite{VJS16}
\begin{equation}
\frac{SU(2)_4 \times SU(2)_{p-6}}{SU(2)_{p-2}} \,,
\label{def:coset_cft}
\end{equation}
where $p$ is related to $Q$ by the usual parametrisation 
$Q = 4 \cos^2(\pi/p)$, and \eqref{def:coset_cft} is defined in principle 
for any integer $p\ge6$, and by analytic continuation for appropriate real 
values of $p$. The central charge for this regime is given by 
\begin{equation}
c \;=\; 2 - \frac{24}{p(p-4)} \,,
\label{eq:c_IV}
\end{equation}
which is the analytic continuation to any real $p>4$ of that of the 
$S_3$-invariant minimal models studied by Fateev and Zamolodchikov 
\cite{FZ1,FZ2}. These models can alternatively be formulated as the 
above conformal cosets \eqref{def:coset_cft}. 
This eigenvalue $g_2$ describes the colouring problem in this Regime~IV; but 
this also holds for the analytic continuation of $g_2$ inside the BK phase, 
at least down to $Q=2^+$ \cite{VJS16}. It is worth mentioning that in 
Regime~IV the thermal operator is irrelevant, as it is in the BK phase.  

Returning now to the phase diagram in the $(Q,v)$-plane, the simplest 
scenario that can account for these findings is that
the AF transition curve must split at a triple point T into a branch 
that goes down to $(Q,v) = (Q_0,-1)$ and
delimits the BK phase, and another branch that extends further to the 
4-colouring point $(Q,v) = (4,-1)$. This simple scenario is the one we 
will find as one of the conclusions of this work (see figure~\ref{fig:tripd}). 

It is also worth mentioning here that the phase diagram for the region 
$Q > 4$ is not as `simple' as the scenario we have just depicted for 
$0\le Q\le 4$. Our numerical TM results for strip graphs of finite
width $2\le L\le 5$ (displayed in figure~\ref{Figure_tri_TM})
show that for $Q>4$ there are involved structures: many `small' phases (most
of them characterised not by a single dominant eigenvalue, but by a pair
of complex conjugate eigenvalues), and outward branches for odd values of 
$L=3,5$. These structures can also be seen in 
figure~\ref{Figure_tri_CP}, where the critical-polynomial data is shown. In 
any case, it is not clear whether these structures converge to anything in the
thermodynamic limit. Additional work on larger values of $L$ would be 
necessary to explore this region of the phase diagram; but this is outside the
scope of this paper. 

%
%
\section{Main technical tools}
\label{sec:tools}

%
%
\subsection{Transfer-matrix computations on the torus}
\label{sec:tools_TM}

The construction of the TM for the $Q$-state Potts model in 
the FK representation on a triangular-lattice strip with toroidal boundary 
conditions%
\footnote{
   We will use the standard notation for boundary conditions of strip graphs:
   free (resp.\/ toroidal) means that the boundary conditions in both 
   directions are free (resp.\/ periodic). Furthermore, cylindrical 
   (resp.\/ cyclic) means that they are periodic (resp.\/ free) in the 
   transverse direction and free (resp.\/ periodic) in the longitudinal
   direction.} 
can be obtained from the general theory developed in detail in 
\cite[and references therein]{SalasSokal01,JacobsenSalasSokal03} 
(in particular, for the chromatic polynomial), the generalisation of the TM
formalism for cylindrical boundary conditions and 
general values of $v$ \cite{ChangJacobsenSalasShrock04}, and the tricks 
needed to deal with periodic boundary conditions in the longitudinal
direction \cite{JacobsenSalas06,JacobsenSalas07,JacobsenSalas13,Salas13-1}. 
We will shortly review the main results here to make the paper as 
self-contained as possible. 

\begin{figure}
\begin{center}
\includegraphics[width=400pt]{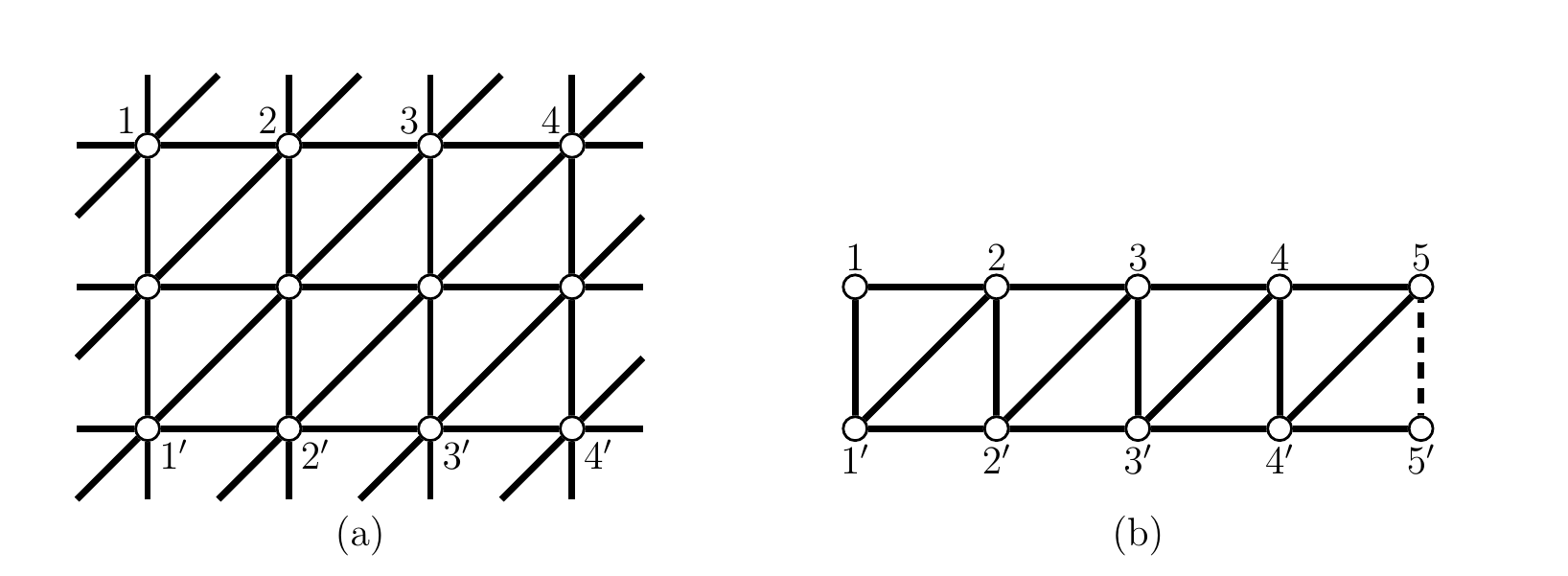}
\end{center}
\caption{Triangular-lattice on a torus. (a) A triangular-lattice strip on a 
torus of size $4\times 3$ with periodic boundary conditions. For simplicity,
we have represented this strip as a square-lattice strip with oblique edges. 
The spins on the bottom row (resp.\/ top row) are labeled as 
$\{1',2',3',4',5'\}$ (resp.\/ $\{1,2,3,4,5\}$). 
The half edges are meant to represent the fully periodic boundary conditions.
Notice that we are using an opposite definition of the oblique edges compared 
to \cite{JacobsenSalas06,JacobsenSalas07}. 
(b) Transfer-matrix construction for a triangular-lattice strip of 
width $L=4$. In this case, we need to start from a triangular-lattice strip
of width $L+1=5$ with free transverse boundary conditions. 
At the end of the computation, we should identify 
the vertices on the leftmost column with the corresponding ones on the 
rightmost column (i.e., $1'=5'$ and $1=5$). The rightmost dashed vertical edge 
is special, as it corresponds to $v=0$. In this way we ensure that the previous
identification gives an edge with the right weight $v$.  
}
\label{fig:tm_tri}
\end{figure}

First of all, if we want to build the TM for a triangular-lattice strip
of width $L$ with periodic boundary conditions in the transverse direction,
one has to start from a triangular-lattice strip of width $L+1$ and 
free transverse boundary conditions \cite{Baxter87,SalasSokal01}. This
is shown in panel~(b) of figure~\ref{fig:tm_tri}. Indeed, at the end of the 
computations, we have to identify this extra column $L+1$ with column $1$. 

%
%

Secondly, dealing with periodic boundary conditions in the longitudinal 
direction in the FK representation forces us to keep track of the 
connectivities among the vertices on the top row $\{1,2,\ldots,L+1\}$ and 
those on the bottom row $\{1',2',\ldots,(L+1)'\}$. Therefore, the TM will 
act on the space of partitions of the set 
$\{1',2',\ldots,(L+1)',1,2,\ldots,L+1\}$.   
Given a partition $\mathcal{P}$ of this set (i.e., a connectivity state
$\bm{v}_\mathcal{P}$), a \emph{link} is a block that contains vertices 
belonging to both the top and bottom rows. 
The number of links in a given partition will be denoted by $\ell$.

The action of the TM $\mathsf{T}_L$ is to add one layer on top of the strip 
graph. As usual, we want to write this matrix as a product of 
`horizontal'  and `vertical' operators:
\begin{subequations}
\label{def_HV}
\begin{align}
\mathsf{Q}_{jk} &\;=\; \mathds{1} + v \, \mathsf{J}_{jk} \,, \\ 
\mathsf{P}_j    &\;=\; v\, \mathds{1} + \mathsf{D}_j \,, 
\end{align}
\end{subequations}
where $\mathds{1}$ is the identity operator, 
$\mathsf{J}_{jk}$ is a `join' operator that amalgamates the blocks 
containing the vertices $j$ and $k$, and $\mathsf{D}_j$ is a `detach' 
operator that removes vertex $j$ from its block and turns it into a singleton 
(if $j$ was already a singleton, it provides a factor $Q$ to that term). 
Then the full TM can be written in a similar way as for the chromatic 
polynomial \cite{JacobsenSalas07}: 
\begin{equation}
\mathsf{T}_L \;=\; 
\left[ 
\left( \prod_{j=1}^L \left( \mathsf{P}_{j} \cdot \mathsf{Q}_{j,j+1} \right) 
\right) \cdot 
\mathsf{P}^{(0)}_{L+1} \right] \cdot 
\left[ \mathsf{J}_{1,L+1} \cdot \left( \prod_{j=1}^L \mathsf{Q}_{j,j+1} 
\right) \right] \,, 
\label{def_TM}    
\end{equation}
where the rightmost operators act first. The first bracket corresponds to
the `vertical' TM $\mathsf{V}_L$, while the second one is related to the 
`horizontal' TM $\mathsf{H}_L$, so that 
$\mathsf{T}_L = \mathsf{V}_L \cdot \mathsf{H}_L$. The oblique edges in the
vertical TM corresponds to the operators $\mathsf{Q}_{j,j+1}$ in the 
$\mathsf{V}_L$ term in \eqref{def_TM}. 
The role of the join operators $\mathsf{J}_{1,L+1}$ in the $\mathsf{H}_L$ 
term is to identify the leftmost and rightmost columns. The operator 
$\mathsf{P}^{(0)}_j = \mathsf{D}_j$ appearing in the
$\mathsf{V}_L$ term is just the evaluation of 
$\mathsf{P}_j$ at $v=0$.\footnote{%
   In the chromatic-polynomial case, there is no need of introducing 
   the operator $\mathsf{P}^{(0)}_j$, as two edges with $v=-1$ in parallel act 
   like a single edge with the same $v=-1$; but this is not true for a 
   generic value of $v$. As a matter of fact, two edges in parallel of 
   weights $v$ and $v'$ are equivalent to a single edge with effective 
   weight $v_\text{eff}=v+v'+vv'$.  
}
Its role is to propagate the spin $(L+1)'$ forwards in time without introducing
an extraneous edge weight between spins $(L+1)'$ and $L+1$, since this is 
already taken into account in the first column.

Then the partition function \eqref{def.Z_FK} for a triangular-lattice 
strip graph $G$ of width $L$ and height $N$ on the torus (represented for 
simplicity by $L_P \times N_P$; see figure~\ref{fig:tm_tri}(a)) is given 
by the expression:
\begin{equation}
Z_{L_P  \times N_P}(Q,v) \;=\; \bm{u}^T \cdot \mathsf{T}_L^N \cdot 
                               \bm{v}_\text{id} \,,
\label{eq_Z_vs_TM} 
\end{equation} 
where the right vector $\bm{v}_\text{id}$ represents the identity connectivity
state $\bm{v}_\text{id}=\{ \{1,L+1,1',(L+1)'\},\{2,2'\},\ldots,\{L,L'\}\}$ 
(in other words, we start with the bottom and top rows identified and with
the rightmost and leftmost sites in a single block). The left vector
$\bm{u}^T$ in \eqref{eq_Z_vs_TM} acts on a connectivity state 
$\bm{v}_\mathcal{P}$ as follows:
\begin{equation} 
\bm{u}^T \cdot \bm{v}_\mathcal{P}  \;=\; Q^{|\mathcal{P}'|} \,,
\end{equation} 
where $|\mathcal{P}'|$ means the number of blocks of the partition 
$\mathcal{P}'$, which is related to $\mathcal{P}$ by the equation 
\begin{equation} 
\bm{v}_\mathcal{P'} \;=\; \mathsf{J}_{1,L+1} \cdot \mathsf{J}_{1',(L+1)'}\cdot  
 \left( \prod_{j=1}^{L+1} \mathsf{J}_{j,j'}\right) \cdot \bm{v}_\mathcal{P}\,. 
\end{equation} 
In words, $\bm{u}^T$ acts on $\bm{v}_\mathcal{P}$ by identifying the top and
bottom rows, and the rightmost vertices $L+1,(L+1)'$ with the leftmost ones
$1,1'$, respectively; and then assigning a factor $Q$ to each block of the
resulting partition $\mathcal{P}'$.  

\medskip

\noindent
{\bf Remark.} The convention used in this paper to build the triangular
lattice (see figure~\ref{fig:tm_tri}) and the corresponding TM 
\eqref{def_TM} are different from those used in previous publications by two
of the authors \cite{JacobsenSalas06,JacobsenSalas07}. The only difference is
that now the oblique edges point North-East, rather than North-West as in 
those papers. The final results are of course independent of this choice. 

\medskip

%
%
\begin{figure}
\begin{center}
\includegraphics[width=200pt]{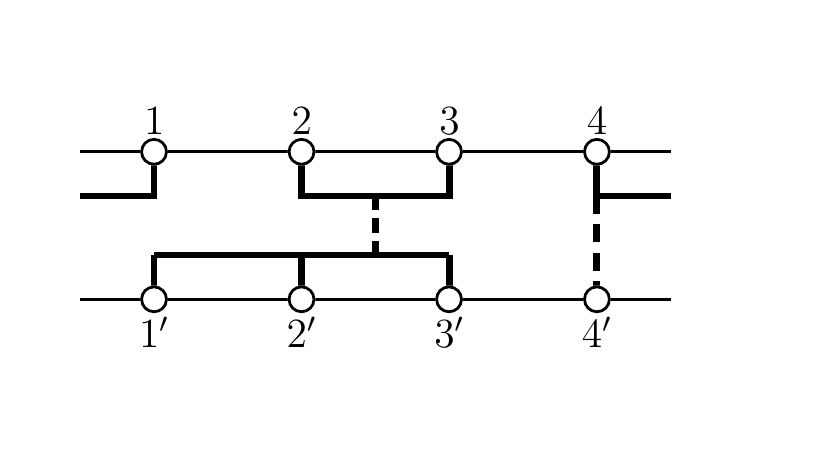}
\end{center}
\caption{Graphical representation of the connectivity state for a 
triangular-lattice strip with periodic transverse boundary conditions of
width $L=4$ (once the `extra' vertices $5,5'$ has been identified  with
$1,1'$, respectively). The picture depicts the connectivity state 
$\mathcal{P}=\{ \{1,4,4'\},\{2,3,1',2',3'\}\}$. 
This state can be though of as a bottom-row connectivity  
$\mathcal{P}_\text{bottom}=\{ \{1',2',3'\},\{4'\}\}$, a top-row 
connectivity $\mathcal{P}_\text{top}=\{ \{1,4\},\{2,3\}\}$, and two links:
one connects the first (resp.\/ second) block of $\mathcal{P}_\text{bottom}$ 
with the second (resp.\/ first) block of $\mathcal{P}_\text{top}$. The 
connectivities $\mathcal{P}_\text{top}$ and $\mathcal{P}_\text{bottom}$
are shown as thick solid lines, while the links are depicted as thick 
dashed lines. 
}
\label{fig:P_tri}
\end{figure}

Let us consider a given connectivity state $\mathcal{P}$ as a connectivity
state of the bottom (resp.\/ top) row $\mathcal{P}_\text{bottom}$ 
(resp.\/ $\mathcal{P}_\text{top}$), and a set of $\ell$ links. We assume that
$1,L+1$ belong to the same block, as well as $1',(L+1)'$. 
As each link connects a single block of the top-row partition to a single 
block of the bottom-row partition, and vice versa, then $0\le \ell \le L$.  
See figure~\ref{fig:P_tri} for an example. 
This computation can be simplified by using several simple observations:

\begin{itemize}
\item As the TM \eqref{def_TM} only acts on the vertices of the top row, 
$T_L$ has a block-diagonal structure, where each block is 
characterised by a given bottom-row connectivity. Indeed, all blocks have the
same eigenvalues (although in some blocks, these eigenvalues might appear 
several times). Therefore, it suffices to choose one representative bottom-row
connectivity. But making this choice prevent us of using any additional 
symmetry in the connectivity space, such as translations along the 
horizontal direction.

\item Neither the detach nor the join operators can increase the number of 
links of a given connectivity state. Therefore, if we order the connectivity 
states in a suitable way (i.e., by decreasing number of links), then 
$\mathsf{T}_L$ takes a block-triangular form. Therefore, its eigenvalues 
are those of the corresponding diagonal sub-blocks $\mathsf{T}_{L,\ell}$.  

\item Given the connectivity space generated by a certain bottom-row 
connectivity and with $\ell$ links, these objects cannot cross, but move 
cyclically around the torus. Therefore the relevant symmetry group for the
links is the cyclic group $C_\ell$. In some sense, toroidal boundary 
conditions appear to be in the middle of two well-known extreme cases. On one 
side, we have planar strip graphs with cyclic boundary conditions. In this 
case, links cannot be permuted at all, and the relevant group is the identity 
$E$ \cite{Richard06}. On the other hand,
for non-planar recursive graphs with periodic longitudinal boundary
conditions \cite{HR05,Salas13-1}, links can be permuted freely, so the 
relevant group is the full symmetric group $S_\ell$. 

The representation theory for the cyclic group $C_\ell$ \cite{Richard07} 
implies that the amplitudes associated to each eigenvalue of the block 
$\mathsf{T}_{L,\ell}$ are certain polynomials in $Q$. In particular, 
these amplitudes are the same as for the chromatic-polynomial case 
\cite{JacobsenSalas07}. 
\end{itemize}

The final goal is to write the Potts-model partition function 
\eqref{def.Z_FK}/\eqref{eq_Z_vs_TM} on an $L\times N$ triangular-lattice strip
graph as a finite sum
\begin{subequations}
\label{def_Z_vs_eigen}
\begin{align}
Z_{L_P \times N_P}(Q,v) &\;=\; \sum\limits_{\ell=0}^L 
     \sum\limits_{j=1}^{d(L,\ell)} \beta^{(\ell)}_j(Q) \,
     \lambda_{\ell,j}(Q,v)^N
\label{def_Z_vs_eigen1} \\
                        &\;=\; \sum\limits_{j=1}^{N_L} \alpha_j(Q) \, 
         \lambda_j(Q,v)^N \,.
\label{def_Z_vs_eigen2}
\end{align}
\end{subequations}
In \eqref{def_Z_vs_eigen1}, $d(L,\ell)$ is the number of distinct 
non-zero eigenvalues $\lambda_{\ell,j}$ of the TM block $\mathsf{T}_{L,\ell}$ 
with the corresponding non-zero amplitudes $\beta^{(\ell)}_j$ given by the 
representation theory of the group $C_\ell$. Notice that there is no reason
why a given eigenvalue cannot appear several times within the same 
block $\mathsf{T}_{L,\ell}$ and/or appear in several blocks simultaneously: 
e.g., in $\mathsf{T}_{L,\ell}$ and $\mathsf{T}_{L,\ell'}$ with $\ell\neq\ell'$. 
Then, it is interesting to rewrite the partition function as  
a `complete' decomposition \eqref{def_Z_vs_eigen2} in terms
of the $N_L$ distinct eigenvalues $\lambda_j$ appearing in the full TM 
$\mathsf{T}_L$ with amplitude $\alpha_j$. These amplitudes should be linear 
combinations with integer coefficients of the former amplitudes
$\beta^{(\ell)}_j$. This form, although not necessary for computing the 
partition function $Z_{L_P \times N_P}$ and its zeros, it very useful to 
compute the accumulation points of these zeros as $N\to\infty$ (e.g., by
verifying that the hypotheses of the BKW theorem \cite{BKW1,BKW2,BKW3} are
met).

The number of connectivity states grows exponentially fast with
the width of the strip $L$, so the actual computation of the partition 
function in the form \eqref{def_Z_vs_eigen2} could only be achieved for a 
small number of values of $L$. However, we could extend our numerical results
to larger values of $L$ by using other techniques. These will be explained
in section~\ref{sec:results_TM}.   

%
%
\subsection{Critical polynomials}
\label{sec:tools_CP}

\begin{figure}
\begin{center}
\begin{tikzpicture}[scale=1.0,>=stealth]
\foreach \xpos in {0,1,2,3}
\foreach \ypos in {0,1,2,3}
 \fill[black!20] (\xpos+0.5,\ypos) -- (\xpos+1,\ypos+0.5) -- 
        (\xpos+0.5,\ypos+1) -- (\xpos,\ypos+0.5) -- cycle;
\foreach \xpos in {0,1,2,3}
\foreach \ypos in {0,1,2,3}
 \draw[blue,ultra thick] (\xpos+0.5,\ypos) -- (\xpos+1,\ypos+0.5) -- 
      (\xpos+0.5,\ypos+1) -- (\xpos,\ypos+0.5) -- cycle;
\foreach \ypos in {0,1,2,3}
{
 \draw[blue,ultra thick] (0,\ypos) -- (4,\ypos);
 \draw[blue,ultra thick] (0,\ypos+0.5) -- (4,\ypos+0.5);
}
\foreach \xpos in {0,1,2,3}
\foreach \ypos in {0,1,2,3}
 \draw[black] (\xpos+0.5,\ypos) -- (\xpos+1,\ypos+0.5) -- 
              (\xpos+0.5,\ypos+1) -- (\xpos,\ypos+0.5) -- cycle;
\draw[very thick,->] (0,-0.5)--(4,-0.5);
\draw (4,-0.5) node[right] {$x$};
\foreach \xpos in {0,1,2,3}
{
 \draw[thick] (\xpos+0.5,-0.6)--(\xpos+0.5,-0.4);
 \draw (\xpos+0.5,-0.5) node[below] {$\xpos$};
}
\draw[very thick,->] (-0.5,0)--(-0.5,4);
\draw (-0.5,4) node[above] {$y$};
\foreach \ypos in {0,1,2,3}
{
 \draw[thick] (-0.6,\ypos+0.5)--(-0.4,\ypos+0.5);
 \draw (-0.5,\ypos+0.5) node[left] {$\ypos$};
}
\end{tikzpicture}
 \caption{$4 \times 4$ basis used to compute the CP on the triangular lattice.}
 \label{fig:triangular}
\end{center}
\end{figure}
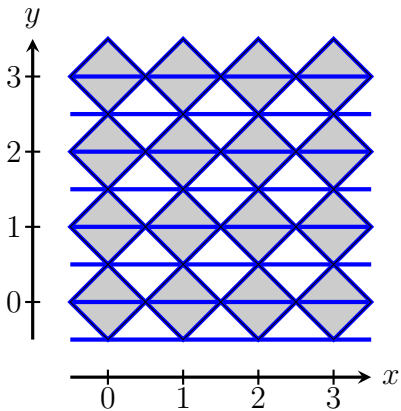

We now discuss a completely different technique for determining the phase 
diagram: the method of CP. There is now a reasonably 
large body of work on this 
\cite{JacobsenScullard12,SJ12,JacobsenScullard13,Jacobsen14,Jacobsen15,JS16},
so we provide only a brief overview here and refer the reader to the literature
for details and justification of the method. The first step is to choose a 
finite graph, which we call the basis, that can be copied and translated to 
cover the entire lattice. In this case, we choose graphs similar to those 
we used for the TM computation, namely squares of size 
$L \times L$. 
The $4 \times 4$ basis is shown in figure~\ref{fig:triangular}. The key idea 
is to compute the weights of two events, $P_{2D}(Q,v)$ and $P_{0D}(Q,v)$, 
in the FK representation. If we view the basis as embedded in 
a torus, then $P_{2D}(Q,v)$ is the weight that a cluster wraps both periodic 
directions, and $P_{0D}(Q,v)$ is the event that there is no wrapping cluster 
of any kind. The CP for the basis $P_B(Q,v)$ is defined to be
\begin{equation}
 P_B(Q,v) \;\equiv\; Q P_{0D}(Q,v)-P_{2D}(Q,v) \label{eq:2D0D}
\end{equation}
and its zeros provide predictions of critical points for arbitrary $Q$. 
These predictions are exact for all solved Potts critical thresholds and 
for unsolved problems they provide accurate estimates that become increasingly 
so for larger and larger bases. Using extrapolation to infinite bases, we 
now have estimates for critical points that far outstrip what can be obtained 
with Monte Carlo techniques \cite{Jacobsen15}. That $P_B(Q,v)$ is a 
polynomial of maximum degree $6 L^2$ follows from the fact that the basis 
is a finite graph with $6L^2$ edges.

The way one computes the CP is not important, and a variety 
of methods have been used in the past 
\cite{ScullardZiff08,Scullard11,Scullard11-2,Scullard12,JacobsenScullard12}. 
But to date, the most efficient technique by far, allowing computation on the 
largest bases, is to use a TM to find $P_{2D}(Q,v)$ and 
$P_{0D}(Q,v)$. Recently, one of the authors \cite{Jacobsen15} was able to 
recast the calculation as an eigenvalue problem, allowing the computation 
of individual CP roots on semi-infinite lattice strips. 
However, 
to get a view of the phase diagram, one really needs to construct the full 
polynomial $P_B(Q,v)$. Polynomials for all the Archimedean lattices have 
previously been computed \cite{Jacobsen14} for bases of size up to 
$5 \times 5$. This was recently extended to $6 \times 6$, using calculations 
on a supercomputer \cite{JS16}, for all except the triangular lattice, which 
is done in the present work. In \cite{JS16}, it is described how the algorithm 
in \cite{Jacobsen14} is parallelised, so the reader is referred to those 
references for the details. 

We ran on Lawrence Livermore National Laboratory's Cab supercomputer which 
has $20,736$ total cores, each 2.6 GHz with 2GB of RAM. To construct 
the entire polynomial, one must do the transfer matrix calculation for a 
variety of $Q$, and our parallel scheme breaks up the problem so that each 
processor handles a piece of the vector of connectivity states and a range 
of $Q$. The maximum size of the vector is $26,332,020$ states and we use 
$144$ $Q$-values, since the degree in $Q$ of $P_B(Q,v)$ is $4L^2$, the number 
of vertices in the basis. Our layout of processors can be pictured as an 
array of $341 \times 12 = 4092$ so that each has $77,220$ states and deals 
with $12$ values of $Q$. The coefficients of the polynomial are very long 
integers and to construct them we use the Chinese remainder theorem. 
This means the computation must be done several times modulo a set of prime 
numbers and in this case we needed 14 primes to get the solution. Each of 
these calculations required about 7 hours with 4092 processors.

%
%

%
%
\section{Numerical results}
\label{sec:results}

%
%
\subsection{Transfer-matrix results}
\label{sec:results_TM}

%
%
\def\kk{\phantom{${}^\dagger$}}
\begin{table}
\centering
\begin{tabular}{rrrrrrrr}
\hline\hline
$L$ &  $d(L,0)$ & $d(L,1)$ & $d(L,2)$ & $d(L,3)$ & $d(L,4)$ & $d(L,5)$ &$N_L$\\
\hline 
2   &       2 &    2\kk &     1\kk &    0\kk &      0 &  0 &    5\kk \\
3   &       5 &   10\kk &    12\kk &    3\kk &      0 &  0 &   30\kk \\
4   &      15 &   33\kk &    55\kk &   25\kk &      3 &  0 &  123\kk \\ 
\hline
5   &      52 &  151${}^\dagger$&   333${}^\dagger$&    
                 178${}^\dagger$&     43 &  5 &  586${}^\dagger$ \\ 
\hline\hline  
\end{tabular}
\caption{Distinct non-zero eigenvalues of the TM associated with a 
triangular-lattice
strip graph of width $L$ and toroidal boundary conditions.   
For each strip graph width $L=2,3,4$, we give the number of distinct 
non-zero eigenvalues $d(L,\ell)$ of the diagonal block $\mathsf{T}_{L,\ell}$.
The number $N_L$ counts the total number of distinct non-zero
eigenvalues of the full TM $\mathsf{T}_L$. This number might be smaller than
the sum $\sum\limits_{\ell=0}^L d(L,\ell)$, as some eigenvalues can appear
in more than one block $\mathsf{T}_{L,\ell}$. For $L=5$, the values with the 
sign ${}^\dagger$ are conjectures based on partial factorisation of the 
corresponding characteristic polynomials and numerical test (see text). 
}
\label{table:d_ell}
\end{table}

In this section we will describe the methods we used within the 
TM  approach to obtain the accumulation points of 
partition-function zeros (i.e., the limiting curves 
$\mathcal{B}_L$ to be defined below), and extract our 
estimates for the AF critical curve $v_\text{AF}(Q)$.   

For the smallest widths $L=2,3,4$ we were able to symbolically compute the
TM blocks $\mathsf{T}_{L,\ell}$ for $0\le \ell \le L$, extract all the
distinct non-zero eigenvalues $\lambda_j$, and compute the corresponding
amplitudes $\alpha_i(Q)$. The practical procedure we used can be summarised 
as follows: Fix a value of $L$, then 
\begin{enumerate}
\item For each value of $\ell$, compute the corresponding diagonal block 
      $\mathsf{T}_{L,\ell}$ of the full TM given a \emph{fixed} bottom-row 
      connectivity (compatible with the value of $\ell$) and use all possible 
      top-row connectivity states.  

\item For each value of $\ell$, compute the characteristic polynomial 
      $\chi_{L,\ell}$ of $\mathsf{T}_{L,\ell}$, and factor it over the 
      integers. Each distinct factor of $\chi_{L,\ell}$ gives distinct 
      eigenvalues $\lambda_{\ell,j}$ [cf.\ \eqref{def_Z_vs_eigen1}]. 
      We drop for simplicity those eigenvalues
      that are identically equal to zero (e.g., for $L=2$ there is one
      zero eigenvalue for $\ell=1$ and another one for $\ell=2$). The total 
      number of distinct non-zero eigenvalues is the quantity $d(L,\ell)$ 
      shown in table~\ref{table:d_ell}. These eigenvalues are
      algebraic functions of $Q$ and $v$.   

\item Once all the non-zero distinct eigenvalues are computed for all
      values of $\ell$, we only take into account those $\lambda_j$
      that are distinct [cf.\ \eqref{def_Z_vs_eigen2}], as we want a 
      `complete' decomposition of the partition function. We finish 
      this part with a list of $N_L$ distinct non-zero eigenvalues 
      $\lambda_j$. (See table~\ref{table:d_ell} for the values of $N_L$.)  

\item We know that the amplitudes $\alpha_j$ are linear combinations of
      the amplitudes $\beta^{(\ell)}_j$ with integer coefficients, and 
      these latter ones are polynomials of degree $\ell$ in $Q$ 
      \cite{JacobsenSalas07,Richard07}. 
      In~\ref{app.coef} we show which amplitudes do appear in the 
      partition function \eqref{def_Z_vs_eigen}, and in the particular
      case of the chromatic-polynomial ($v=-1$). 
      Therefore we expect that each 
      eigenvalue $\lambda_j$ is associated with an amplitude $\alpha_j$ which
      is a polynomial in $Q$ whose degree is the maximum value of $\ell$
      with which it appears in the spectrum of $\mathsf{T}_{L,\ell}$. 
      Indeed, all 
      eigenvalues coming from the same algebraic equation should have the 
      same amplitude. These amplitudes are fixed by computing the first 
      partition-function polynomials $Z_{L_P \times N_P}(Q,v)$ for small
      values of $N\ge 1$, and solving the corresponding linear equations.  
\end{enumerate}  
This procedure has been carried out symbolically using {\sc Mathematica}
for $L=2,3,4$. We have made several cross-checks: 
a) $Z_{L_P \times N_P} = Z_{N_P \times L_P}$ for any values of $L,N$; and
b) $Z_{L_P \times N_P}(Q,-1)=P_{L_P \times N_P}(Q)$, where these
chromatic polynomials were already computed in \cite{JacobsenSalas07}. 

For $L=5$ this procedure could only be performed successfully in part:
for $\ell=1,2,3$ we could find a block structure of the corresponding 
blocks $\mathsf{T}_{L,\ell}$; but we were unable to symbolically compute
the characteristic polynomials $\chi_{L,\ell}$ of the largest sub-block for 
each of these values of $\ell$. (The dimension of these sub-blocks
are $126,240,135$ for $\ell=1,2,3$, respectively.) To ensure that these
blocks have non-zero distinct eigenvalues, and that these eigenvalues only
appeared in the corresponding blocks, we numerically evaluate all the 
eigenvalues for several pairs of $(Q,v)$, and concluded that our conjecture
was right. In table~\ref{table:d_ell}, these numbers are marked with the
superscript ${}^\dagger$. In this case, we could not obtain the final
form for the partition function \eqref{def_Z_vs_eigen2}; but we at least 
computed all the relevant eigenvalues of $\mathsf{T}_5$, and therefore, the
corresponding limiting curve $\mathcal{B}_5$. 

For $6\le L\le 9$, we obtained numerical results for the limiting curve 
$\mathcal{B}_L$ using a {\sc C} code. We checked this program by computing 
these curves for $2\le L\le 5$, and the agreement with the results obtained
from the symbolic TM was excellent.  

\begin{figure}
  \centering
  \begin{tabular}{cc}
  \includegraphics[width=200pt]{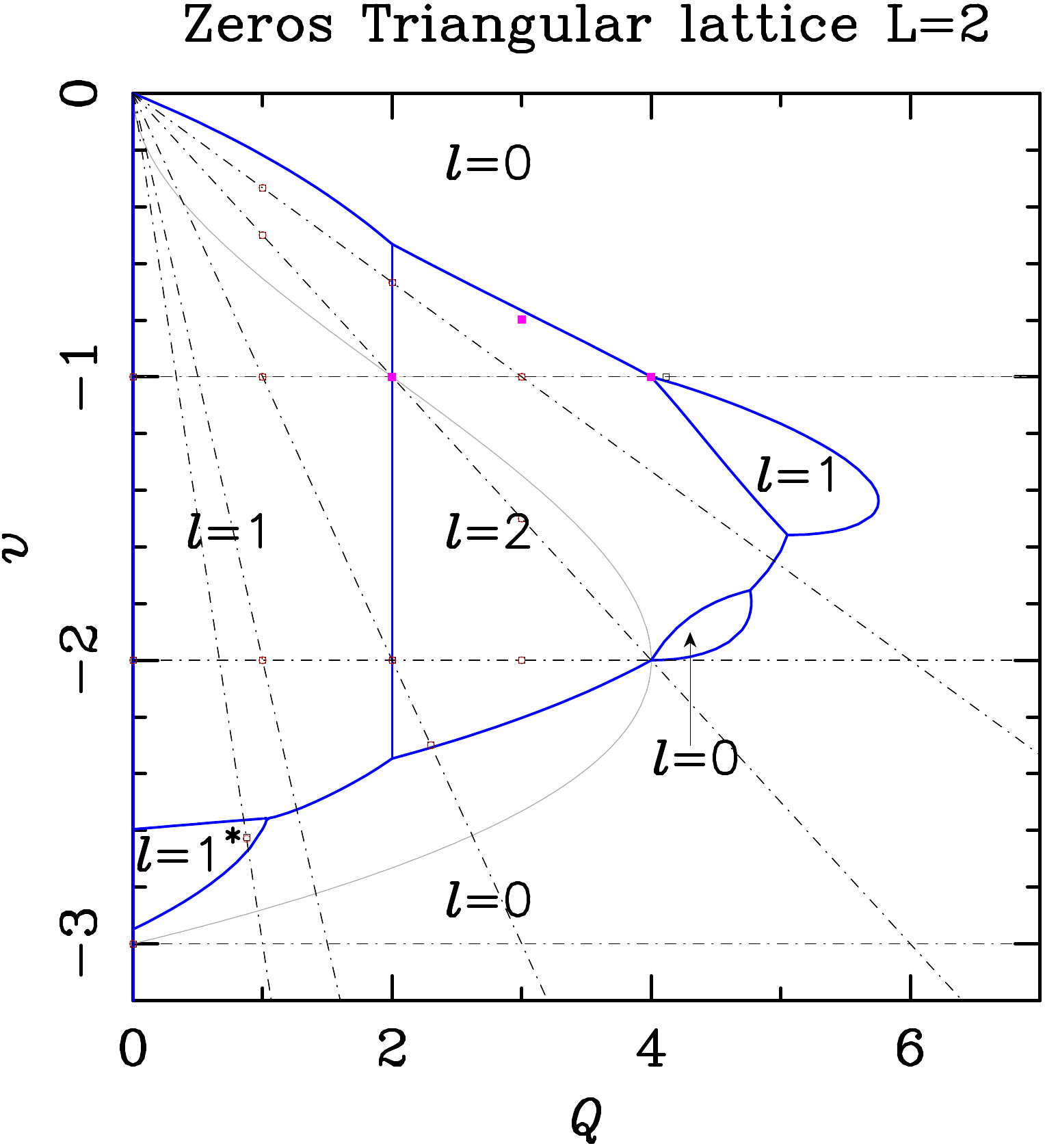} &
  \includegraphics[width=200pt]{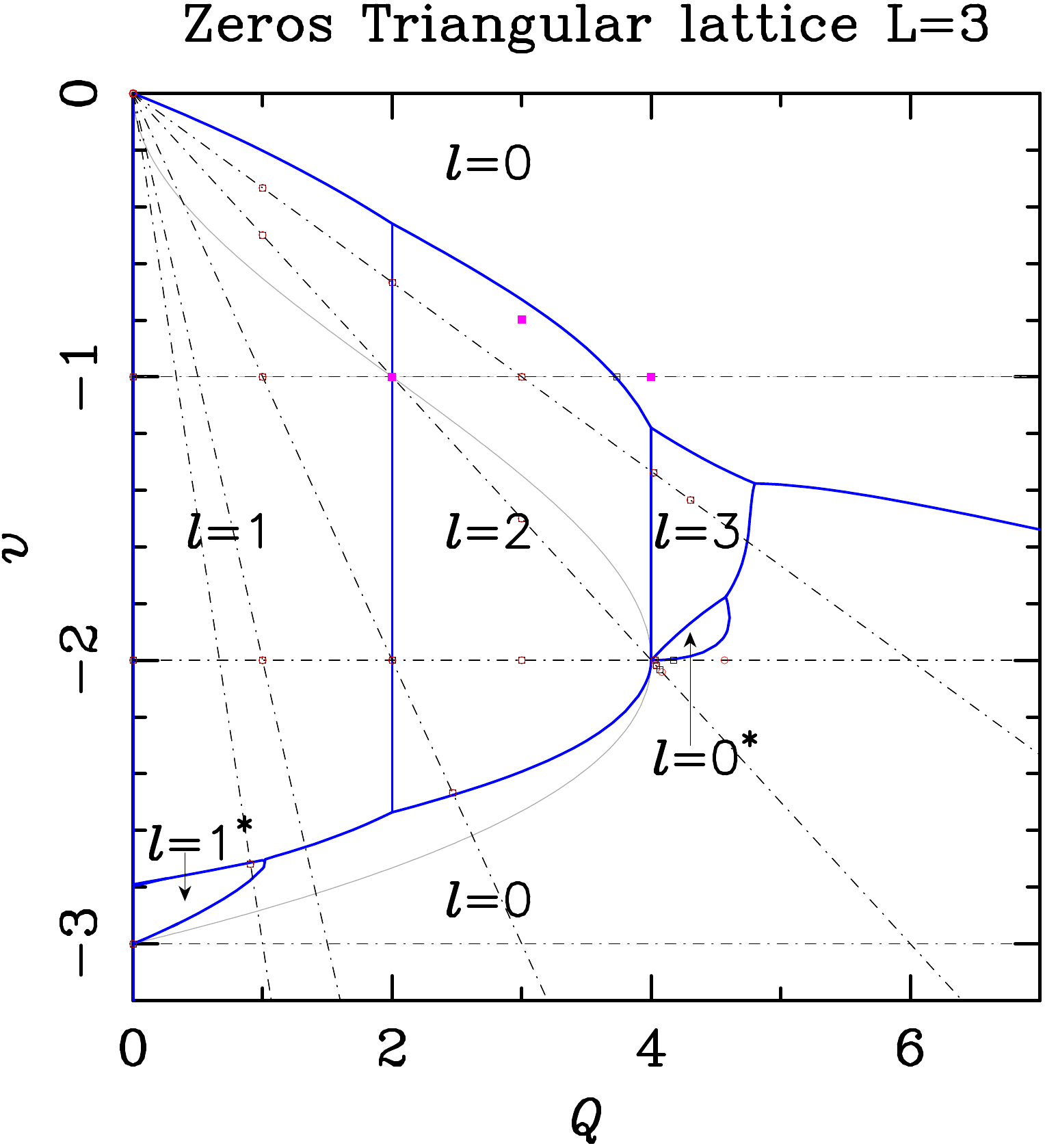} \\
  \qquad (a) & \qquad (b) \\[2mm]
  \includegraphics[width=200pt]{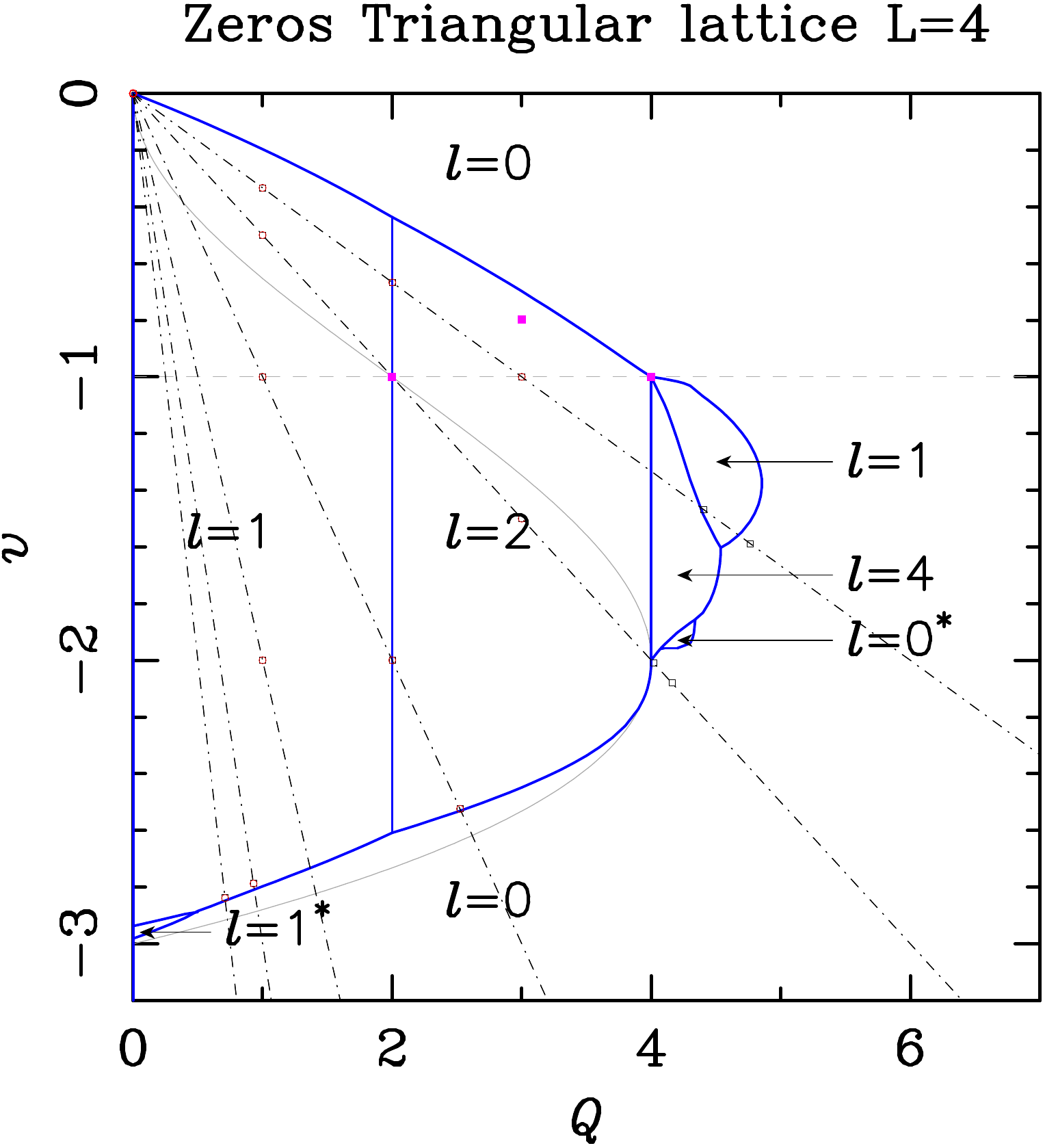} &
  \includegraphics[width=200pt]{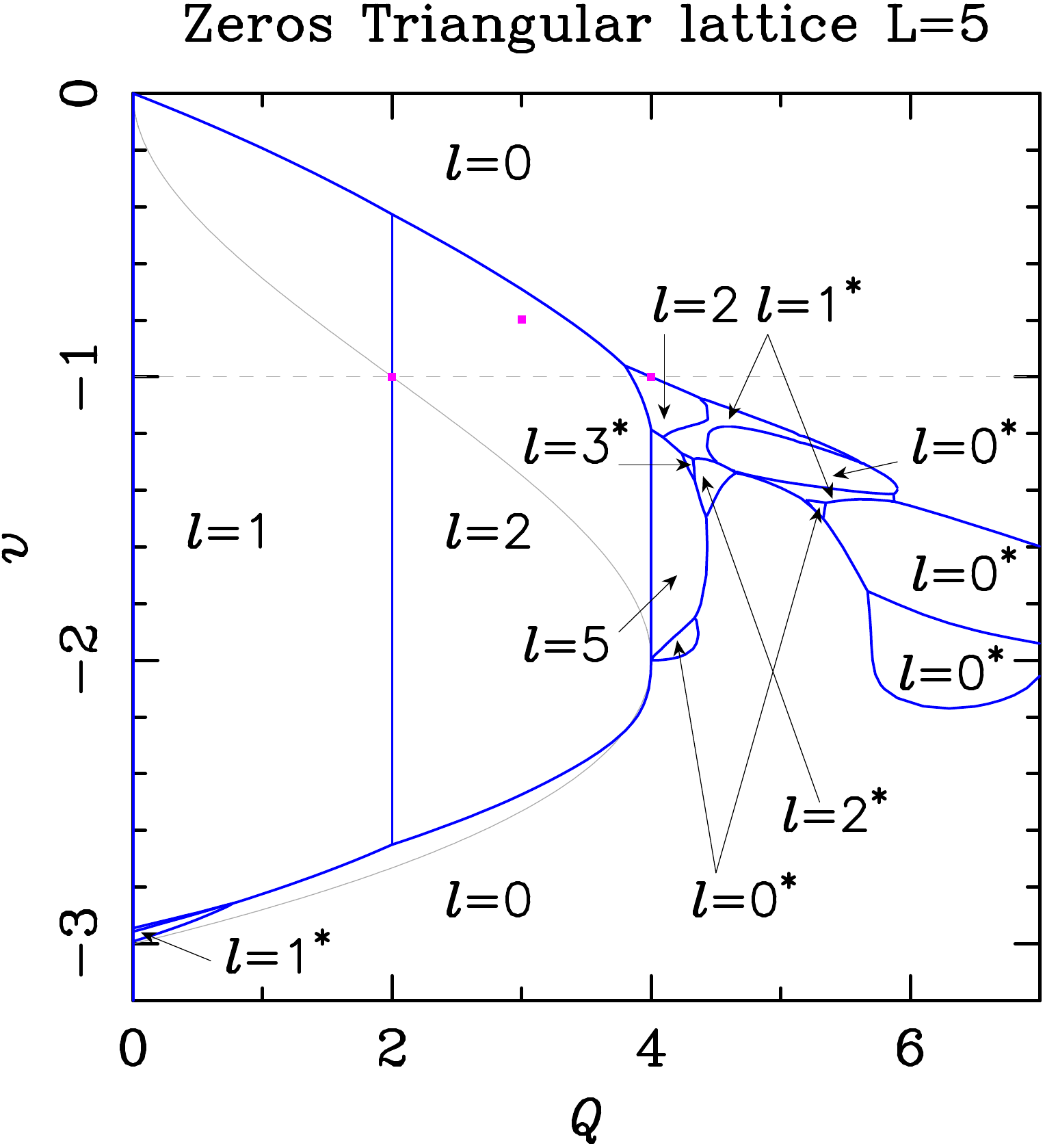} \\
  \qquad (c) & \qquad (d)
  \end{tabular}
  \caption{
  Limiting curves $\mathcal{B}_L$ (solid blue curves) in the real 
  $(Q,v)$-plane for 
  the $Q$-state Potts model on triangular-lattice strip graphs embedded in 
  tori of widths $L=2$ (a), $L=3$ (b), $L=4$ (c), and $L=5$ (d). 
  The solid grey curve corresponds to the two lower branches of the 
  cubic \eqref{cubic_tri}.   
  We also show, for each width, the corresponding real zeros for finite 
  strips of length $N=19$ (black $\square$) and  $N=20$ 
  (red $\color{red} \circ$)
  along several lines (depicted as dot-dashed lines) of the type $Q=-p v$ 
  ($p=1,2,3$), $v=-p Q$ ($p=1,2,3$), and $v=-1,-2,-3,-4$.
  The pink full squares ($\color{magenta} \blacksquare$) show the known 
  critical values for $Q=2,3,4$.
  Each region is labeled with the sector to which the dominant eigenvalue 
  belongs  
  (e.g., $\ell = 1$). An asterisk in the sector label (e.g., $\ell = 1^*$) 
  means that there is a pair of complex-conjugate dominant eigenvalues 
  in that region.
  }
\label{Figure_tri_TM}
\end{figure}

As explained above, for $2\le L\le 4$, we could compute the exact symbolic 
form for the partition function $Z_{L_P \times N_P}$ \eqref{def_Z_vs_eigen2}.
As explained in section~\ref{sec:potts}, we look for the zeros of this
partition function in the real $(Q,v)$-plane for $Q \ge 0$ and $v\le 0$. 
In order to achieve 
this goal, we computed the zeros along several lines in this quadrant:  
$Q=-p v$ with $p=1,2,3$ (the case $p=1$ essentially corresponds to the 
flow polynomial \cite{JacobsenSalas13}), $v=-p Q$ with $p=1,2,3$, 
and $v=-1,-2,-3,-4$ (the case $v=-1$ is the chromatic polynomial 
\cite{JacobsenSalas07}). We have represented in figure~\ref{Figure_tri_TM},
for each value of $L$, these zeros for $N=19$ (black squares) and $N=20$ 
(red open circles). Unfortunately, there are not many zeros along any of
these lines.  
 
%
%

So far, we have dealt with finite strips, so our results suffer from large
finite-size-scaling corrections. Therefore, we should study the thermodynamic 
limit. As shown in previous papers 
\cite{SalasSokal01,JacobsenSalas01,ChangSalasShrock02,JacobsenSalasSokal03,%
ChangJacobsenSalasShrock04,JSS05,JacobsenSalas06,JacobsenSalas07,transfer5,%
transfer6,JacobsenSalas13,Salas13-1}, 
a very efficient way to take the thermodynamic limit is the following. We 
first take the limit $N\to\infty$, so we have semi-infinite strips with 
an infinite number of vertices. Then according to the BKW theorem 
\cite{BKW1,BKW2,BKW3} 
(see also \cite{Sokal_04} for a slight generalisation),
the partition-function zeros accumulate around two
types of accumulation points:
\begin{enumerate}
 \item Isolated limiting points, when there is a single dominant eigenvalue 
       $\lambda_\star$ (with $|\lambda_\star|>|\lambda_i|$ for all 
       $i\neq \star$), but its amplitude vanishes identically, 
       $\alpha_\star=0$. 
 \item Limiting curves $\mathcal{B}_L$ formed by points where two or more
       eigenvalues are dominant in modulus $|\lambda_1|=|\lambda_2|=\ldots$. 
       Notice that if all the amplitudes $\alpha_k$ vanish,
        then we obtain isolated limiting points \cite{Sokal_04}.
\end{enumerate}  
By taking this first limit $N\to\infty$, we can compute both the isolated 
limiting points and the limiting curves $\mathcal{B}_L$. If the  
free energy $f_{L_P \times N_P}$ [cf. \eqref{eq.free_energy}] for a finite
strip graph of size $L\times N$ is given by 
\begin{equation}
f_{L_P \times N_P}(Q,v) \;=\; \frac{1}{L\, N} \, \log Z_{L_P \times N_P}(Q,v)
\,,
\label{eq.free_energy_finite}
\end{equation}
the corresponding limiting free energy $f_L(Q,v)$ reads
\begin{equation}
f_{L}(Q,v) \;=\; \lim_{N\to \infty} f_{L_P \times N_P}(Q,v) \;=\; 
                 \frac{1}{L} \, \log |\lambda_\star(Q,v)| \,.
\label{eq.free_energy_L}
\end{equation}
Therefore, for each value of $L$ we have a `phase diagram' given by
the limiting curves $\mathcal{B}_L$. We empirically know that when the strip
graph has periodic longitudinal boundary conditions (as in this case), the
corresponding limiting curves bound closed regions 
\cite{JacobsenSalas06,JacobsenSalas07} (see figure~\ref{Figure_tri_TM}).
Each of these regions is characterised by the sector to which the 
leading eigenvalue $\lambda_\star$ belongs. In the chromatic-polynomial case 
\cite{JacobsenSalas06,JacobsenSalas07} the dominant eigenvalue within each
region was unique; but here we find that some (small) regions are
characterised by a pair of complex-conjugate eigenvalues. These regions are
denoted by a sector label with an asterisk in figure~\ref{Figure_tri_TM}.  

By inspecting the panels of figure~\ref{Figure_tri_TM} we arrive at the
following empirical conclusions:
\begin{itemize}
  \item The `phase diagram' has two parts. The `regular' one corresponds
        to the interval $Q\in[0,4]$, and the relevant sectors are $\ell=0,1,2$.
        The other piece of the phase diagram corresponds to $Q>4$, and it 
        contains curves that extend to infinity, and very involved features.

  \item In the regular part, there is an outer phase characterised by $\ell=0$,
        and generically two other phases bounded by two curves. 
        One corresponds to 
        $Q\in[0,2)$ where $\ell=1$ is dominant, and the other one corresponds
        to $Q\in (2,4)$, where $\ell=2$ is dominant (for $L\ge 3$). 
        The upper boundary curve $v_{+,L}$ can be interpreted as a finite-size
        approximation to the AF critical curve $v_\text{AF}$, while the 
        lower one $v_{-,L}$ approaches the lower branch 
        of Baxter's cubic \eqref{cubic_tri} (as one would expect). 
        Notice that for $L=5$ there is a third phase close
        to $Q=4$, where $\ell=2$ is also dominant. This region can be 
        interpreted as a finite-size approximation to Regime~IV. $L=5$ is the
        smallest width for which this phase can be directly observed.

  \item In this regular part, we find three perfectly vertical lines at 
        $Q=0$, $Q=2$ and $Q=4$. These lines are limiting curves. While the
        lines at $Q=0,2$ cover all the points between $v_{\pm,L}$,  
        the one at $Q=4$ does not in general:  
        this line goes from $v=-2$ to a value $v_1(L) < -1$. 
        See table~\ref{table.v1} for our estimates for $2 \le L \le 12$.

  \item The curve $v_{+,L}$ contains a T-point $(Q_T(L),v_T(L))$ for 
        $L=5,7,8$. In these cases, one branch goes from the T-point to
        $(4,-1)$, and the other one goes to $(4,v_1(L))$. This latter branch 
        crosses the chromatic line $v=-1$ at $Q=Q_2(L)$ (see 
        \cite[table~7]{JacobsenSalas07} for their numerical values).  
        The values of $Q_T(L)$ and $v_T(L)$ are displayed in 
        table~\ref{table.v1}. 
        The curve $v_{+,L}$ for $L=2,3,4,6,9,12$ does not show such
        T-point; for $L=2,4$ it goes through the point $(4,-1)$; but for
        $L=3,6,9,12$ it crosses the line $v=-1$ at a value $Q=Q_0(L)$. The 
        values of $Q_0(L)$ are also displayed in 
        \cite[table~7]{JacobsenSalas07}. From these results and those found
        on the chromatic line \cite{JacobsenSalas07}, we can conjecture that 
        the curve $v_{+,L}$ contains a T-point and goes through the point 
        $(4,-1)$ for any $L$ not multiple of 3.

  \item The lines $Q=1$ and $Q=3$ can be considered \emph{lines} of 
        isolated limiting points, as the dominant eigenvalues in each of 
        the corresponding regions have amplitudes vanishing at these values. 
        Although this seems to be some sort of contradiction, every point 
        $(Q,v)$ with $v_{-}(Q) < v < v_{+}(Q)$ and $Q=1,3$ satisfies 
        the conditions of the BKW theorem, and the convergence of the zeros
        of $P_{L\times N}(Q)$ is exponentially fast in $N$, rather than 
        algebraically fast as for the ordinary limiting points.  

  \item In the non-regular part of the phase diagram, we find a parity effect:
        for odd $L$, there are outward lines going to infinity, while for
        even $L$, this part is compact. We also find that for $L\ge 3$ the 
        sector dominant to the right of the vertical line at $Q=4$ (for
        $-2 < v < v_1(L)$) is $\ell=L$. Furthermore, we find many 
        regions with dominant complex-conjugate pairs of eigenvalues.  
\end{itemize}

%
%
\begin{table}[h]
\centering
\begin{tabular}{rlll}
\hline\hline\\[-2mm]
\multicolumn{1}{c}{$L$} & \multicolumn{1}{c}{$v_1(L)$} & 
                          \multicolumn{1}{c}{$Q_T(L)$} & 
                          \multicolumn{1}{c}{$v_T(L)$} \\ 
\hline \\
2  & $-2$             &                &     \\
3  & $-1.1798383409$  &                &     \\
4  & $-1$             &                &     \\
5  & $-1.1862100032$  & $3.7967440229$ & $-0.9606237874$\\
6  & $-1.245919843$   &                &     \\
7  & $-1.19085474895$ & $3.826(4)$     & $-0.971(2)$  \\
8  & $-1.25306265476$ & $3.772(4)$     & $-0.9541(9)$ \\
9  & $-1.27354$       &                &     \\
12 & $-1.28069$       &                &     \\ 
\hline\hline 
\end{tabular}
\caption{%
  Estimates of the values $v_1(L)$, $Q_T(L)$, and $v_T(L)$ as a function
  of the width of the triangular-lattice strip graph. The precision of each
  non-integer entry is of order $10^{-n}$, where $n$ is the position of the 
  last quoted digit, unless a larger error is explicitly indicated. 
}
\label{table.v1}
\end{table}

%
%
\begin{figure}
  \centering
  \includegraphics[width=200pt]{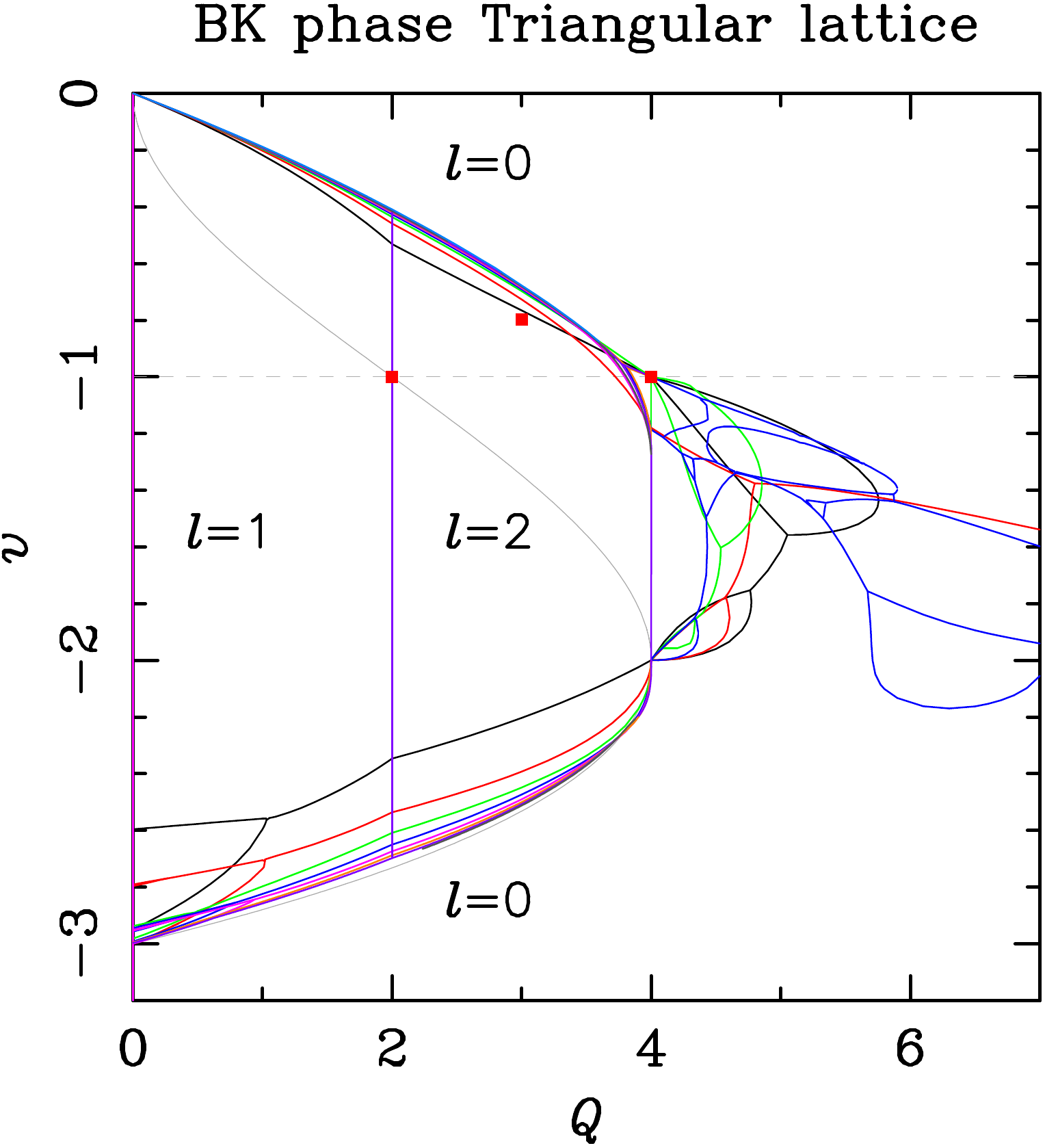} 
  \caption{
  Limiting curves from figure~\ref{Figure_tri_TM} plotted
  in the same figure. These curves $\mathcal{B}_L$ are depicted as
  solid lines with colour code $L=2$ (black), $L=3$ (red), $L=4$ (green),
  $L=5$ (navy blue), $L=6$ (pink), $L=7$ (orange), $L=8$ (violet),
  and $L=9$ (dark grey).
  The upper blue curve in the range $Q\in [0,3.6]$ corresponds to the
  extrapolation of the TM limiting curves.
  The grey solid curve corresponds to the cubic \eqref{cubic_tri}, and 
  the dashed horizontal line, to the chromatic-polynomial subspace ($v=-1$).
  In each region, we show the dominant link sector. The solid red squares 
  ($\color{red} \blacksquare$) show the known critical points for integer 
  values of $Q=2,3,4$.  
  }
\label{Figure_tri_TM_all}
\end{figure}

%
%
\begin{figure}
  \centering
  \begin{tabular}{cc}
  \includegraphics[width=200pt]{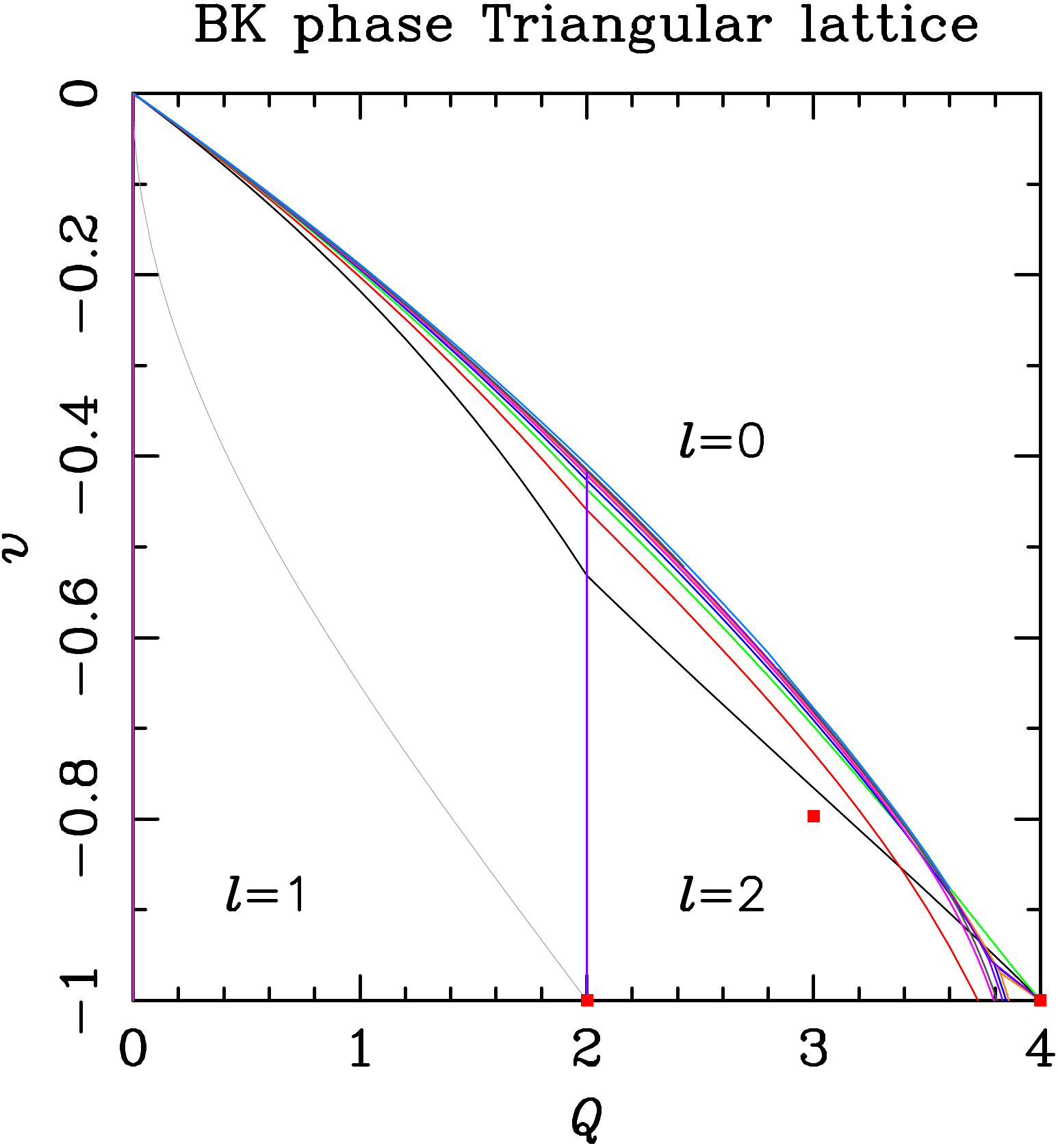} &
  \includegraphics[width=200pt]{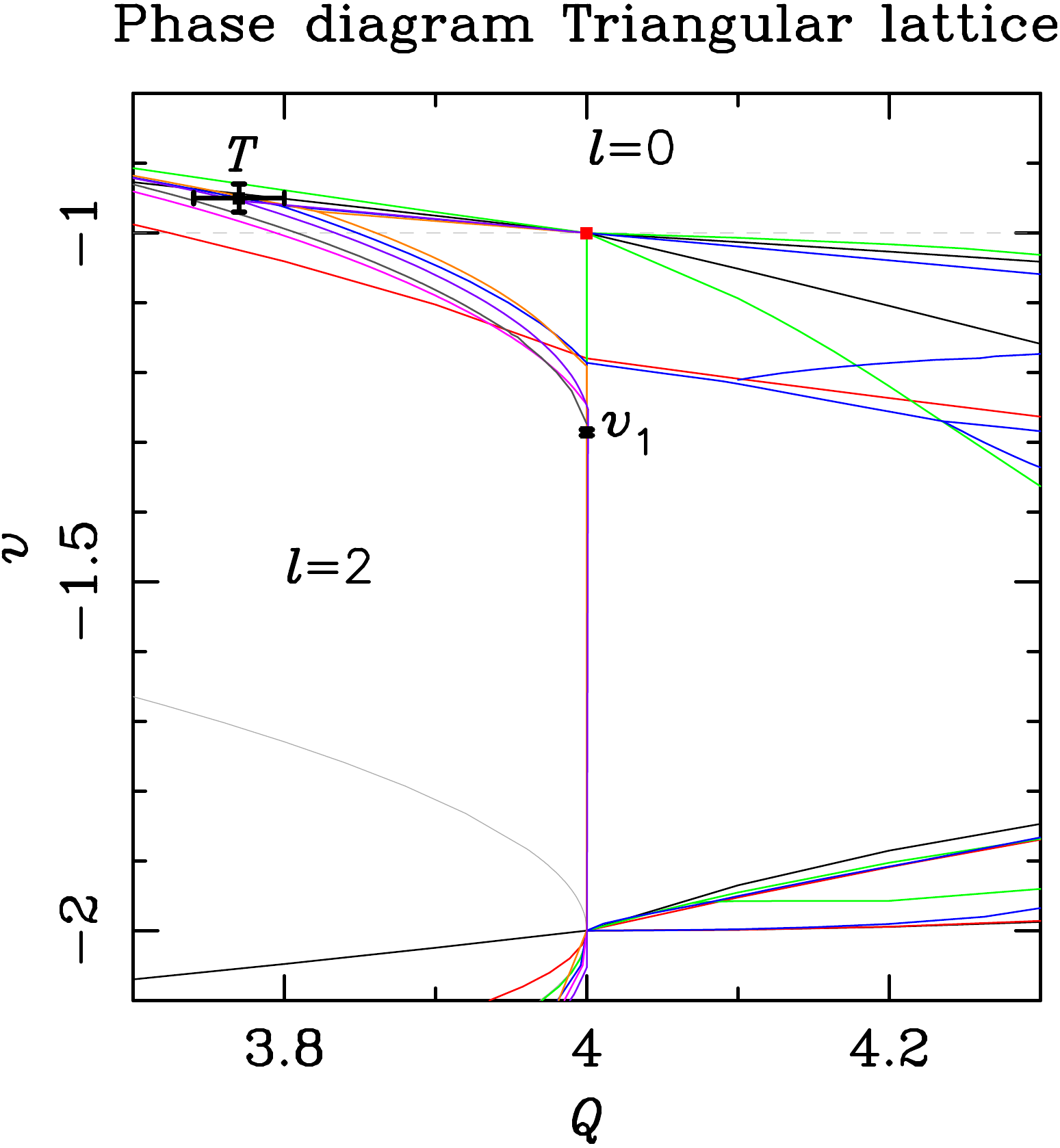} \\
  \qquad (a) & \qquad (b) \\
  \end{tabular}
  \caption{
  (a) Zoom of figure~\ref{Figure_tri_TM_all} showing the different 
      approximants to the antiferromagnetic critical curve $v_\text{AF}$.
  (b) Zoom of figure~\ref{Figure_tri_TM_all} around the line $Q=4$. 
      The point labeled $T$ is
      our best estimate for the T-point $(Q_T,v_T)$; and the point labeled
      $v_1$ is our best estimate for $v_1$. These points are represented by 
      black squares ($\blacksquare$). 
  In both panels we use the same symbols and colour codes as in 
  figure~\ref{Figure_tri_TM_all}.
  }
\label{Figure_tri_TM_all_zoom}
\end{figure}

The final goal is to extract information about the true thermodynamic limit 
from the above finite-$L$ data. We then take the limit $L\to \infty$, so that
the infinite-volume free energy becomes:
\begin{equation}
f(Q,v) \;=\; \lim_{L\to \infty} f_L(Q,v) \,.  
\label{eq.free_energy_infinity}
\end{equation}
In our case, we should try to extract patterns from the above finite-width 
data. It is interesting to plot together all our limiting curves, at least for
the `regular' part of the phase diagram 
(see figures~\ref{Figure_tri_TM_all} and~\ref{Figure_tri_TM_all_zoom}).
The non-regular part of the phase diagram is very involved, with many tiny
features, and, as observed above, its structure suffers from large 
finite-size corrections. Therefore, we refrain from making any further study
of this part of the phase diagram, and focus on its regular part.

In figure~\ref{Figure_tri_TM_all} we show all the limiting curves (or parts of 
them) for $2\le L\le 9$; they define the full phase diagram.
In figure~\ref{Figure_tri_TM_all_zoom}(a), we focus on the AF regime 
$v\in[-1,0]$ of figure~\ref{Figure_tri_TM_all}, where the physical part of the
AF critical curve is expected to be. It is clear from the figure that,
at least on the interval $Q\in[0,3]$, the finite-$L$ estimates for $v_{+,L}$ 
do not suffer from parity effects and they seem to converge rather 
quickly to a stable value, that we will identify as the AF critical point 
$v_\text{AF}$. For this reason, we have tried to fit the numerical data to a 
power-law Ansatz
\begin{equation}
v_{+,L} \;=\; v_\text{AF} + A \, L^{-\Delta} \,,
\label{def_power_law}
\end{equation} 
where the $Q$-dependence has been dropped for simplicity. The fits are 
rather stable, although the smaller the value of $Q$ the more stable the fit.
The estimate for the power increases slowly from $\Delta \approx 1.85$ for 
$Q=0.1$ to $\Delta \approx 2.76$ for $Q=3$. These estimates are depicted
in figure~\ref{Figure_tri_TM_all} as a solid blue curve. If we fit the
estimates for the smallest values of $Q$ to a polynomial Ansatz, we can
obtain a slightly more accurate estimate for the derivative of 
$v_\text{AF}$ at $Q=0$ [cf. \eqref{def_der_Q=0}]:
\begin{equation}
\left. \frac{{\rm d} v_\text{AF}}{{\rm d}Q} \right|_{Q=0} \;=\; 
-0.17526 \pm 0.00001\,.
\label{def_der_Q=0_fit}
\end{equation}

To extend the above results beyond $Q=3$ and to prevent systematic errors
due to parity effects, we have considered the data for $L=3,6,9,12$ and 
made a numerical fit to the same Ansatz \eqref{def_power_law}.
In this way, we have obtained a reasonably smooth curve 
$v_{+,L}$ up to $Q \lesssim 3.6$. This is shown
in figures~\ref{Figure_tri_TM_all} and~\ref{Figure_tri_TM_all_zoom}(a).  
Beyond the value $Q\gtrsim 3.6$, we should expect numerical difficulties
due to the existence of the T-point $T$, whose position is given below 
[cf. \eqref{def_T}].

\noindent
Our main conclusions are:

\begin{itemize}

\item In the region $Q\in[0,4]$ there is a disordered phase (belonging to the
sector $\ell=0$ when $v$ is close enough to $v=0$ ($=$ infinite temperature) or
$v$ is large enough in modulus and negative (belonging to the unphysical 
regime).

\item There are two boundary curves $v_{\pm,L}$ such that the upper one
$v_{+,L}$ (resp.\/ lower one $v_{-,L}$) converges to the AF critical curve
$v_\text{AF}$ (resp.\/ to the lower branch $v_{-}$ of the cubic 
\eqref{cubic_tri}).
The AF critical curve has a T-point denoted $T$ in 
figure~\ref{Figure_tri_TM_all_zoom}(b), and two branches: one of them, 
$v_\text{AF,1}$ goes to the point $(Q_c,v)=(4,-1)$, and the other one, 
$v_\text{AF,2}$, to $(Q_0,-1)$, with $Q_0$ given by \eqref{eq.Q0}, 
and continues to the point $(4,v_1)$, where  
\begin{equation}
v_1 \;=\; -1.286 \pm 0.004 \,.  
\label{def_v1}
\end{equation}
This T-point $T=(Q_T,v_T)$ appears for $L=5$ and $L=7,8$. Its approximate 
position is 
\begin{equation}
Q_T \;=\; 3.77 \pm 0.03 \,, \qquad v_T \;=\; -0.95 \pm 0.02 \,.
\label{def_T}
\end{equation} 
With our current data it is only possible to give rather
rough estimates for the error bars in \eqref{def_v1}/\eqref{def_T}, as these 
quantities show strong parity effects $\pmod{3}$. The above estimates for 
$T$ and $v_1$ are displayed in figure~\ref{Figure_tri_TM_all}(b).

\item There is a critical phase in the region $Q\in (0,2)$ and 
$v_\text{AF} < v < v_{-}$ characterised by the sector $\ell=1$. 
This phase has a line of isolated limiting points at $Q=1$. 

\item There is a critical phase in the region bounded by the curves
$v_\text{AF}$ for $Q\in(2,Q_T)$, $v_\text{AF,2}$ for $Q\in(Q_T,4)$, 
the line $Q=4$ for $v\in(-2,v_1)$, the curve $v_{-}$ for $Q\in(2,4)$,
and the line $Q=2$ for $v\in (v_{-}(2),v_\text{AF}(2))$. (See  
figure~\ref{Figure_tri_TM_all_zoom}(b) for a graphical representation.)
This phase is characterised by the sector with $\ell=2$, and contains a line
of isolated limiting points at $Q=3$. 

\item There is another critical phase (corresponding to the regime~IV
discussed in section~\ref{sec:setup_tri}), which is bounded by the 
point $(4,-1)$, the $T$ point \eqref{def_T}, the points $(Q_0,-1)$ 
[cf. \eqref{eq.Q0}], and the point $(4,v_1)$ [cf. \eqref{def_v1}]. This phase
might possibly extend to $Q>4$, but we have not considered this issue here.
Finally, we expect that this phase belongs to the sector 
$\ell=0$, at least for $L\equiv 0\pmod{3}$ (see also \cite{JacobsenSalas07}).
\end{itemize}

Note that the facts that vertical segments at $Q=0,2,4$ are parts of the 
limiting curve $\mathcal{B}_\infty$, and that the vertical segments at 
$Q=1,3$ are lines of isolated limiting points are in full agreement with the
conclusions of \cite{Salas13-1}: the non-negative integers play the same 
role for non-planar graphs, as the one played by the Beraha numbers
for planar graphs. Our results also agree (when restricted to $v=-1$) with 
those found for the chromatic polynomial of the triangular lattice 
\cite{JacobsenSalas07}.  

%
%
\subsection{Critical-polynomial results}
\label{sec:results_CP}

\begin{figure}
  \centering
  \begin{tabular}{cc}
  \includegraphics[width=200pt]{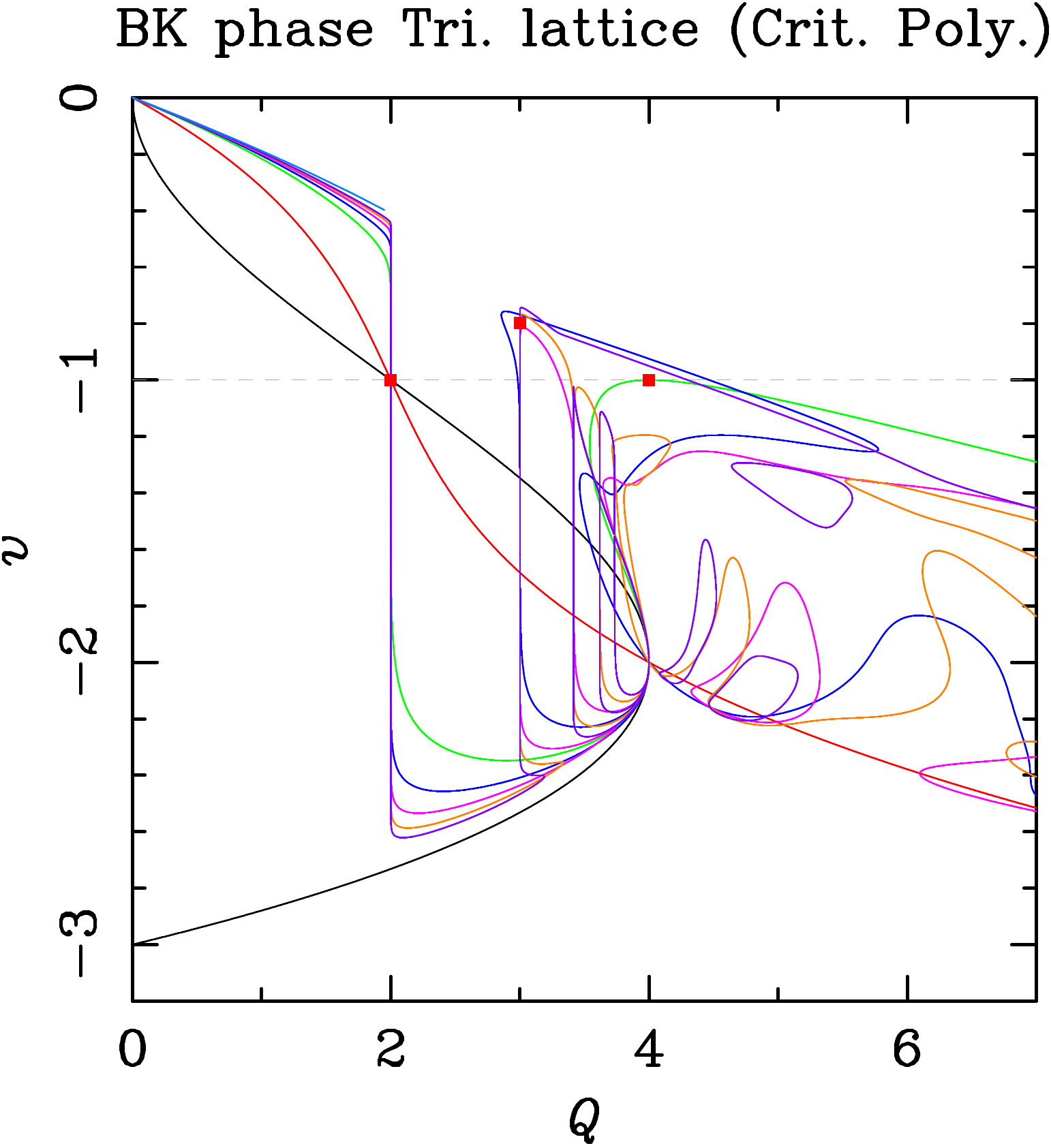} &
  \includegraphics[width=200pt]{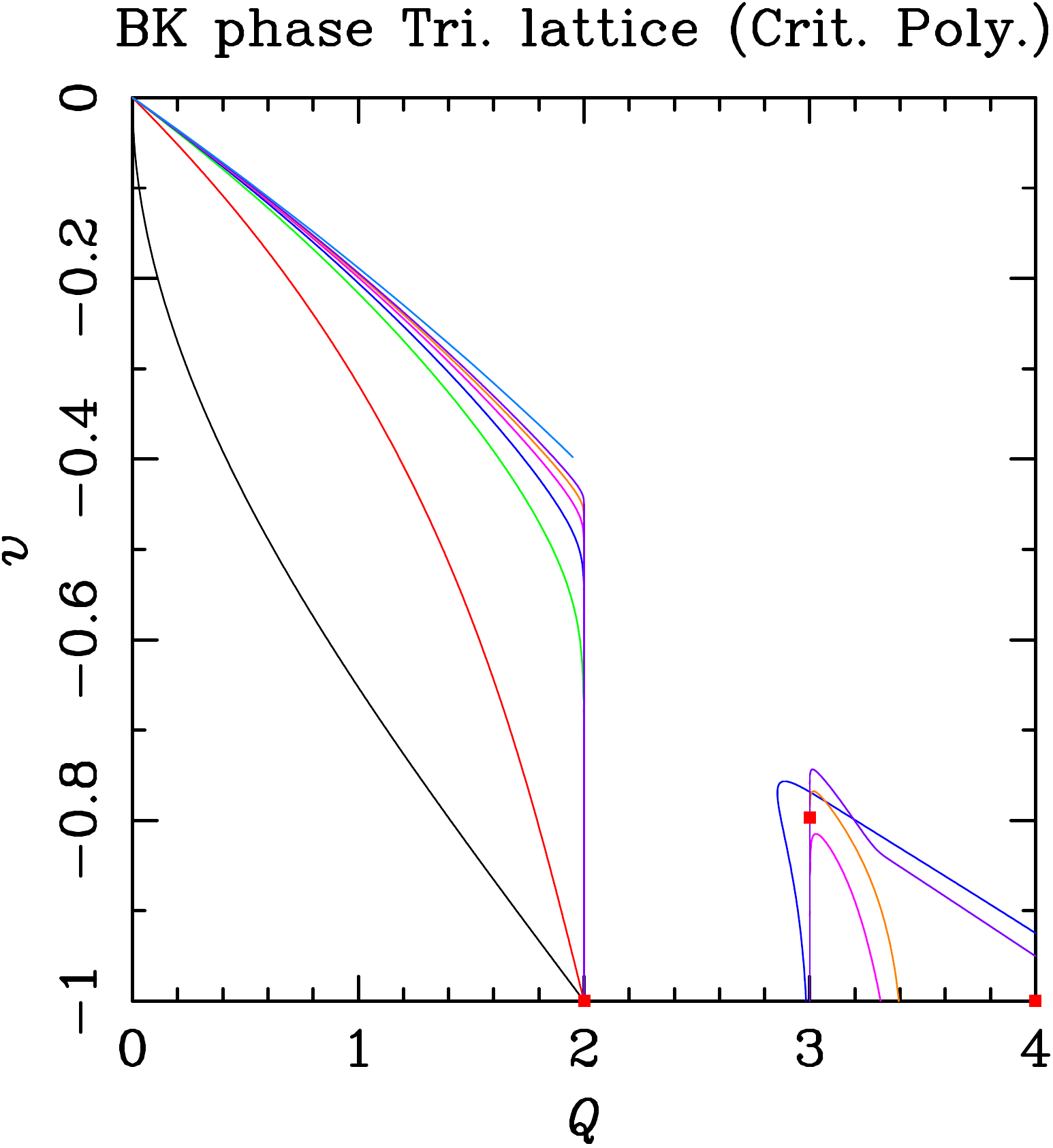} \\
  \qquad (a) & \qquad (b) \\
  \end{tabular}
  \caption{
  (a) Zeros of the CP for basis size 3 (black),
      6 (red), 24 (green), 54 (navy blue), 96 (pink), 150 (orange),
      and 216 (violet). Notice that the CP for the basis 
      of size 3 is a common factor for the rest of the polynomials, and its 
      zeros coincide with the cubic \eqref{cubic_tri}. 
      The upper blue curve in the range $Q\in [0,1.95]$ corresponds to the
      extrapolation of the previous curves. The solid red squares 
      ($\color{red} \blacksquare$) show the known critical points for 
      integer values of $Q=2,3,4$.
  (b) Zoom of the previous panel showing the different approximants to the
      AF critical curve $v_\text{AF}(Q)$.
  }
\label{Figure_tri_CP}
\end{figure}

In figure \ref{Figure_tri_CP}(a), we show the locus of zeros of the CP 
for $L \times L$ bases with $6L^2$ edges (see 
figure~\ref{fig:triangular}) for $L=1,2,\ldots,6$, as well as the simple 
three-edge basis described in \cite{JacobsenScullard12}.
As in our preceding work \cite{SJ12,JacobsenScullard13,Jacobsen14} 
the polynomials are available in electronic form
as supplementary material to this paper.\footnote[1]{%
  This text file {\tt triangular.m} can be processed by 
  {\sc Mathematica} or---perhaps after minor changes of formatting---by any 
  symbolic computer algebra program of the reader's liking.
}
Note that the part of 
the phase diagram which is known exactly, the cubic polynomial 
\eqref{cubic_tri}, is a prediction of the polynomials on every basis. That is, 
the cubic factors out of every higher-order polynomial. As we increase the 
size of the basis, the predicted phase diagram becomes more intricate. In 
particular, vertical rays are clear at the Beraha numbers $Q=B_p$ but 
apparently only for even $p$. More of these become visible with increasing $L$,
up to $B_{12}$ for $L=6$. It therefore seems reasonable to believe that all 
$B_p$, for $p$ even, would be present in the limit $L \to \infty$. For $Q$ 
not an integer, there is no counterpart 
of these rays in the partition function zeros. However, as the 
$p \rightarrow \infty$ limit is reached, the partition function
zeros in figure~\ref{Figure_tri_TM_all} seem to indicate that there is a 
vertical ray at $Q=4$, a feature not evident in the CP.

%
%

While the picture is fairly clear for $Q<4$, for $Q \ge 4$ one can 
discern some features that may be coming into focus, such as a small loop 
that emerges from the vicinity of $(Q,v)=(4,-2)$, but it generally is not 
really clear what exactly the CP is predicting in this region.
The limiting curves obtained by the TM method seem
to provide a more coherent view here, and do include 
the aforementioned loop, but even so there seems to be more variation 
between the different values of $L$ than for $Q<4$.

The AF transition curve $v_+$ appears to be well converged for $L=6$, at 
least where $Q<2$. This curve is shown in closeup in 
figure~\ref{Figure_tri_CP}(b). The gap in $v_+$ between $Q=2$ and $Q=3$ is 
possibly an artifact of our choice of basis, and might be 
filled in by modifying it, perhaps by introducing a twist into the toroidal 
boundary conditions \cite{JacobsenScullard12} 
(although this may introduce gaps elsewhere). The 
phenomenon of individual bases bringing out certain features of the 
phase diagram while being blind to others is a common feature of critical 
polynomials on Archimedean lattices \cite{JS16}, and is presently not well 
understood. Nevertheless, comparing figure~\ref{Figure_tri_CP}(b) with 
\ref{Figure_tri_TM_all}(b) we find very good agreement between the CP  
prediction of $v_+$ for $Q<2$ and the limiting curves computed by 
the TM method. To be more precise, if we consider the interval $Q\in [0,1.9]$
in steps of size $10^{-1}$, the absolute difference between the TM and CP 
estimates increases as $Q$ does: it grows from $\approx 9.9 \times 10^{-6}$ 
at $Q=0.1$ to $4.5 \times 10^{-4}$ at $Q=1.5$, where this difference attains 
its maximum value. For larger values of $Q\le 1.9$, this difference becomes
somewhat smaller, but is still of order $10^{-4}$. We obviously expect identical
results of the two methods in the thermodynamical limit.

We also employ the generalisation of the CP method by one of us 
\cite{Jacobsen15} to bases of size $L \times M$ in the limit $M \to \infty$. 
In this approach there is no actual polynomial, but for any given $(Q,v)$ the 
$M \to \infty$ limit of each of $P_{2D}(Q,v)$ and $P_{0D}(Q,v)$ in 
\eqref{eq:2D0D} can be related to the largest eigenvalue of a 
corresponding TM. The value of $v$ is then adjusted by a root-finding 
algorithm, such as Newton-Raphson or Householder, until $P_B(Q,v)=0$ to 
within some tolerance, giving the estimate of the critical point. Once again, 
the reader is referred to \cite{Jacobsen15} for background details. The main 
difference here is that the numerical diagonalisation is done using the 
Arnoldi method as implemented in the ARPACK \cite{Lehoucq} library, rather 
than the simpler power method of \cite{Jacobsen15}. This makes it possible 
to explore the region $3<Q<4$, where the power method suffers from severe 
convergence problems. As we wished to do the calculations in high numerical 
precision we needed to make major modifications to ARPACK to work with the 
{\sc C}++ CLN arbitrary precision library \cite{Haible1996}. This required 
translating the original {\sc Fortran} code of ARPACK into {\sc C}++ 
and creating template specialisations to handle CLN numbers. 

\begin{figure}
  \centering
  \begin{tabular}{cc}
  \includegraphics[width=200pt]{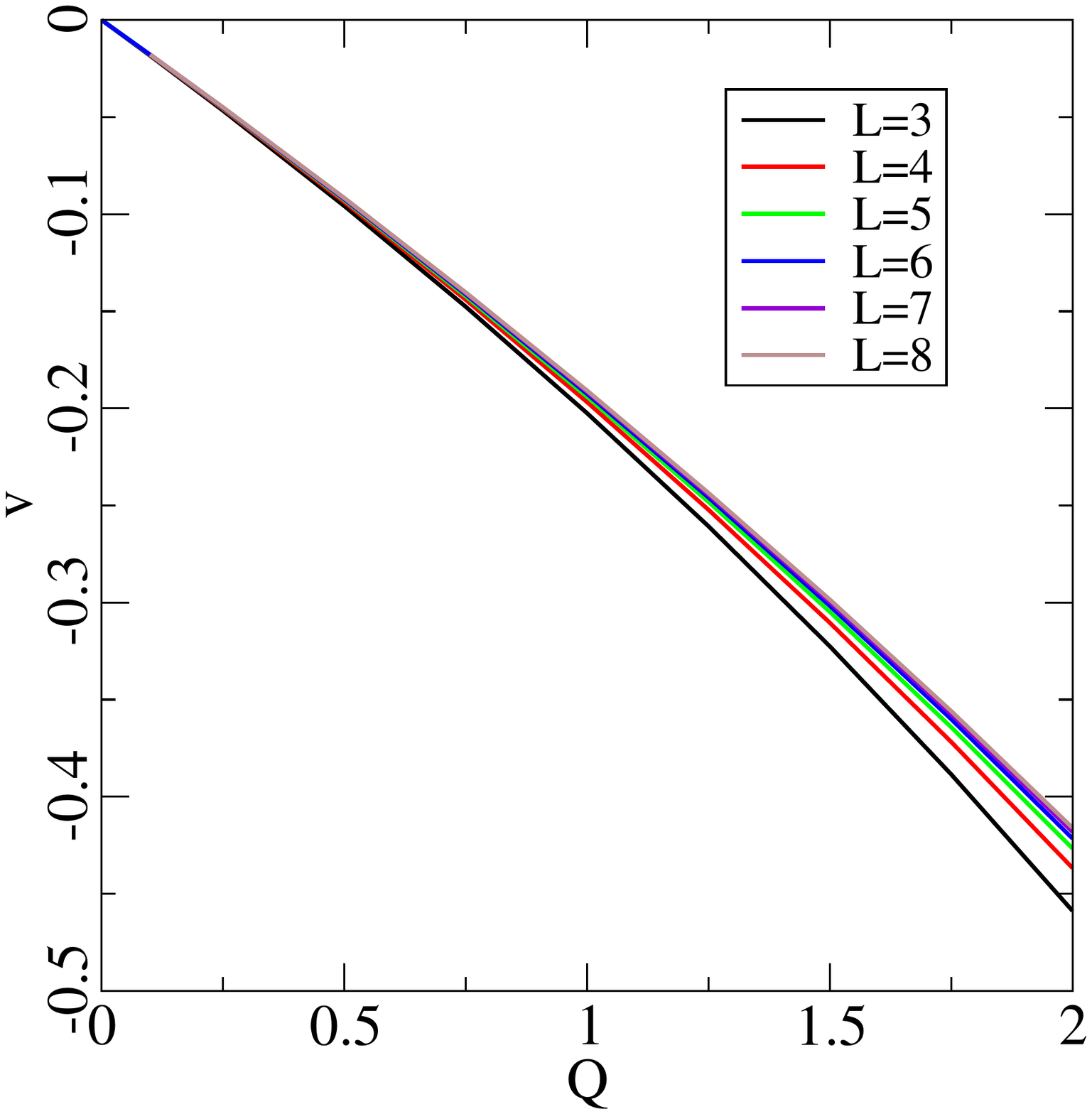} &
  \includegraphics[width=163pt, trim=0 -20 0 0]{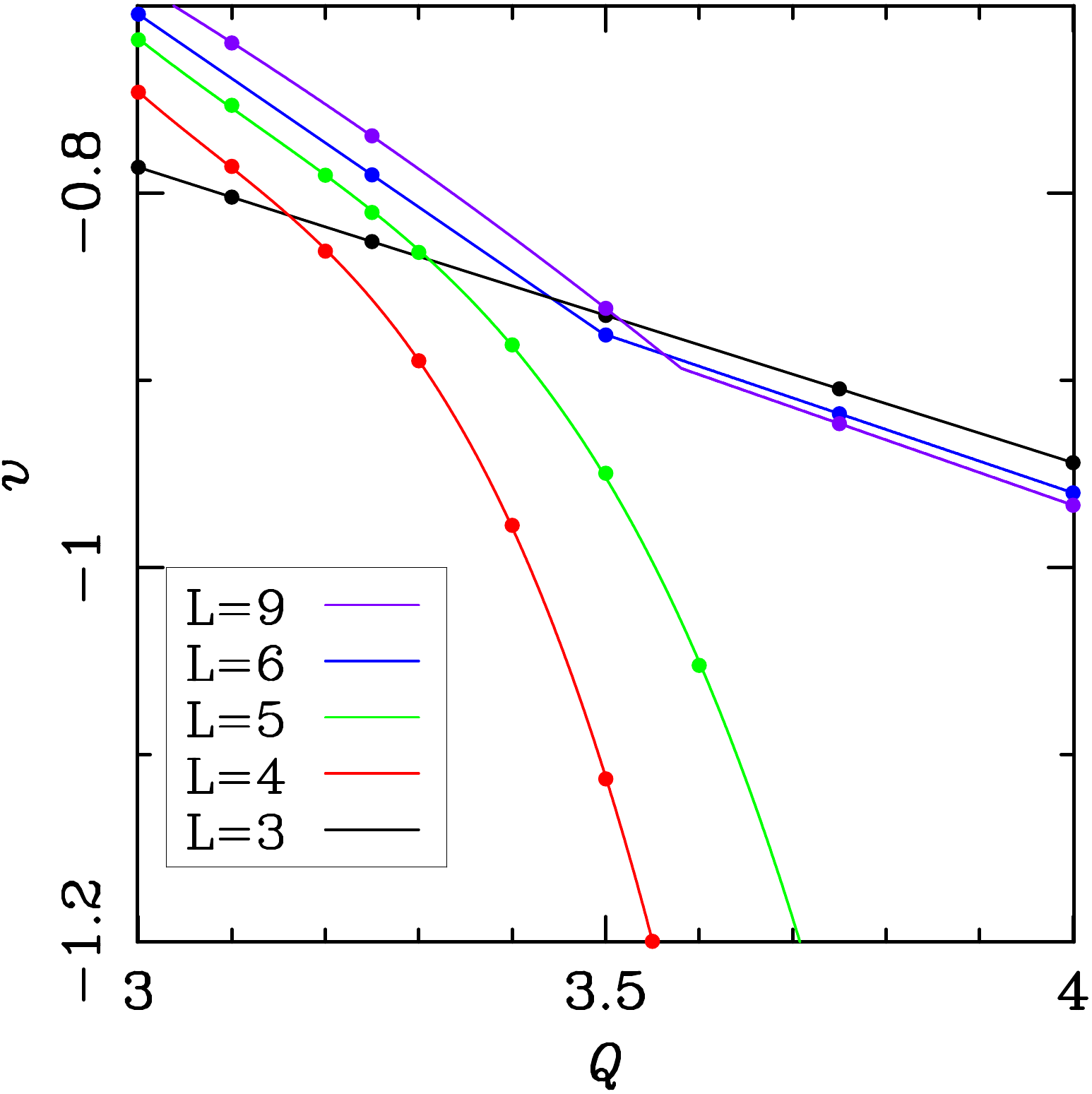} \\
  \quad\, (a) & \quad\, (b) \\
  \end{tabular}
  \caption{The AF curve calculated using the extension of the CP method to 
   bases of size $L \times \infty$; (a) the region $0<Q<2$; (b) the region 
   $3<Q<4$. This method gives no information for $2<Q<3$.
  }
\label{Figure_CP_Arnoldi}
\end{figure}

To find the AF curve, we chose a set of $Q$ values and found the corresponding 
$v$ satisfying $P_B(Q,v)=0$ using the above algorithm with an intial guess for 
$v$ that was close to the data points we had already obtained for this curve. 
Interestingly, the gap between $Q=2$ and $Q=3$ is still present even for 
these $L \times \infty$ bases. That is, when we choose an initial $v$ near 
the AF curve for $2<Q<3$, the iteration converges only to the cubic curve 
\eqref{cubic_tri}. The AF curve obtained in this way for $0<Q<2$ is shown in 
figure~\ref{Figure_CP_Arnoldi}(a) for $L=3,4,5,6,7,8$. 
The curve for $Q>3$, given 
in figure~\ref{Figure_CP_Arnoldi}(b), exhibits a mod $3$ parity effect; 
the computed curves for $L=3,6,9$ cross the region uninterrupted by 
Beraha numbers, much as they do in the $3 \times 3$ and $6 \times 6$ 
polynomial curves of figure~\ref{Figure_tri_CP}(a). The $L=4$ and $L=5$ 
curves bend downwards, probably on their way to form a vertical ray at some 
Beraha number as they also do in the polynomial. It thus appears that the 
results from $L \times \infty$ and $L \times L$ bases are qualitatively 
very similar. Another feature of both is a `kink' in the $L=6$ curve around 
$(Q,v) \approx (3.50,-0.88)$, visible in figure~\ref{Figure_tri_CP}(a) and 
more obvious in \ref{Figure_CP_Arnoldi}(a). A similar behaviour is also seen
for $L=9$. We have estimated the position of such kinks by fitting the points
for $L=6$ (resp.\/ $L=9$) with $Q \le 3.3$ (resp.\/ $Q\le 3.5$) 
to a linear (resp.\/ quadratic) {\em Ansatz} in $Q$, and then repeat the 
linear fits for $Q \ge 3.7$. The crossing point for these two fits provide
estimates of the T-point \eqref{def_T}: i.e., $(Q_T,v_T)\approx (3.497,-0.875)$
for $L=6$, and $(3.581,-0.894)$ for $L=9$. A linear fit in $1/L$ 
to these estimates 
leads to $(Q_T,v_T)\approx (3.75,-0.93)$, which agrees within errors 
with \eqref{def_T}. We hypothesise that 1) this 
feature will develop further for higher $L$ and converge to the T-point 
\eqref{def_T}, and 2) the CP curve will converge to $v_{\rm AF,1}$ for 
$Q \in [Q_T,4]$.

%
%

%
%
\section{The RSOS model on the torus}
\label{sec:RSOS}

There exists another representation of the $Q$-state Potts model on a 
planar graph $G$ when $Q$ is a Beraha number $B_p$ \eqref{eq.Bp}, namely as an
RSOS model of the $A_{p-1}$ type 
\cite{Pasquier87,SaleurBauer89,PasquierSaleur90}, where $p$ denotes an integer,
such that $p \ge 3$. 
It is well-known that, if we build the TM for the $Q$-state Potts model 
in the FK representation, when evaluated at $Q=B_p$ there are massive 
eigenvalue and amplitude cancellations, so that the physical ground state
is deeply buried in the original spectra. 
This RSOS representation is very useful in these cases, as it contains only 
the eigenvalues that do not cancel. 

\medskip

\noindent
{\bf Remark.} By eigenvalue cancellation we mean more precisely the following 
phenomenon. Let us suppose we compute the TM $\mathsf{T}_L$ for a strip width 
$L$, and use its eigenvalues to express the partition function on an 
$L \times N$ torus as  
$Z_{L\times N}(Q,v) = \sum_j \alpha_j(Q) \, \lambda_j(Q,v)^N$, where the
$\alpha_j$ (resp.\/ $\lambda_j$) are the amplitudes (resp.\/ eigenvalues)
for $1\le j\le \dim \mathsf{T}_L$. Let us further assume that there are 
$m$ eigenvalues $\lambda_k(Q,v)$ $(1\le k\le m$) which are distinct for 
generic values of $Q$, but which for a particular value $Q = B_p$ become all 
identical, $\lambda_k(B_p,v)=\lambda_\circ(v)$, and the sum of 
their corresponding amplitudes vansishes $\sum_{k=1}^m \alpha_k(B_p) =0$. 
Then we say that $\lambda_\circ(v)$ cancels from $Z_{L \times N}(Q,v)$, 
since its net contribution to the partition function is zero. 

\medskip

These cancellations imply that the behaviour of the Potts model at Beraha 
numbers can be quite different from its behaviour at generic values of $Q$, 
accessible in the FK representation. In particular, the vertical rays in 
figures~\ref{fig:sq-tripd} and~\ref{Figure_tri_CP} will turn out to conceal 
a rich physics, including several interesting critical points and critical 
phases, once the cancelling eigenvalues have been stripped off
via the RSOS construction.

The RSOS model can be defined on any graph embedded in an 
orientable surface; and in particular on a torus. Therefore, it makes perfect
sense to define a RSOS model on a strip graph with toroidal boundary 
conditions. The equivalence between the RSOS and the Potts models 
holds true in great generality if the graph is planar, in the sense that if 
we take appropriate boundary conditions on both models the corresponding 
partition functions are equal
\cite{Kostov89,RichardJacobsenSalas,JS_in_prep}.
However, this relation is more subtle when the graph is not planar 
\cite{Pasquier87,Pasquier87b,JacobsenRichard05}. In this section we will
investigate such relation when both models are defined on a torus.
 
When considering the RSOS model of the $A_{p-1}$ type, it is convenient 
to use the temperature variable 
\begin{equation}
 x \;=\; \frac{v}{\sqrt{Q}}
\label{eq.x}
\end{equation} 
instead of $v$. It is also useful to define the weights
\begin{equation}
S_h \;=\; \frac{ \sin(h\, \pi/p)}{ \sin(\pi/p)}  \,. 
\label{eq.Sh}
\end{equation}
 
%
%
\subsection{TM construction for the periodic RSOS model}
\label{sec:TM-perio-RSOS}

\begin{figure}
\begin{center}
\includegraphics[width=200pt]{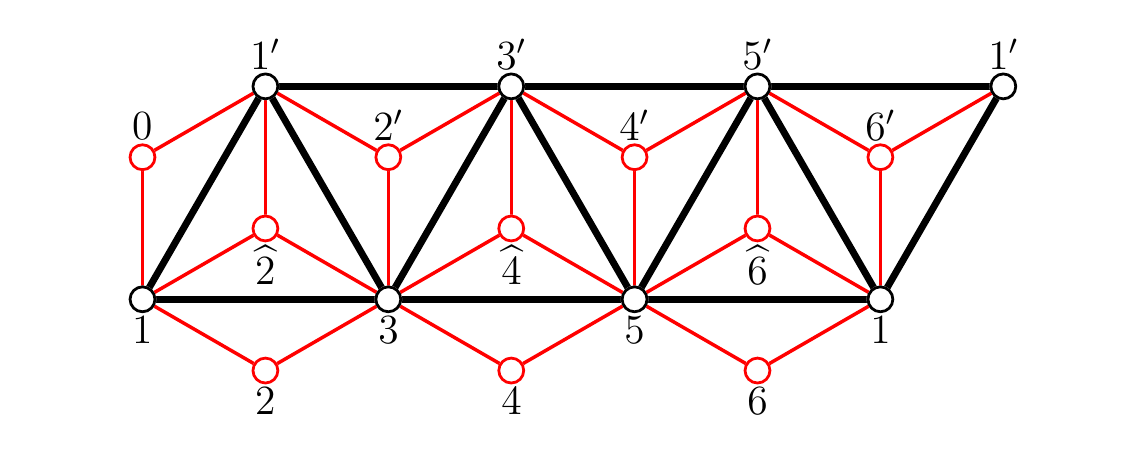}
\end{center}
\caption{Construction of the TM for a triangular-lattice strip of width $L=3$
and periodic transverse boundary conditions at $Q=B_p$ with 
$p\ge 3$. 
A row of this triangular-lattice strip is depicted with black thick edges and
black vertices. We also show the relevant part of the dual strip graph (which
is a hexagonal-lattice strip graph with transverse periodic boundary 
conditions). This dual graph is depicted as thin red lines and red vertices.
The RSOS model lives on the vertices of both lattices with interactions 
around each quadrangular face. In this case, the basis of the space the TM 
acts on is formed by vectors of the type 
$(h_1,h_2,\ldots,h_{2L})$ where $0< h_i < p$ and $|h_i - h_{i+1}|=1$ for 
each $1\le 1\le 2L$ (assuming that $h_{2L+1}=h_1$).
The action of TM gives the final vector $(h'_1,\ldots,h'_{2L})$ starting  
from the initial one $(h_1,\ldots,h_{2L})$.
}
\label{fig:TM_RSOS_tri}
\end{figure}

Let us consider a triangular-lattice strip of width $L$ and periodic 
transverse boundary conditions. In order to define an RSOS-type model,
we need to consider also its dual lattice, which in this case is a 
hexagonal-lattice strip graph with periodic transverse boundary conditions. 
(Notice that given a graph embedded in a surface, not necessarily planar, 
its dual graph is well-defined and it is also embedded in the same surface.)
See figure~\ref{fig:TM_RSOS_tri}.   

The variables in the TM formulation of this RSOS model are vectors of 
dimension $2L$ of the type
$(h_1,h_2,\ldots,h_{2L})$ such that the odd (resp.\/ even) labeled entries 
live on the triangular (resp.\/ dual hexagonal) lattice.%
\footnote{This parity convention is just one of two possible choices, but
we will stick to it hereafter.}
Periodic boundary conditions mean that $h_{2L+1}=h_1$ and $h_0=h_{2L}$.
The variables $h_j$, defined for $1\le j \le 2L$, take values in the set 
$\{1,2,\ldots,p-1\}$, and we shall think of them as `heights'. 
They are required to satisfy the constraint $|h_j - h_{j+1}|=1$, which will 
be extended to any pair of nearest neighbours
through the definition of the TM (see below), hence explaining the name 
Restricted Solid-On-Solid (RSOS) model. The interactions of the RSOS model 
live on the quadrangular faces; these faces have two vertices of each lattice 
type in opposite corners. There are two types of interaction terms 
(we assume that the TM acts upwards; see 
\cite{JacobsenRichard05}): 
$$
\includegraphics[width=200pt]{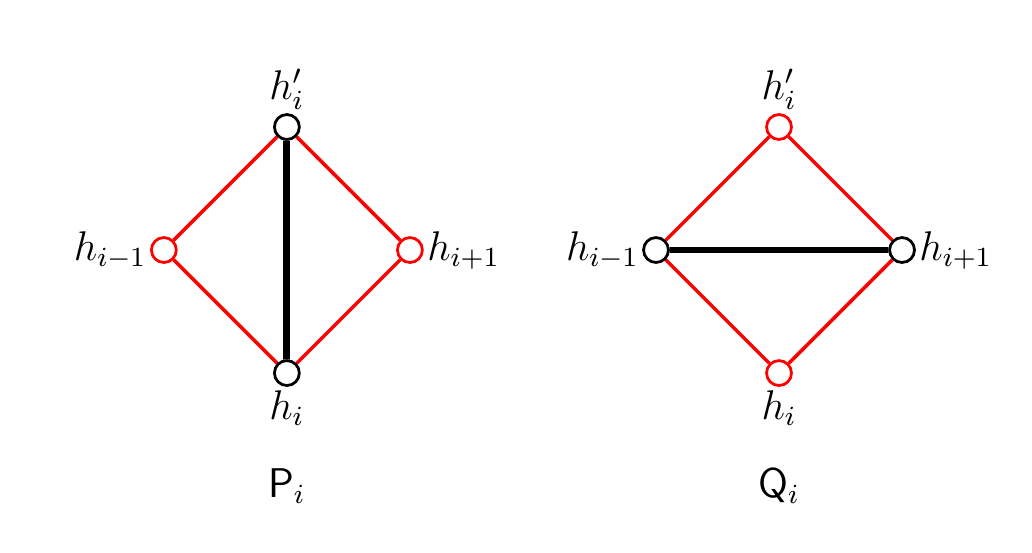}
$$

If we have a vector $\bm{v}=(h_1,\ldots,h_i,\ldots,h_{2L})$, then the
action of the first of the above operators is the following:
\begin{equation}
\frac{1}{\sqrt{Q}}\, \mathsf{P}_i 
      \cdot \bm{v} \;=\; x \, \bm{v} + \delta(h_{i+1},h_{i-1}) \, 
\frac{ \sqrt{S_{h_i}\, S_{h'_i}} }{S_{h_{i-1}}} \, \bm{v}^{(i)} \,,
\label{def.RSOS.H}
\end{equation}  
where $x$ is defined in \eqref{eq.x}, $\delta(a,b)$ stands for the usual
Kronecker delta, the factors $S_h$ are defined in \eqref{eq.Sh}, and the
vector $\bm{v}^{(i)}$ is given by 
$\bm{v}^{(i)}=(h_1,\ldots,h'_i,\ldots,h_{2L})$. 
This operator propagates a height $h_i \to h'_i$ standing on a vertex 
of the original triangular lattice, and it is just another representation 
of the `vertical' operator $\mathsf{P}_i$ in the FK picture based on 
join/detach operators \cite{SalasSokal01}, already discussed in 
section~\ref{sec:tools_TM}. The normalisation on the left-hand side
matches the definition \eqref{eq.x}. Notice also that with the parity 
convention for indices made above, the operator $\mathsf{P}_i$ 
acts only on odd vertices $i$ in figure~\ref{fig:TM_RSOS_tri}. 

The action of the second operator is
\begin{equation}
\mathsf{Q}_i \cdot \bm{v} \;=\; \bm{v} + 
x \, \delta(h_{i+1},h_{i-1}) \, 
\frac{ \sqrt{S_{h_i}\, S_{h'_i}} }{S_{h_{i-1}}} \, \bm{v}^{(i)} \,.
\label{def.RSOS.V}
\end{equation}
This operator propagates a height $h_i \to h'_i$ standing on a vertex
of the dual hexagonal lattice, and hence, it corresponds to a `horizontal'
operator in the FK picture based on join/detach operators \cite{SalasSokal01}.
In this case, the operator $\mathsf{Q}_i$ acts only on even sites in 
figure~\ref{fig:TM_RSOS_tri}.   

%
%

We can build the TM in the RSOS representation for a triangular-lattice
strip graph with toroidal boundary conditions by going through the following 
steps: Let us start with a vector $(h_1,h_2,\ldots,h_{2L})$, where the 
variables $h_i$ with $1\le i\le 2L$ satisfy the required constraints; then,  

\begin{enumerate}
 \item Build the `horizontal' TM given by  
       $\mathsf{H}_L = \mathsf{Q}_2\cdot \mathsf{Q}_4 \cdots  \mathsf{Q}_{2L}= 
       \prod\limits_{j=1}^{L} \mathsf{Q}_{2j}$, to obtain the new vector 
       $(h_1,\widehat{h}_2,h_3,\ldots,h_{2L-1},\widehat{h}_{2L})$,
       where the $\widehat{h}_{2j}$ with $1\le j\le L$ are 
       intermediate variables that we need to keep track of.     
 \item Insert an extra site which should be neighbour to site $1$. The 
       corresponding variable $h_0$ can take any of the allowed values of  
       $h_0=h_1 \pm 1$. Then, our vector has been enlarged and 
       takes the form:
       $(h_0,h_1,\widehat{h}_2,h_3,\ldots,h_{2L-1},\widehat{h}_{2L})$.

 \item Build the `vertical' TM given by  
       $\mathsf{V}_L = \mathsf{P}_1\cdot \mathsf{Q}_2 \cdot \mathsf{P}_3
        \cdot \mathsf{Q}_4 
        \cdots \mathsf{P}_{2L-1}\cdot \mathsf{Q}_{2L} = 
        \prod\limits_{j=1}^{L} (\mathsf{P}_{2j-1}\cdot \mathsf{Q}_{2j})$,
       to obtain the new enlarged vector
       $(h_0,h'_1,h'_2,h'_3,\ldots,h'_{2L-1},h'_{2L})$.
 
 \item If $h_0 = h'_{2L}$, then this vector is valid, so we can drop the first
       entry and get the final vector 
       $(h'_1,h'_2,\ldots,h'_{2L})$ of the right dimension $2L$. If 
       $h_0 \neq h'_{2L}$, then this vector is not valid, and its  
       contribution should be deleted. 
\end{enumerate}
 
Due to the normalisation of the operators $\mathsf{P}_i$ 
[cf.~\eqref{def.RSOS.H}], the transfer matrices in the FK and RSOS 
representations are now consistently normalised. In more algebraic terms, 
we can in fact write
\begin{subequations}
\label{def_TLcorr}
\begin{align}
 \frac{1}{\sqrt{Q}} \, \mathsf{P}_i &\;=\; x \, \mathds{1} + \mathsf{E}_i \,, 
 \label{def_TLcorr1} \\
 \mathsf{Q}_i &\;=\; \mathds{1} + x \, \mathsf{E}_i \,, \label{def_TLcorr2}
\end{align}
\end{subequations}
where $\mathds{1}$ is the identity operator, and the $\mathsf{E}_i$ are 
the generators of the Temperley-Lieb (TL) algebra \cite{TemperleyLieb} obeying 
the abstract relations
\begin{subequations}
\label{def_TLrels}
\begin{align}
 \mathsf{E}_i^2 &\;=\; \sqrt{Q} \, \mathsf{E}_i \,, \label{def_TLrels1} \\
 \mathsf{E}_i \, \mathsf{E}_{i \pm 1}\,  \mathsf{E}_i &\;=\; \mathsf{E}_i \,, 
 \label{def_TLrels2} \\
 \mathsf{E}_i \, \mathsf{E}_j &\;=\; \mathsf{E}_j \, \mathsf{E}_i \quad  
 \text{for $|i-j| > 1$.} \label{def_TLrels3} 
\end{align}
\end{subequations}
The FK and RSOS expressions given earlier are but two different 
representations of this abstract algebra.%
\footnote{In the case of periodic boundary conditions, all indices $i$ are 
   considered mod $2L$. In addition, \eqref{def_TLrels} must be supplemented 
   by further relations to make the algebra finite-dimensional, giving
   rise to the so-called affine TL algebra. We refer to \cite{BGJSV17} 
   for further discussion.
}
The dimensions of these representations are not identical, so obviously the 
spectra of the two TM cannot be identical. There are however many shared 
eigenvalues, and in particular the `ground state' eigenvalue (i.e., 
the dominant eigenvalue for $x > 0$, and its analytic continuation for 
$x < 0$) is identical in both representations. This is of course
not enough to ensure the equality of the corresponding partition functions on a finite-size torus. Moreover, each representation does not necessarily admit 
the same kind of boundary conditions. In particular, it can be shown 
\cite{BGJSV17} that taking periodic boundary conditions in the RSOS 
representation (i.e., taking $h_{i+2L} \equiv h_i$)
will produce a superposition of several different (twisted) boundary conditions
in the FK representation.
We shall admit for the time being that it is of interest to study
the RSOS model with periodic boundary conditions, and pursue and clarify 
some of these subtle aspects below.

In particular, we shall compute the TM in the periodic RSOS representation 
for several values of $p=3,4,5,6,7,8$; but we will not attempt to compute the 
full partition function $Z_{L_P \times N_P}^\text{RSOS}$. Therefore, our 
comparison will be made at the level of eigenvalues (or free energies). 
We stress again that the above normalisation allows the 
direct comparison among eigenvalues in different representations of the 
TL algebra. In particular, we shall compare the RSOS representation with the 
FK representation for integer $p \ge 3$, and with the Potts spin representation
for integer $Q = 2,3$ (hence $p=4,6$).

%
%
\subsection{Symmetries and sector decomposition}
\label{sec:TM-syms}

If we consider the RSOS model described above, we can try to use the
full configuration space for moderate values of $L$. However, since its
dimension grows exponentially fast with $L$, it is also of interest to diminish
its size using lattice and target space symmetries. To this end, we first 
observe that given a configuration vector $(h_1,h_2,\ldots,h_{2L})$, it is 
obvious from the definition of $\mathsf{P}_i$ and $\mathsf{Q}_i$ that the 
height variables $h_i$ having
labels $i$ of the same parity will take values with the
same parity. This condition is preserved under the action of 
the full TM. Therefore, we can define two sectors:  
\begin{itemize}
    \item {\bf Even sector}: the heights on the triangular lattice take
    even values (and those on the dual hexagonal lattice take odd values).
    This means that the odd (resp.\/ even) entries of the state vector
    $(h_1,h_2,\ldots,h_{2L})$ take even (resp.\/ odd) values: e.g.,
    $(2,1,2,3,4,3)$ for $L=3$. The transfer matrix computed on this sector
    will be denoted by $\mathsf{T}_\text{RSOS}^\text{(even)}$.

    \item {\bf Odd sector}: the heights on the triangular lattice take
    odd values (and those on the dual hexagonal lattice take even values).
    This means that the odd (resp.\/ even) entries of the state vector
    $(h_1,h_2,\ldots,h_{2L})$ take odd (resp.\/ even) values: e.g.,
    $(1,2,3,4,3,2)$ for $L=3$. The transfer matrix computed on this sector
    will be denoted by $\mathsf{T}_\text{RSOS}^\text{(odd)}$.
\end{itemize}

The RSOS model with periodic transverse boundary conditions has two  
symmetries that might help to reduce the size of the configuration space:
\begin{itemize}

\item {\bf Spin-reversal symmetry}: the RSOS partition function is invariant
  under the transformation $h_k \to p-h_k$ for $0\le k\le 2L$. If we
  denote by $\bm{v}'$ the vector obtained from $\bm{v}$ by applying the
  latter transformation, we can make a change from the initial
  basis $\mathcal{B} = \{ \bm{v}_j, \bm{v}'_j\}$ to
  $\mathcal{B}' = \{ \bm{v}_j + \bm{v}'_j, \bm{v}_j - \bm{v}'_j \}$. Indeed,
  the spectra of the corresponding transfer matrices should coincide.
  We can now use the subspace of those states that are invariant under
  spin reversal $\mathcal{B}'' = \{ \bm{v}_j + \bm{v}'_j\}$, and build the
  corresponding transfer matrix $\mathsf{T}_\text{RSOS}^\text{(S)}$ on it.
  We can do the same procedure on the even and odd sectors defined
  above.\footnote{%
     This can only be done for even values of $p$. For odd values
     of $p$, $\bm{v}_j$ and $\bm{v}'_j$ belong to distinct sectors.
  }
  The corresponding transfer matrices are denoted
  $\mathsf{T}_\text{RSOS}^\text{(even,S)}$ and
  $\mathsf{T}_\text{RSOS}^\text{(odd,S)}$.
  
\item {\bf Rotational symmetry}: due to the periodic boundary conditions
  in the transverse direction, the RSOS partition function should be
  invariant under any translation along that direction. Note that,
  since the odd (resp.\/ even) entries in a state vector correspond to vertices
  on the triangular (resp.\/ dual hexagonal) sublattices, these translations
  should be done in steps of two units. Instead of 
  building a basis that is merely invariant under rotations, we 
  consider a basis that is invariant under both rotation \emph{and}
  spin-reversal symmetries: now each element of the basis
  contains a given state, its spin-reversal counterpart, and all those
  vectors obtained from the previous ones by translations of even length.
  The corresponding transfer matrix will be denoted by
  $\mathsf{T}_\text{RSOS}^\text{(F)}$. 
  (The superscript F stands for `Fully symmetric'.)
  If we perform the same procedure
  on each of the two even/odd sectors for even values of $p$ we obtain
  transfer matrices that we denote
  $\mathsf{T}_\text{RSOS}^\text{(even,F)}$ and
  $\mathsf{T}_\text{RSOS}^\text{(odd,F)}$.
\end{itemize}

Finally, we have implemented the construction of these TM in two independent 
ways: a) a symbolic approach using {\sc Mathematica} (similar to the one
used in section~\ref{sec:results_TM}); and b) a numerical approach using 
code written in {\sc C} and using Arnoldi's algorithm. Indeed, we have 
cross-checked the results for the smaller values of $L$, and the agreement 
is perfect (within the precision of the corresponding numerical procedures).

%
%
\subsection{Dimension of the TM}

The dimension of the TM in the FK representation is independent of $Q$, 
but depends on the width $L$ and on the boundary conditions. For periodic 
boundary conditions in the longitudinal direction, the TM can be decomposed 
in sectors corresponding to the number $\ell = 0,1,\ldots,L$ of marked 
clusters that propagate along the time direction. Each sector
can be equivalently realised, for $\ell \ge 1$, as a standard module in the 
representation theory of the affine TL algebra, and as an appropriate quotient 
for $\ell = 0$. The dimensions $d_\ell^{\rm FK}$ read
\begin{equation}
 d_0^{\rm FK} = \frac{1}{L+1} \, \binom{2L}{L} \,, \qquad
 d_\ell^{\rm FK} = \binom{2L}{L-\ell} \quad \text{for $1 \le \ell \le L$.} 
\end{equation}

The dimension of the TM in the RSOS representation can be inferred from the 
number of paths $M_{ba}$ on the Dynkin diagram $A_{p-1}$, going from node 
$h_0 = a$ to node $h_{2L} = b$ in precisely
$2L$ steps. This is given by \cite[eq.~(4.17)]{SaleurBauer89}
\begin{equation}
 M_{ba} = \frac{2}{p} \, \sum_{j=1}^{p-1} 
 \sin\left( \frac{\pi a j}{p} \right) \, 
 \sin \left( \frac{\pi b j}{p} \right) \, 
 \left[ 2 \, \cos \left( \frac{\pi j}{p} \right) \right]^{2L} \,.
 \label{dimMba}
\end{equation}
To have periodic boundary conditions we must impose $h_{2L} = h_0$. It 
follows that
\begin{equation}
 {\rm dim}\, \mathsf{T}^{\rm (even)}_{\rm RSOS} \;=\; 
 {\rm dim}\, \mathsf{T}^{\rm (odd)}_{\rm RSOS} \;=\; 
 \frac{1}{2} \, \sum_{a=1}^p M_{aa} \,.
 \label{dimT-RSOS}
\end{equation}
These dimensions of course depend explicitly on $p$.

%
%
\subsection{Boundary conditions}
\label{sec:braidtranslation}

It was mentioned towards the end of section~\ref{sec:TM-perio-RSOS} that the 
periodic boundary condition imposed on the RSOS model
corresponds to a superposition of several boundary conditions in the RSOS 
model. We now review this construction in sufficient
details to be able to correctly interpret the results that will be given in 
section~\ref{sec:RSOS.results}.

Consider the space ${\cal W}_a$ of RSOS states $(h_0,h_1,\ldots,h_{2L})$
in which the first and last heights have been constrained to take equal values, $h_0 = h_{2L} = a$, where $a \in \{1,2,\ldots,p-1\}$ is fixed.
The space ${\cal W}_a$ has dimension $M_{aa}$ given by \eqref{dimMba}. 
Within ${\cal W}_a$ we now consider the {\em non-periodic} TL
algebra in the RSOS representation, with generators 
$\{\mathsf{E}_i\}_{i=1}^{2L-1}$. Combining the concepts of braid translation
and blobbed boundary conditions, one may manufacture another 
operator $\mathsf{E}_0$ (it is given by an explicit, albeit highly non-local,
expression in terms of the non-periodic generators $\mathsf{E}_i$) so that 
the set $\{\mathsf{E}_i\}_{i=0}^{2L-1}$ obeys the algebraic
relations of the {\em periodic} TL algebra \cite{BGJSV17}. Any operator 
${\cal O}$ within the periodic TL algebra can now be replaced by a
corresponding expression $\widetilde{\cal O}_a$ in the non-periodic one, 
by using $\mathsf{E}_0$ as the periodic generator; we shall refer to 
this procedure as `periodicisation'. In particular, one might periodicise 
the TM of the triangular-lattice Potts model, by exploiting its expression 
in terms of the TL generators and using the RSOS representation for the latter,
as given in section~\ref{sec:TM-perio-RSOS}.

It turns out that the union over $a$ of the spectra of each periodicised 
operator $\widetilde{\cal O}_a$ is equal to the spectrum of the original
operator ${\cal O}$ in the genuine periodic RSOS representation. Moreover, 
the parameter $a$ can be interpreted as a certain twist which amounts,
in the corresponding loop model representation of the TL algebra, to giving 
the weight
\begin{equation}
 n_a \;=\; 2 \cos \left( \frac{\pi a}{p} \right)
 \label{eq:na}
\end{equation}
to each loop that wraps around the periodic direction (i.e., a loop which is 
non-homotopic to a point). Alternatively, in the FK cluster representation,
each wrapping cluster will get the weight $Q_a = n_a^2$ because it is 
surrounded by a pair of wrapping loops; notice in particular that the sign of
$n_a$ is immaterial. The case $a=1$ thus gives rise
to $Q_a = Q$, meaning that wrapping clusters get the same weight as 
non-wrapping ones---in other words, this reproduces the effect of periodic
boundary conditions in the FK representation.

Summarising, the periodic RSOS representation can be decomposed into several 
sectors $a$, of which $a=1$ corresponds to periodic boundary
conditions in the corresponding FK model, whereas other values of $a$ are 
equivalent to giving a modified weight $Q_a = n_a^2$ to wrapping
clusters. We shall use this result below to analyse the TM 
$\mathsf{T}_\text{RSOS}$ in the periodic RSOS representation. However, we 
shall make no attempt on constructing the operators 
$(\widetilde{\mathsf{T}}_\text{RSOS})_a$ for each sector $a$ individually.

%
%
\section{Numerical results for the RSOS model on the torus} 
\label{sec:RSOS.results}

In this section we discuss our numerical results for the RSOS model
of type $A_{p-1}$ for $3\le p \le 8$ on a triangular-lattice strip graph 
with toroidal boundary conditions. We shall consider only strips of 
widths that are a multiple of $3$: i.e., $L=3,6,9,12,15$, 
in order to avoid the
mod~3 parity effects which we have observed above in both the limiting curves
and the roots of the CP.
Because the computations for $L\ge 12$ are very
demanding, we will focus in this case (except for $p=4$) on the even RSOS 
sector with all symmetries taken into account (i.e., the spectra of 
$\mathsf{T}_\text{RSOS}^\text{(even,F)}$).

We first consider, in section~\ref{sec:RSOS_integerQ}, the values
$p=3,4,6$ for which $Q=B_p$ is an integer. In these cases, we have three 
representations of the corresponding $Q$-state Potts model: i.e., the 
spin, FK, and RSOS representations. 
Next, in section~\ref{sec:RSOS_non-integerQ}, we will consider the
cases $p=5,7,8$ for which $Q$ is non-integer. In these cases, only the
FK and the RSOS representations are available.

For each fixed value of $p$, we will study in detail the corresponding RSOS 
model by considering the different sectors and symmetries 
described in section~\ref{sec:TM-syms}; the main focus will be on the even 
and odd RSOS sectors with all symmetries taken into account (i.e., 
$\mathsf{T}_\text{RSOS}^\text{(even,F)}$ and 
$\mathsf{T}_\text{RSOS}^\text{(odd,F)}$). 
We will compare the resulting RSOS spectra with the eigenvalue structure of the 
spin \cite[and references therein]{SalasSokal01} and FK representations 
of the associated $Q$-state Potts model (with $Q=B_p$). Our goal will be to 
find, for each $p$, values of the variable $x$ for which
the RSOS model displays new physical properties.

In particular, we aim at identifying critical points and critical phases.
This is most easily done by computing the effective central charge $c$ from
the finite-size corrections to the free energy. In some cases $c$ can be
identified with that of a known CFT, and so the whole operator content is 
fixed by inference. In other cases the identification is ambiguous or 
uncertain, due to an insufficient numerical precision or the lack of an 
obvious candidate CFT. In
those cases we content ourselves by determining a numerical value of $c$, and
we leave the precise identification of the operator content for future work.

Two known CFT will play a special role below. The first one is the usual 
unitary minimal model of central charge
\begin{equation}
 c_\text{FM} \;=\; 1 - \frac{6}{p(p-1)} \,.
 \label{cFM}
\end{equation}
As indicated by the subscript, this CFT is expected to be related with the 
ferromagnetic (FM) transition in the Potts model. The other one is the 
$sl(2,\mathbb{R})/u(1)$ Euclidean black hole sigma model with
\begin{equation}
 c_\text{PF} \;=\; 2 - \frac{6}{p} \,,
 \label{cAF}
\end{equation}
which is the continuum limit of the AF transition in the square-lattice Potts 
model \cite{JacobsenSaleur06,IkhlefJacobsenSaleur08,IkhlefJacobsenSaleur12}. 
Numerical evidence has previously been given in 
\cite[figure~27]{JacobsenSaleur06} that this theory also describes, 
for real $p \in [2,\infty)$, the transition at $v_{-}(Q)$ of the
triangular-lattice model in the FK representation. We shall see here that 
the restriction of this theory to integer $p \ge 3$---that is, 
the $sl(2)/u(1)$ parafermion CFT, still with
central charge \eqref{cAF}---will account as well for the behaviour of the 
RSOS model at $v_{-}(Q)$. The subscript PF in \eqref{cAF} is chosen to
remind us about this connection with parafermions.

%
%
\subsection{Results for integer $Q$} 
\label{sec:RSOS_integerQ}

We have made two independent 
cross-checks to ensure the correctness of our numerical 
techniques.

The first one is to reproduce the trivial result for $Q=1$ ($p=3$):
the TM in the RSOS representation has a unique eigenvalue 
$\mu = (1 + x)^{3L} = (1+v)^{3L}$. This is indeed the expected result, as the 
partition function for a triangular-lattice strip on a torus of dimensions 
$L\times N$ is $Z_G^\text{Potts}(1,v) = (1+v)^{3LN}$. 
We have also computed the free energy \eqref{eq.free_energy_L} of a 
triangular-lattice $Q$-state Potts model on the torus of width $L=3$ for 
$Q=2,3$ using the TM in the spin representation. These free energies 
coincide (within the numerical precision) with those obtained using the TM
in the FK representation when taking into account the vanishing amplitudes
and the eigenvalue cancellations. Notice that in these three cases 
$Q=1,2,3$, we do find such vanishing/cancellation features in the FK 
amplitudes/eigenvalues.

%
%
\subsubsection{$Q=B_4=2$.}
\label{sec:RSOS_p=4}

We show the dimensions of the TM in the spin, FK, and RSOS representations
in table~\ref{table.dim.p=4-6}. For $p=4$, we see that
$\dim \mathsf{T}_\text{spin}=\dim \mathsf{T}_\text{RSOS}^\text{(even)}=
 \dim \mathsf{T}_\text{RSOS}^\text{(odd)}$
for all widths $L$. This equality can be interpreted as follows: 
the RSOS variables on one sublattice all take the value $2$, 
while those on the other sublattice can take the other two possible
values $1,3$. Hence, there is a bijection between RSOS configurations in the 
even or odd sectors, and Ising configurations on one sublattice.

The dimensions of $\mathsf{T}_\text{RSOS}^\text{(even)}$ and 
$\mathsf{T}_\text{RSOS}^\text{(odd)}$
of course also agree with \eqref{dimT-RSOS}, when the sum is constrained 
to the specified parity of $a$.

%
%
\begin{table}[h]
\centering
\begin{tabular}{rrrrrr}
\hline\hline\\[-2mm]
$p$ & $L$ & $\dim \mathsf{T}_\text{FK}$
          & $\dim \mathsf{T}_\text{spin}$
          & $\dim \mathsf{T}_\text{RSOS}^\text{(even)}$
          & $\dim \mathsf{T}_\text{RSOS}^\text{(even,F)}$\\[2mm]
\hline \\
4  & $3$ & 30 &   8 &    8 &   2 \\
   & $6$ &    &  64 &   64 &   8 \\
   & $9$ &    & 512 &  512 &  30 \\
   & $12$&    &4096 & 4096 & 180 \\
\hline \\[-2mm]
6  & $3$ & 30 &  27  &   28&    6 \\
   & $6$ &    & 729  &  730&   68 \\
   & $9$ &    &19683 &19684& 1098 \\
   & $12$ &  &           & 531442 & \\
\hline\hline
\end{tabular}
\caption{%
  Dimension of the TM for the Potts model with $Q=B_p$ ($p=4,6$) 
  on a triangular-lattice strip graph on the torus in the FK representation 
  $\mathsf{T}_\text{FK}$, in the spin representation $\mathsf{T}_\text{spin}$, 
  and in the RSOS representation in the \emph{even} sector 
  $\mathsf{T}_\text{RSOS}^\text{(even)}$, and in the same sector with all
  symmetries $\mathsf{T}_\text{RSOS}^\text{(even,F)}$. 
  Note that
  $\dim \mathsf{T}_\text{RSOS}^\text{(even)}=
   \dim \mathsf{T}_\text{RSOS}^\text{(odd)} = 
   \dim \mathsf{T}_\text{RSOS}^\text{(S)}= 
  2\dim \mathsf{T}_\text{RSOS}^\text{(even,S)} = 
  2\dim \mathsf{T}_\text{RSOS}^\text{(odd,S)}$,   
   and
  $\dim \mathsf{T}_\text{RSOS}^\text{(F)}=  
  2\dim \mathsf{T}_\text{RSOS}^\text{(odd,F)} = 
  2\dim \mathsf{T}_\text{RSOS}^\text{(even,F)}$.
}
\label{table.dim.p=4-6}
\end{table}

The free energy (per site) for a given sector of the RSOS model is 
expressed in terms of the dominant eigenvalue $\mu_*$
of the corresponding TM by the analogue of \eqref{eq.free_energy_L}.
For each pair $(p,L)=(4,L)$ with $L=3,6,9,12$, we have computed 
the free energy for each RSOS 
sector $f^\text{(RSOS)}_L(p,x)$ in the interval $x\in[-3,1]$ in steps of 
$10^{-2}$. We can compare these free energies with those of the corresponding 
$Q$-state Potts model with periodic boundary conditions in the spin 
representation, $f_L^\text{(spin)}(Q,v)$, with
$Q=B_p=2$, and $v=x\sqrt{B_p}$.

According to section~\ref{sec:braidtranslation}, the quantities 
$f^\text{(RSOS)}_L(p,x)$ and $f_L^\text{(spin)}(Q,v)$
need not agree for any value of $x$. Indeed, the former quantity contains 
two different boundary conditions,
corresponding to giving the weight $Q_1 = 2$ or $Q_2 = 0$ to each wrapping 
cluster. These can be identified with,
respectively, periodic ($a=1$) and antiperiodic ($a=2$) boundary conditions 
for the Ising spins, where the label $a$
has the same meaning as in \eqref{eq:na}.  For $x > 0$,
the periodic boundary condition is dominant by probabilistic reasons, 
implying that $f^\text{(RSOS)}_L(p,x) = f_L^\text{(spin)}(Q,v)$ indeed,
but this equality might not hold true for any $x < 0$.

To give a quantitative measure of the 
difference between these two free energies, we introduce the quantity   
\begin{equation}
\Delta_f(p,L) \;=\; \max_{x\in[-3,1]} 
\left| f_L^\text{(spin)}(B_p,v)-f^\text{(RSOS)}_L(p,x) \right| \,.
\label{def.disc}
\end{equation}
In table~\ref{table.dis.p=4-6} we show this quantity for each pair 
$(p,L)=(4,L)$ and each RSOS model we have studied. 
 
%
%
\begin{table}[h]
\centering
{
\small
\begin{tabular}{rrccccccccc}
\hline\hline\\[-2mm]
$p$ & $L$ & RSOS
          & RSOS
          & RSOS
          & RSOS
          & RSOS
          & RSOS
          & RSOS
          & RSOS  \\
    &     & (even)
          & (odd)
          & (even,S)
          & (odd,S)
          & (S)
          & (even,F)
          & (odd,F)
          & (F)       \\[2mm]
\hline \\
4  & $3$ & $3\cdot 10^{-4}$
         & $0$
         & $0$
         & $0$
         & $0$
         & $0$
         & $0$
         & $0$ \\
   & $6$ & $10^{-5}$
         & $0$
         & $0$
         & $0$
         & $0$
         & $0$
         & $0$
         & $0$ \\
   & $9$ & $10^{-6}$
         &
         &
         &
         &
         & $0$
         & $0$
         & $0$ \\
   & $12$& $10^{-7}$
         &
         &
         &
         &
         &
         &
         & $0$ \\
\hline \\[-2mm]
6  & $3$ & $6\cdot 10^{-2}$
         & $6\cdot 10^{-2}$
         & $4\cdot 10^{-4}$
         & $0$
         & $4\cdot 10^{-4}$
         & $8\cdot 10^{-2}$
         & $8\cdot 10^{-2}$
         & $8\cdot 10^{-2}$ \\
   & $6$ & $9\cdot 10^{-6}$
         & $6\cdot 10^{-6}$
         & $9\cdot 10^{-6}$
         & $0$
         & $9\cdot 10^{-6}$
         & $9\cdot 10^{-6}$
         & $0$
         & $9\cdot 10^{-6}$ \\
   & $9$ & $6\cdot 10^{-3}$
         &
         &
         &
         &
         & $3\cdot 10^{-2}$
         & $3\cdot 10^{-2}$
         & \\
\hline\hline
\end{tabular}
}
\caption{%
  Difference $\Delta_f(p,L)$ \eqref{def.disc} between the free energy
  obtained in the periodic spin representation for $Q=2,3$, and the free energy 
  computed for the different sectors of the periodic RSOS model with $p=4,6$, 
  respectively.
}
\label{table.dis.p=4-6}
\end{table}

As expected from the bijection between states in the RSOS and spin
representation, we find that for \emph{all} (but one) sectors of the
RSOS model, the free energy $f^\text{(RSOS)}_L$ coincides exactly with 
$f_L^\text{(spin)}$.

\begin{figure}
\begin{center}
  \begin{tabular}{cc}
  \includegraphics[width=200pt]{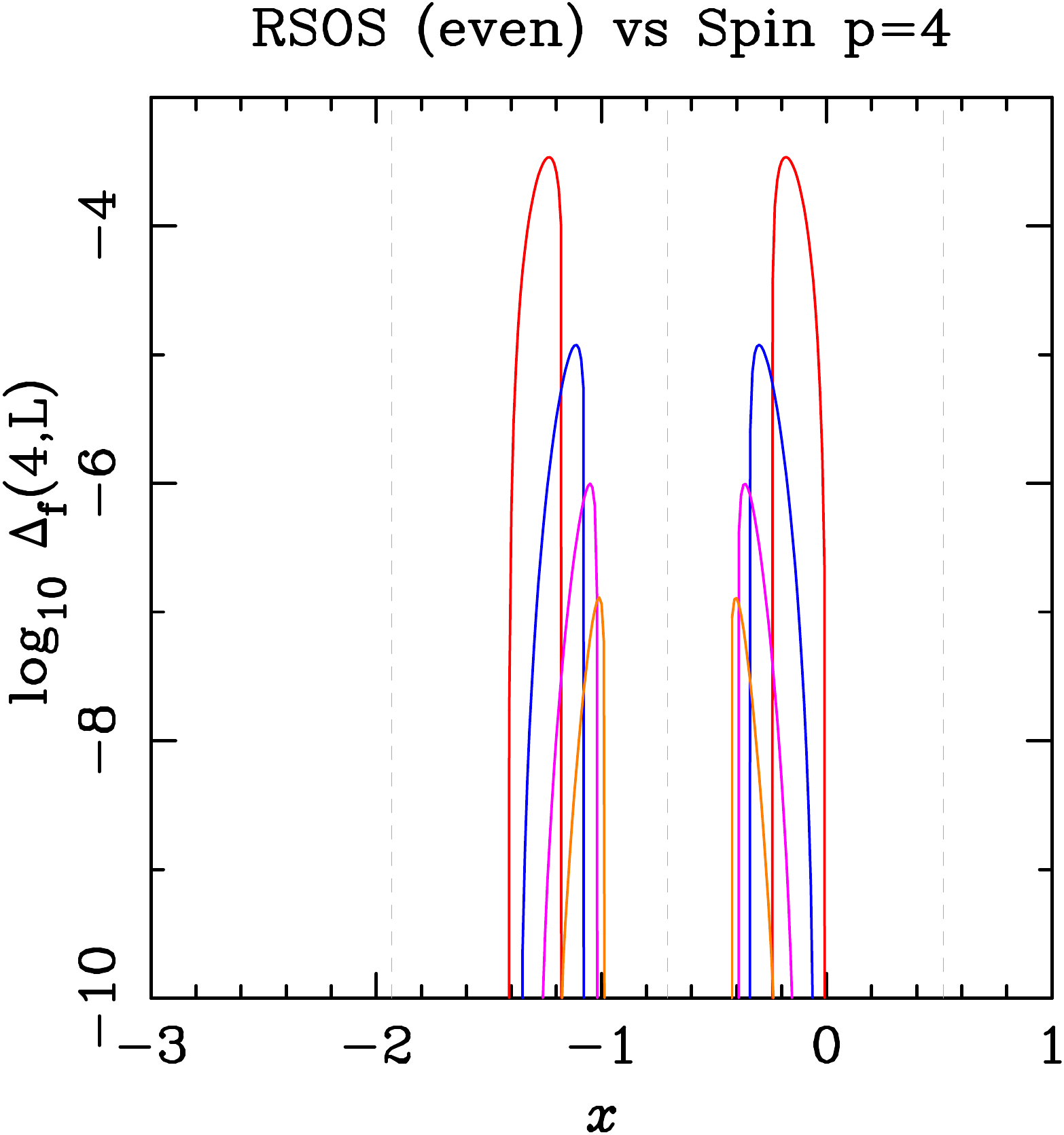} & 
  \includegraphics[width=200pt]{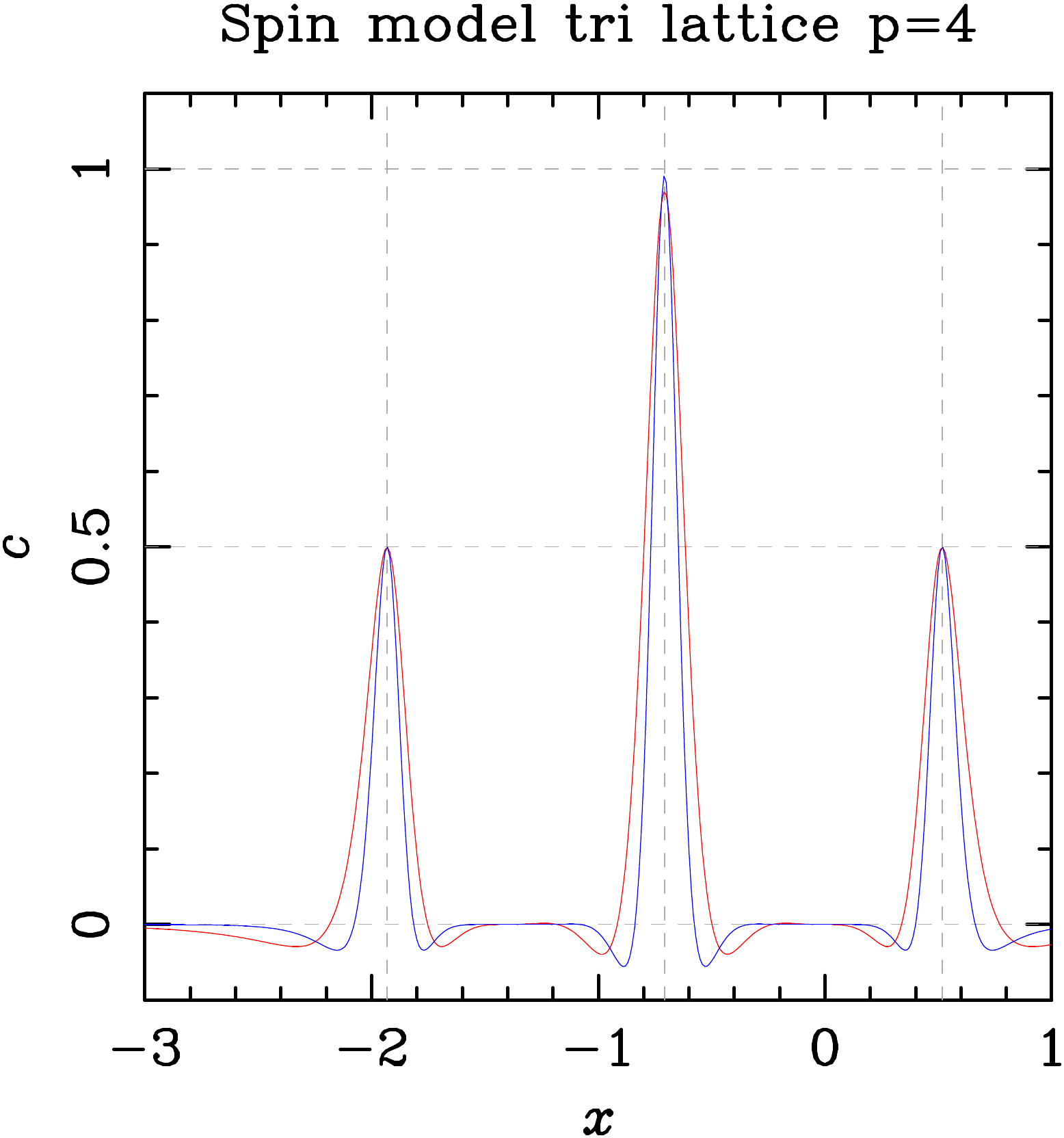} \\ 
  \qquad (a) & \qquad (b) \\
  \end{tabular}
\end{center}
\caption{
  Free energy and central charge for the Ising model on a triangular
  lattice with toroidal boundary conditions and widths $L \equiv 0 \bmod{3}$ 
  in the RSOS representation. 
  (a) Absolute difference $\Delta_f(4,L)$ [cf. \eqref{def.disc}] 
      between $f_L^\text{(spin)}$ and the free energy 
      $f_L^\text{(RSOS,even)}$ of the even sector of the RSOS model. (For the
      other RSOS sectors $\Delta_f(4,L)=0$.) We show the data for widths 
      $L=3$ (red), $L=6$ (blue), $L=9$ (pink), and $L=12$ (orange). 
  (b) Central charge obtained from the Ansatz \eqref{eq.CFT_Ansatz}. We
      display the results for $L_\text{min}=3$ (red) and 
      $L_\text{min}=6$ (blue). 
  The vertical dashed grey lines in both panels show the roots of Baxter's 
  cubic \eqref{cubic_tri}, and the horizontal lines in (b) mark the expected
  results $c=1/2$ and $c=1$.}
\label{fig:p=4}
\end{figure}

The single exception to this generic behaviour is the free energy 
coming from the even sector of the RSOS model 
(i.e., $\mathsf{T}_\text{RSOS}^\text{(even)}$). Notice that the configuration 
space in this RSOS sector is equivalent to that of an Ising model on the dual
hexagonal sublattice, and no symmetry has been taken into account in both
models. In any case, the RSOS free energy in this sector differs slightly 
from the one computed in the spin representation, and the
difference $\Delta_f(4,L)$ decreases with $L$. Roughly speaking,
as $L$ increases by $3$ units, the difference decreases by
one order of magnitude. (See figure~\ref{fig:p=4}(a) and
table~\ref{table.dis.p=4-6}.) It is worth mentioning that
$f_L^\text{(spin)}$ and $f_L^\text{(RSOS, even)}$ differ precisely at values
of $x$ where the model is not critical, as it is clear from 
figure~\ref{fig:p=4}.

We can go further and compare the full spectra of these various 
transfer matrices. We find that
\begin{equation}
 {\rm sp}\, \mathsf{T}^\text{(RSOS)}_L(x) \;\subset\; 
 \bigcup_{a=1,2} {\rm sp}\, \mathsf{T}^\text{(spin, $a$)}_L(x) \,,
\end{equation}
as has been checked explicitly for $L=3$ and $L=6$. This is again in 
agreement with section~\ref{sec:braidtranslation}. We note that the inclusion 
is strict (as expected by comparing the dimensions): the spin TM contains 
some eigenvalues which are not found in the RSOS representation.

We now compute the central charge from our finite-$L$ free-energy estimates
by using the standard CFT Ansatz
\begin{equation}
f_L(x)  \;=\; f_\text{bulk}(x) + \frac{\pi\, G\, c(x)}{6\, L^2} + 
                                 \frac{A(x)}{L^4} \,,
\label{eq.CFT_Ansatz} 
\end{equation}
where $G=\sqrt{3}/2$ is the geometric factor for a triangular lattice on a 
torus. To exhibit the residual finite-size-scaling (FSS) corrections, we 
will fit the data to the Ansatz \eqref{eq.CFT_Ansatz} for 
$L=L_\text{min},L_\text{min}+3,L_\text{min}+6$, and display the results as 
a function of $L_\text{min}$. These results are shown in 
figure~\ref{fig:p=4}(b).

%
%

The results for the central charge are not unexpected: the critical points
coincide with the solutions of the cubic \eqref{cubic_tri} for $Q=2$. There 
is a critical point in the ferromagnetic regime with $c_\text{FM}=1/2$. 
We also find a zero-temperature critical point in the AF regime 
\cite{Stephenson64}
with central charge $c=1$ \cite{BloteNightingale93} 
(see also \cite{BloteHilhorst82,NienhuisHilhorstBlote84}).
Finally, we also find a $c_\text{PF}=1/2$ critical point on the curve $v_{-}$, 
which lies in the unphysical region. Recall that this is expected, as there 
is numerical evidence \cite{JacobsenSaleur06} that $v_{-}(Q)$ for the 
triangular lattice belongs to the same universality class as $v_\text{AF}(Q)$ 
for the square lattice \cite{IkhlefJacobsenSaleur08,IkhlefJacobsenSaleur12}.%
\footnote{Our use of the subscripts FM and AF for the two $c=1/2$ points does
   not imply that the two CFT are different in any way. But we shall soon 
   see that the identification of $c_\text{FM}$ and $c_\text{PF}$ extends 
   to other values of $p$ as well.
}

%
%
\subsubsection{$Q=B_6=3$}
\label{sec:RSOS_p=6}

The dimensions of the TM in the spin, FK, and RSOS representations are 
displayed in table~\ref{table.dim.p=4-6}, where again those of the RSOS
case can be inferred from \eqref{dimT-RSOS}. In this case, we observe that
there is an extra state in $\mathsf{T}_\text{RSOS}^\text{(even)}$:
$\dim \mathsf{T}_\text{RSOS}^\text{(even)}-\dim \mathsf{T}_\text{spin}=1$.
The origin of this extra state can be found as follows: we split the RSOS 
variables in groups of two $g_i=h_{2i-1} h_{2i}$ for $1\le i\le L$. Then in 
the \emph{even} sector, each `word' $g_i$ should start with $2$ or $4$, 
finish with $1,3,5$, and respect the RSOS rule. Therefore there are $4$ 
possible RSOS words, and we assign each of them to a spin value: 
$21 \to 1$, $23\to 2$, $43\to 3$, \emph{and} $45\to 1$. 
There is a one-to-one relation between RSOS and spin configurations, 
with a \emph{single exception}: the spin configuration $11\ldots 1$ has two
RSOS counterparts $(2121\ldots 21)$ and $(4545\ldots 45)$. Indeed, we need
to assign $45\to 1$ in order to be able to express e.g. spin configurations
like $133\to (454343)$.

%
%
\begin{figure}[ht]
\begin{center}
  \begin{tabular}{cc}
  \includegraphics[width=200pt]{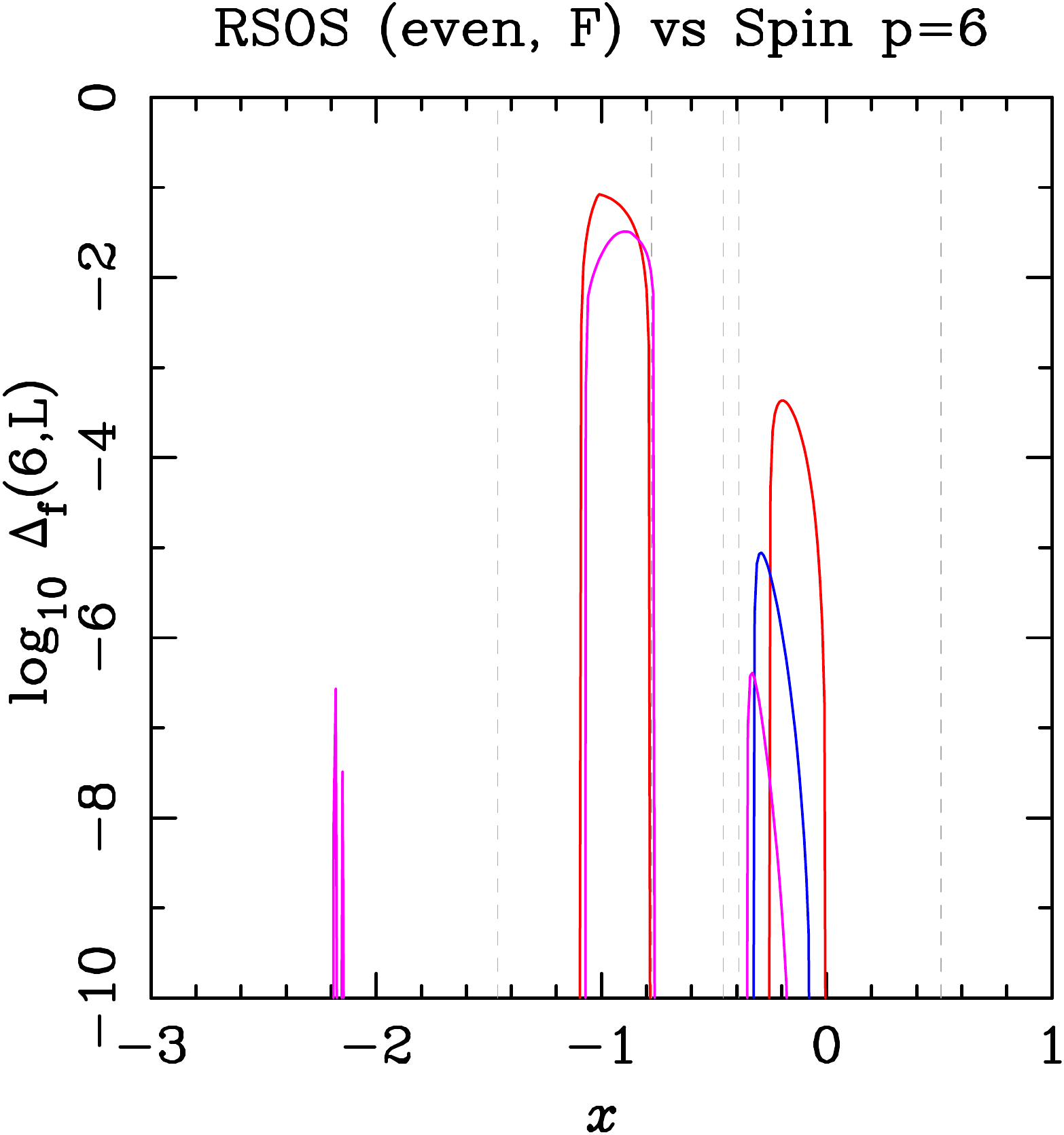} & 
  \includegraphics[width=200pt]{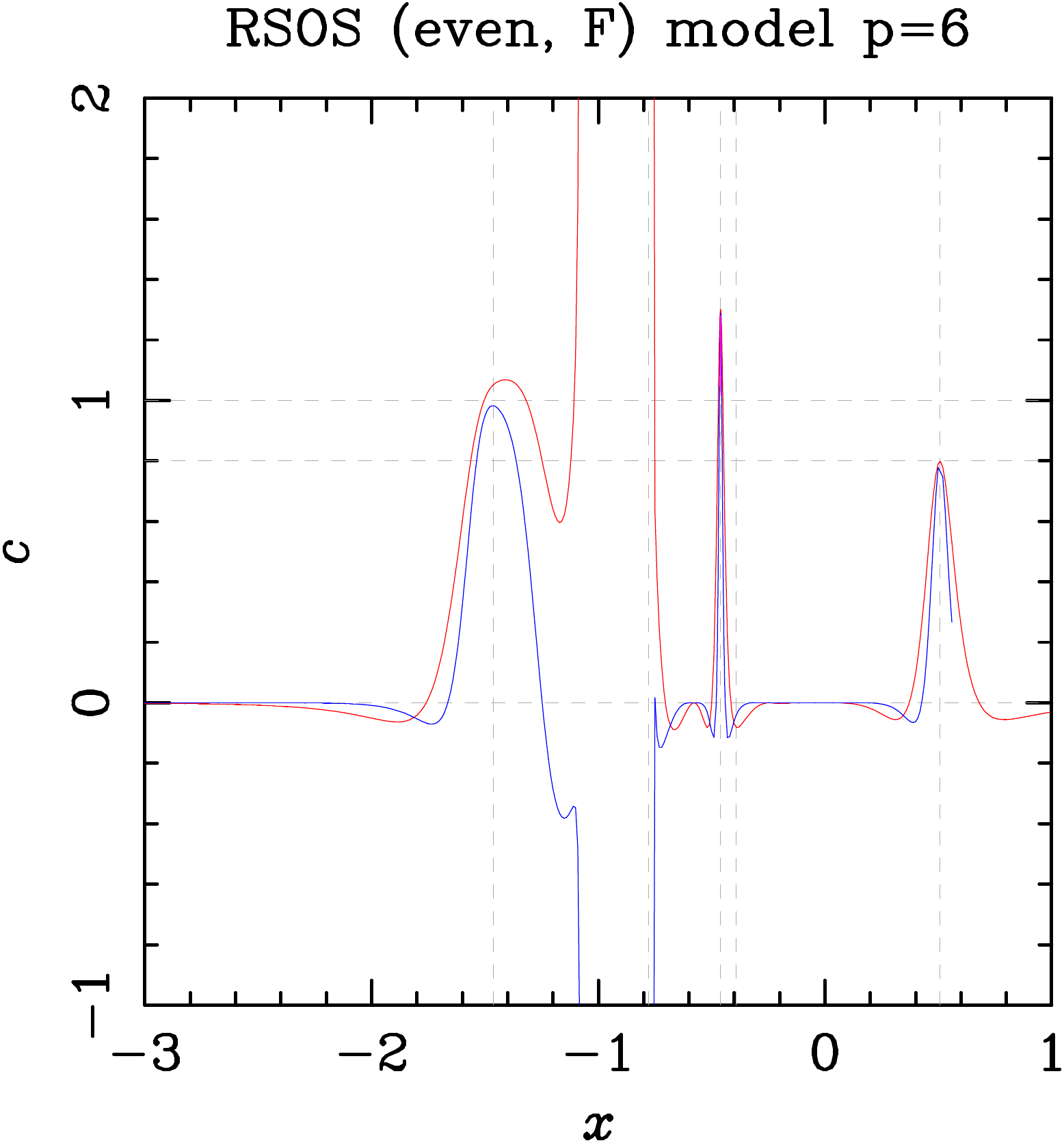} \\ 
  \qquad (a) & \qquad (b) \\
  \end{tabular}
\end{center}
\caption{
  Free energy and central charge for the 3-state Potts model on a triangular
  lattice with toroidal boundary conditions and widths $L \equiv 0 \bmod{3}$ 
  in the RSOS representation. 
  (a) Absolute difference $\Delta_f(6,L)$ [cf. \eqref{def.disc}] 
      between $f_L^\text{(spin)}$ and the free energy 
      $f_L^\text{(RSOS,even,F)}$ of the even sector of the RSOS model with
      all symmetries taken into account. We show the data for widths 
      $L=3$ (red), $L=6$ (blue), and $L=9$ (pink). 
  (b) Central charge obtained from the Ansatz \eqref{eq.CFT_Ansatz} and  
      the RSOS data in the (even,F) sector. We display the results for 
      $L_\text{min}=3$ (red), $L_\text{min}=6$ (blue), and 
      $L_\text{min}=9$ (pink).  
  The vertical dashed grey lines in both panels show the roots of Baxter's 
  cubic \eqref{cubic_tri}, the position of the first-order phase 
  transition at $x=-0.46011(12)$ \cite{ChangJacobsenSalasShrock04}, 
  and the position of the AF critical curve $x_\text{AF}\approx -0.39125$.
  The horizontal lines in (b) mark the expected results, 
  $c_\text{FM}=4/5$ and $c_\text{PF}=1$.
}
\label{fig:p=6}
\end{figure}

We have computed the free energy (per site) for the different RSOS sectors
defined in section~\ref{sec:RSOS}, as well as the free energy in the spin 
representation, for each pair $(p,L)=(6,L)$ with $L=3,6,9,12,15$.
Also in this case, the quantities $f_L^\text{(RSOS)}(p,x)$ and 
$f_L^\text{(spin)}(Q,v)$
can only be guaranteed to agree for $x > 0$, since the RSOS model contains 
now three different boundary conditions, corresponding to giving the 
weight $Q_1 = 3$, $Q_2 = 1$ or $Q_3 = 0$ to each wrapping cluster. Only the 
first of those corresponds to periodic boundary conditions in the spin 
representation.

The absolute differences $\Delta_f(6,L)$, defined by \eqref{def.disc},
are listed in table~\ref{table.dis.p=4-6}. This is now non-zero for all
symmetry sectors of the RSOS models. Notice
that for $L=12,15$ we only have data for $f^\text{(RSOS,even,F)}_L$, and in 
this case, only in the interesting ranges $x\in [-1.7,-0.4]$ and   
$x\in [-0.463,-0.456]$, respectively.
In figure~\ref{fig:p=6}(a) we show the differences between $f_L^\text{(spin)}$
and $f^\text{(RSOS,even,F)}_L$ for $L=3,6,9$. 

We have also compared the full spectra of the RSOS and spin transfer matrices,
finding the (sharp) inclusion
\begin{equation}
 {\rm sp}\, \mathsf{T}^\text{(RSOS)}_L(x) \;\subset\; 
 \bigcup_{a=1,2,3} {\rm sp}\, \mathsf{T}^\text{(spin, $a$)}_L \,.
\label{spectraQ3}
\end{equation}
On the right-hand side, $a=1$ corresponds to periodic
boundary conditions on the spins. To get the $a=2$ sector we must impose 
boundary conditions on the spins so that wrapping clusters in the 
corresponding FK model get the weight $Q_2 = 1$. We know
that the spin is constant throughout a cluster, implying that its weight is 
equal to the number of possible spin labels that it can support. If we 
permute two of the spin labels when crossing the
periodic direction, e.g., $(1,2,3) \to (2,1,3)$, a wrapping cluster can 
only support spins of the remaining third value, hence it will get the 
weight $1$ indeed. We can therefore identify the $a=2$ sector
with $\mathbb{Z}_2$-twisted boundary conditions. Similarly, if we permute 
all three spin labels cyclically, e.g., $(1,2,3) \to (2,3,1)$ across the seam, 
then wrapping clusters are disallowed altogether.
Thus we can identify the $a=3$ sector with $\mathbb{Z}_3$-twisted boundary 
conditions. We have constructed the corresponding TM in the spin 
representation and verified \eqref{spectraQ3} explicitly for sizes 
$L=3$ and $L=6$.

If we fit the RSOS free energy $f^\text{(RSOS,even,F)}_L$ to the
Ansatz \eqref{eq.CFT_Ansatz} and use the same protocol as for $Q=2$, 
we obtain the results displayed in figure~\ref{fig:p=4}(b). 
As expected, there is a peak at the ferromagnetic critical point with the
central charge $c_\text{FM}=4/5$ given by \eqref{cFM}. At $x=-0.46010(2)$
we observe a sharp peak attaining the value $c = 1.265(15)$.%
\footnote{These central values and error bars are estimated from the 
   following results for the position $x_*$ and the value $c_*$ of the peak 
   in the three-point fit for the central charge:
   For $L=3$, $x_* = -0.46019$ and $c_* = 1.3011$.
   For $L=6$, $x_* = -0.46012$ and $c_* = 1.2886$.
   And for $L=9$, $x_* = -0.46011$ and $c_* = 1.2796$.
}
This estimate for the central charge appears rather stable, but does not 
coincide with any known CFT possessing an $S_3$ symmetry. If this theory 
were conformal indeed, it might seem to challenge the works 
\cite{Adler95,ChangJacobsenSalasShrock04},
according to which the transition is (weakly) first order. But actually, 
having a discontinuity in the first derivative of the free energy (the 
hallmark of a first-order transition) is not incompatible with the 
transition being simultaneously second-order (and conformal):
indeed, this scenario was found for the AF transition in the 
square-lattice Potts model \cite{JacobsenSaleur06}. In any case, determining 
the exact nature of this transition would require further work; 
Density-matrix-renormalisation-group (DMRG) simulations for larger sizes 
would be one option.

In the (unphysical) region in between the central and lower branches of the 
cubic \eqref{cubic_tri}, we find large absolute values of the central charge, 
which are not expected to be physical at all. As a matter of fact, the fits
of the spin free energy $f^\text{(spin)}_L$ to the Ansatz \eqref{eq.CFT_Ansatz}
for $L=3,6,9$ give very similar results to those depicted in 
figure~\ref{fig:p=6}(b) for $L_\text{min}=3$, even in that region. 
These observations may imply that none of these models are well defined in 
this interval, or that we need data for larger widths in order to obtain 
meaningful physical results there.  

Finally, at the lower branch $x_{-}=v_{-}/\sqrt{3} \approx -1.4619$, we find 
a peak with a central charge $c=0.98(2)$. This is compatible with the expected
result $c_\text{PF} = 1$.

%
%
\subsection{Results for non-integer values of $Q$} 
\label{sec:RSOS_non-integerQ}
 
In this section we will discuss the cases $p=5,7,8$. As $Q$ is not an 
integer, we do not have immediate access to the spin representation (but
see also \cite{DJS10-dw}). We can however still compare our results 
with those of the FK representation.
Notice that for any of the three values of $p$, we do not find any vanishing
amplitude and/or eigenvalue cancellation for the FK model, at least not 
for $L=3$, and so the non-local aspects of FK clusters would seem to remain 
present in the corresponding partition function on the torus. Therefore, it 
is not clear to us to which extent the FK and RSOS models can be considered 
equivalent for such values of $p$. In any case, we shall henceforth pursue 
the study of the RSOS model in its own right.

In table~\ref{table.dim.p=578} we show the dimensions of the TM in the FK
representation and in the RSOS representation in the even sector, and in
the same sector with all symmetries taken into account. Notice that when 
$p$ is odd, then the symmetry $h\to p-h$ mixes both sectors, so in this case
we only consider the rotational symmetry. On the contrary, when $p$ is even,
we consider all (i.e., spin reversal and rotational) symmetries. 
The dimensions of $\mathsf{T}^\text{(even)}_\text{RSOS}$ and
$\mathsf{T}^\text{(odd)}_\text{RSOS}$ agree of course
with \eqref{dimT-RSOS}.

%
%
\begin{table}[h]
\centering
\begin{tabular}{rrrrr}
\hline\hline\\[-2mm]
$p$ & $L$ & $\dim \mathsf{T}_\text{FK}$
          & $\dim \mathsf{T}_\text{RSOS}^\text{(even)}$
          & $\dim \mathsf{T}_\text{RSOS}^\text{(even, F)}$\\[2mm]
\hline \\[-2mm]
5  & $3$ & 30 &   18 &    8 \\
   & $6$ &    &  322 &   60 \\
   & $9$ &    & 5778 &  648 \\
   &$12$ &    & 103682 &      \\
\hline \\[-2mm]
7  & $3$ & 30 &   38 &   16 \\
   & $6$ &    & 1186 &  210 \\
   & $9$ &    &40169 & 4475 \\
   & $12$ & & 1373466 & \\
\hline \\[-2mm]
8  & $3$ & 30 &   48 &   10 \\
   & $6$ &    & 1648 &  148 \\
   & $9$ &    &63552 & 3538 \\
   & $12$ & & 2513024 & \\
\hline\hline
\end{tabular}
\caption{%
  Dimension of the TM for the Potts model with $Q=B_p$ ($p=5,7,8$) 
  on a triangular-lattice strip graph on the torus in the FK representation 
  $\mathsf{T}_\text{FK}$, and in the RSOS representation in the \emph{even} 
  sector $\mathsf{T}_\text{RSOS}^\text{(even)}$, and in the same 
  sector with all symmetries $\mathsf{T}_\text{RSOS}^\text{(even, F)}$. 
  Note that $\dim \mathsf{T}_\text{RSOS}^\text{(even)} =
             \dim \mathsf{T}_\text{RSOS}^\text{(odd)}$.
}
\label{table.dim.p=578}
\end{table}

%
%
\subsubsection{$Q=B_5$.} \label{sec:RSOS_p=5}

\begin{figure}[ht]
\begin{center}
  \begin{tabular}{cc}
  \includegraphics[width=200pt]{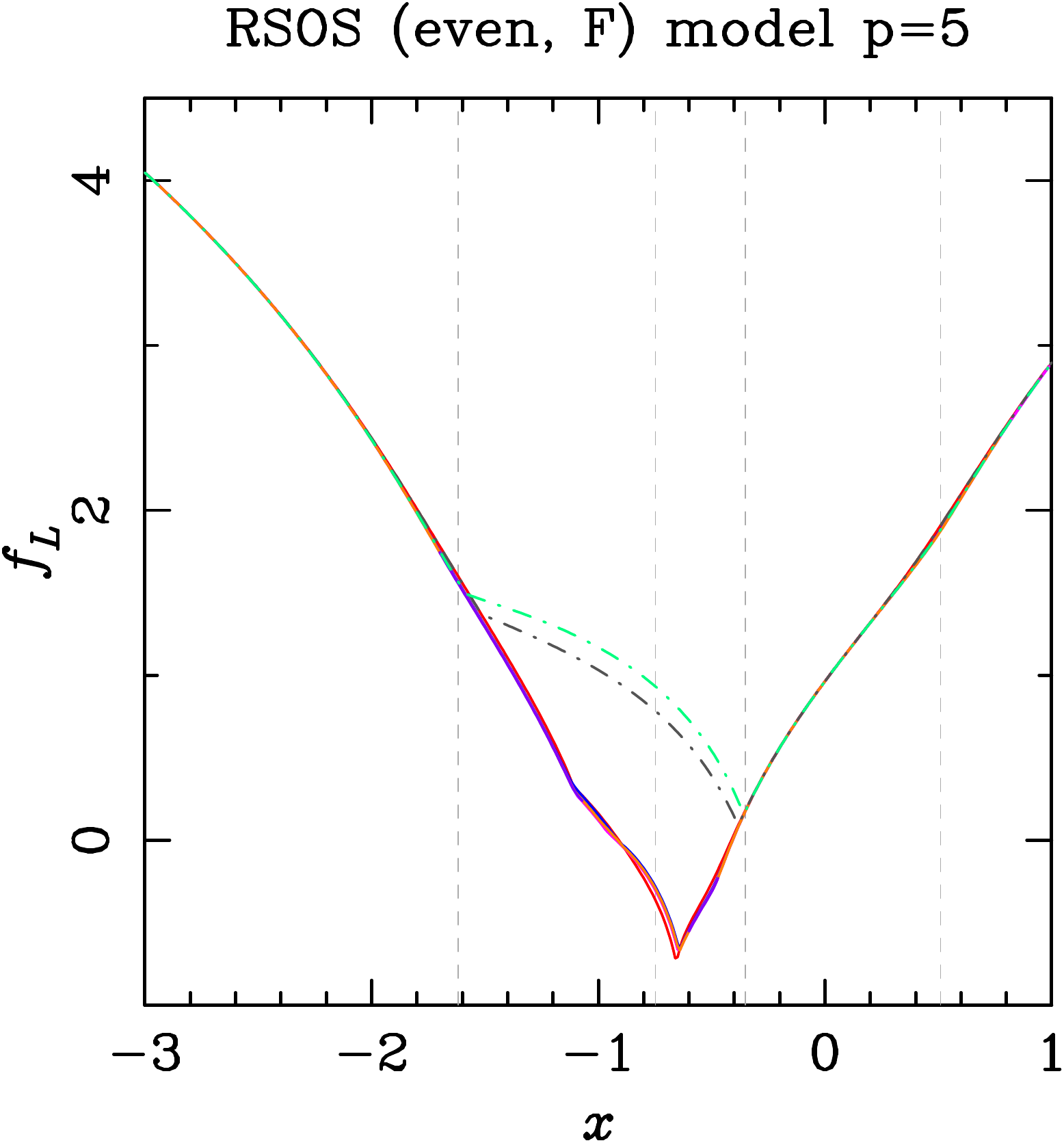} & 
  \includegraphics[width=200pt]{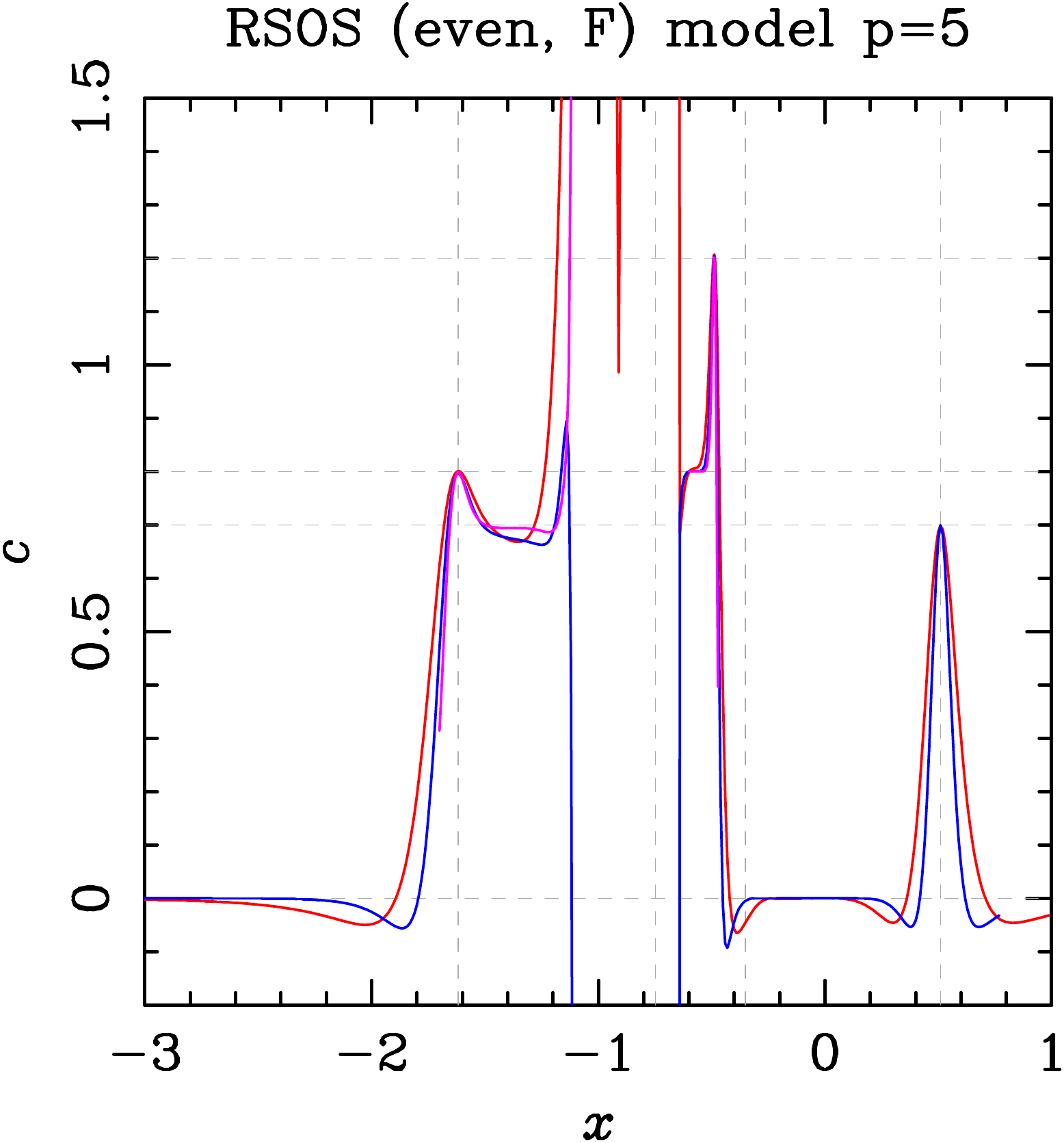} \\ 
  \qquad (a) & \qquad (b) \\
  \end{tabular}
\end{center}
\caption{
  Free energy and central charge for the RSOS model with $p=5$ on a 
  triangular lattice with toroidal boundary conditions and widths 
  $L \equiv 0 \bmod{3}$. 
  (a) The solid curves depict the free energy $f_L^\text{(RSOS,even,F)}$ 
      of the even sector of the RSOS model with all symmetries taken into 
      account. We show the data for widths $L=3$ (red), $L=6$ (blue),  
      $L=9$ (pink), $L=12$ (orange), and $L=15$ (violet).
      The dot-dashed curves show the 
      free energy in the FK representation $f_L^\text{(FK)}$ for 
      widths $L=3$ (violet), and $L=6$ (dark grey). 
  (b) Central charge obtained from the Ansatz \eqref{eq.CFT_Ansatz} and  
      the RSOS data in the (even, F) sector. We display the results for 
      $L_\text{min}=3$ (red), $L_\text{min}=6$ (blue), and $L=9$ (pink).  
  The vertical dashed grey lines in both panels show the roots of Baxter's 
  cubic \eqref{cubic_tri}, and the position of the AF critical curve 
  $x_\text{AF}\approx -0.35065$. The horizontal lines in (b) mark the expected
  results $c_\text{FM}=7/10$ and $c_\text{PF}=4/5$, 
  as well as our conjectured new value, $c=6/5$.
}
\label{fig:p=5}
\end{figure}

In this case we find that the free energies satisfy 
$f^\text{(RSOS,even,F)}_L=f^\text{(RSOS,odd,F)}_L$, so we will consider the
former in our analysis. If we compare $f^\text{(RSOS,even,F)}_L$ with the
free energy of the Potts model with $Q=B_5$ in the FK representation
$f^\text{(FK)}_L$, we see in figure~\ref{fig:p=5}(a) that they disagree
strongly in an interval that starts close to the lower branch of the
cubic $v_{-}$ \eqref{cubic_tri} and finish 
at the position of the AF critical curve $v_\text{AF} \approx -0.35065$.
In the remainder
of the interval $x\in [-3,1]$, the agreement between these two free energies
is perfect (within the numerical precision). Indeed, for $x\ge 0$ this is
expected, since then the FK representation has a probabilistic interpretation,
and we can argue (like in \cite{JacobsenRichard05}) that these two 
models should be equivalent in the ferromagnetic regime.   

%
%

\begin{figure}[ht]
\begin{center}
  \begin{tabular}{cc}
  \includegraphics[width=200pt]{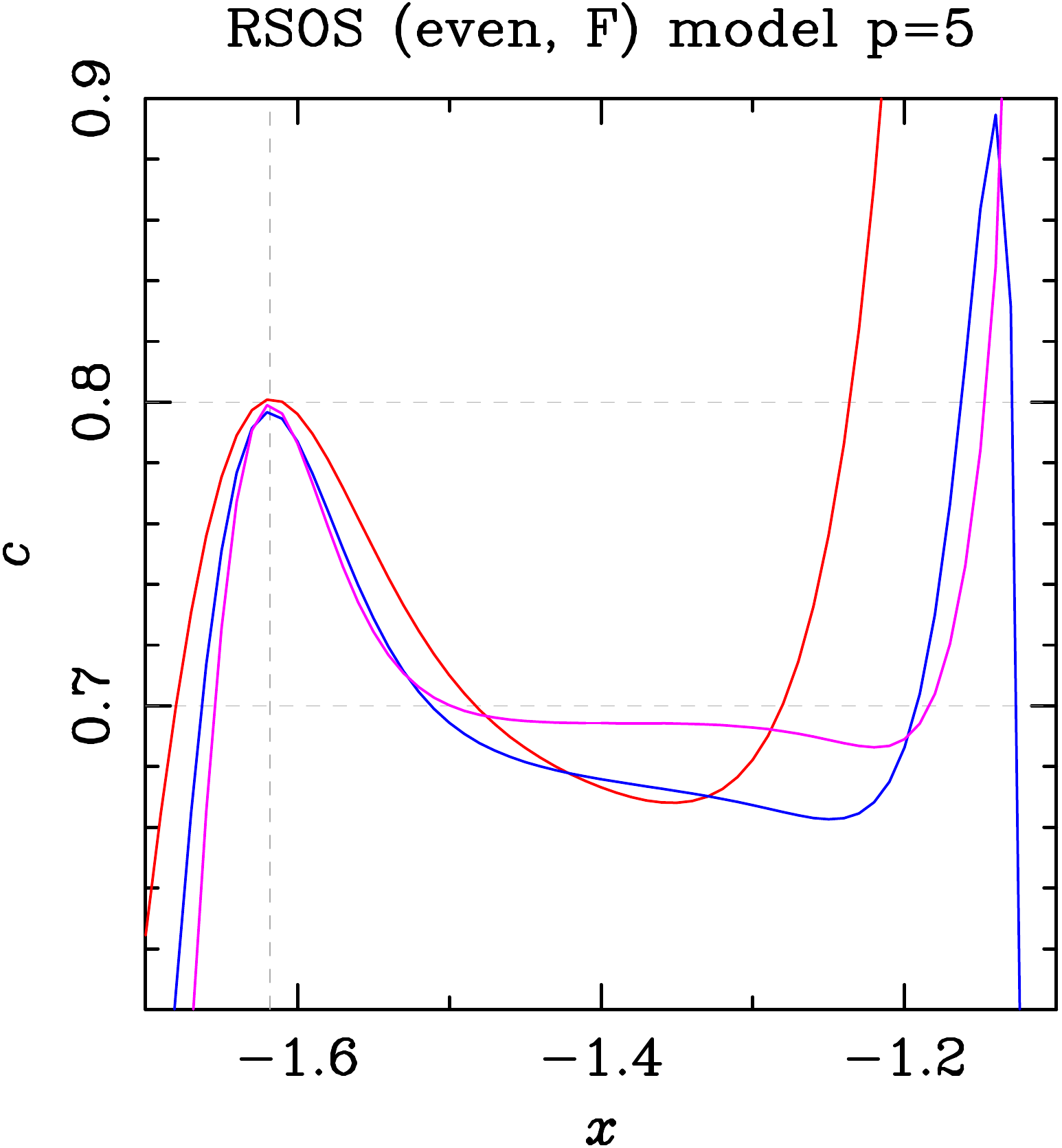} & 
  \includegraphics[width=213pt]{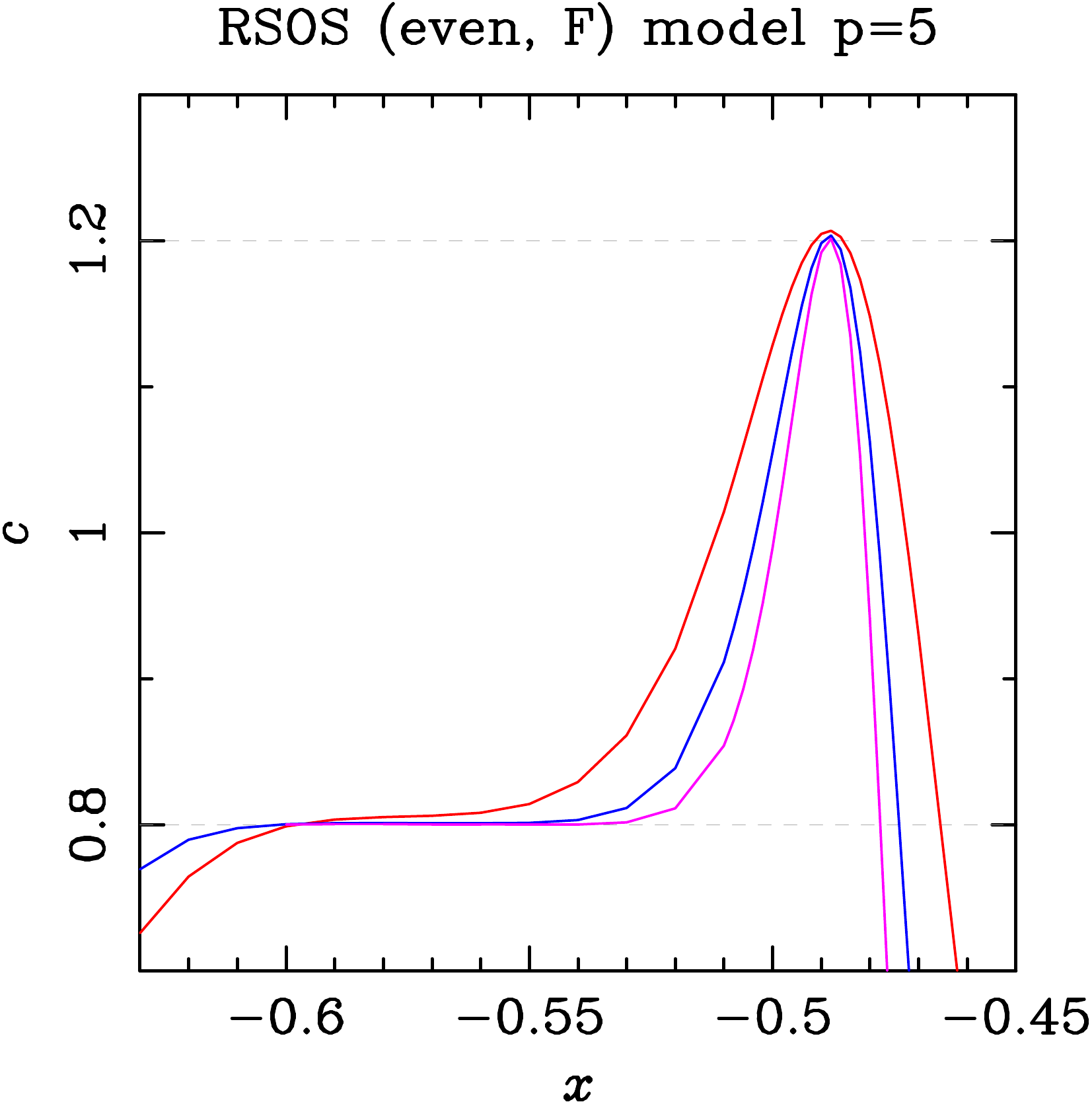} \\ 
  \qquad (a) & \quad (b) \\
  \end{tabular}
\end{center}
\caption{
  Zooms of figure~\ref{fig:p=5}(b) close to the two most interesting regions:
  $x \in [-1.7,-1.1]$ (a) and $x \in [-0.5,-0.45]$ (b). In each panel, we 
  show the central charge obtained from the Ansatz \eqref{eq.CFT_Ansatz} 
  with $L_\text{min}=3$ (red), $L_\text{min}=6$ (blue), 
  and $L_\text{min}=9$ (pink).    
  Vertical and horizontal dashed grey lines have the same meaning as in 
  figure~\ref{fig:p=5}.
}
\label{fig:p=5_zoom}
\end{figure}

We have fitted the free-energy data $f^\text{(RSOS,even,F)}_L$ to the
CFT Ansatz \eqref{eq.CFT_Ansatz} and used the same protocol as for the 
other cases. The results are displayed in figure~\ref{fig:p=5}(b).
As expected, we find a peak at the ferromagnetic critical point with the
central charge $c_\text{FM}=7/10$. A closer look at this
figure reveals four other interesting regions:
\setcounter{footnote}{0}
\begin{enumerate}
  \item A broad peak close to $x_{-}= v_{-}/\sqrt{B_5} \approx -1.6180$ with 
        central charge $c=0.797(3)$. This is in perfect agreement with
        the expected result $c_\text{PF} = 4/5$. 
        [See figure~\ref{fig:p=5_zoom}(a)].
  \item A broad plateau in the interval $x \in (x_{-},-1.14]$ with a constant
        central charge $c=0.70(1)$.
        [See figure~\ref{fig:p=5_zoom}(a)].
  \item A narrow plateau in the interval $x\in [-0.61,-0.57]$ with
        a constant central charge $c=0.800(2)$.
        [See figure~\ref{fig:p=5_zoom}(b)].
  \item A narrow peak at $x=-0.4883(1)$ with central charge $c=1.198(2)$.
        [See figure~\ref{fig:p=5_zoom}(b)].%
\footnote{%
   These central values and error bars are estimated from the 
   following results for the position $x_*$ and the value $c_*$ of the peak 
   in the three-point fit for the central charge:
   For $L=3$, $x_* = -0.48833$ and $c_* = 1.2069$.
   For $L=6$, $x_* = -0.48828$ and $c_* = 1.2038$.
   And for $L=9$, $x_* = -0.48829$ and $c_* = 1.2016$.
}
\end{enumerate} 

%
%

We conjecture that the exact values of the central charge on the two 
plateaux, (ii) and (iii), are $c=7/10$ and $c=4/5$ respectively. Note 
that these values coincide with
$c_\text{FM}$ and $c_\text{PF}$, respectively, and hence we can presumably
identify them with the corresponding CFT. Each plateau
appears to be controlled by an attractive fixed point of the indicated nature, 
at it would be interesting to examine further why the temperature variable 
$x$ does not couple to any of the relevant operators present in those 
fixed-point theories. It would also be interesting to study further the precise
nature of the flow from the $c=1.198(2)$ theory (iv) towards the $c=4/5$
plateau (iii), and the flow from the $c_\text{PF}=4/5$ theory (i) towards 
the $c=7/10$ plateau (ii).

We further conjecture that the exact value of the central charge at the 
peak (iv) is $c = 6/5$. Perhaps we can split this result as $6/5 = 1/2 + 7/10$,
so that this model corresponds to a model belonging to the same universality 
class as the ferromagnetic one plus an Ising fermion. The flow from the peak 
(iv) towards the fixed point controlling the plateau (iii) could then be 
characterised by this fermion becoming massive.

Notice that the result $c_\text{PF} = 4/5$ at the peak (i) can also be 
produced by evaluating $c_\text{FM}$ with $p \to p+1$. Therefore we can 
speculate that the flow from (i) to (ii) is just the usual flow between 
minimal models, ${\cal M}_p \to {\cal M}_{p-1}$.

Finally, it should be remarked that the zero-temperature point in 
the AF regime corresponds to $x=-1/\sqrt{B_5}\approx -0.6180$, which is very 
close to the lower bound of the interval defining the plateau (ii). It is 
possible that this value is indeed the termination point of the plateau.
In addition, the region containing the interval $x\in [-1.1,-0.8]$ shows
estimates for the central charge that are rather large in absolute value, and
hence presumably unphysical or at least non-critical, similar to the behavior 
we found for $p=6$ in approximately the same region. 

%
%
\begin{figure}[ht]
\begin{center}
  \begin{tabular}{cc}
  \includegraphics[width=200pt]{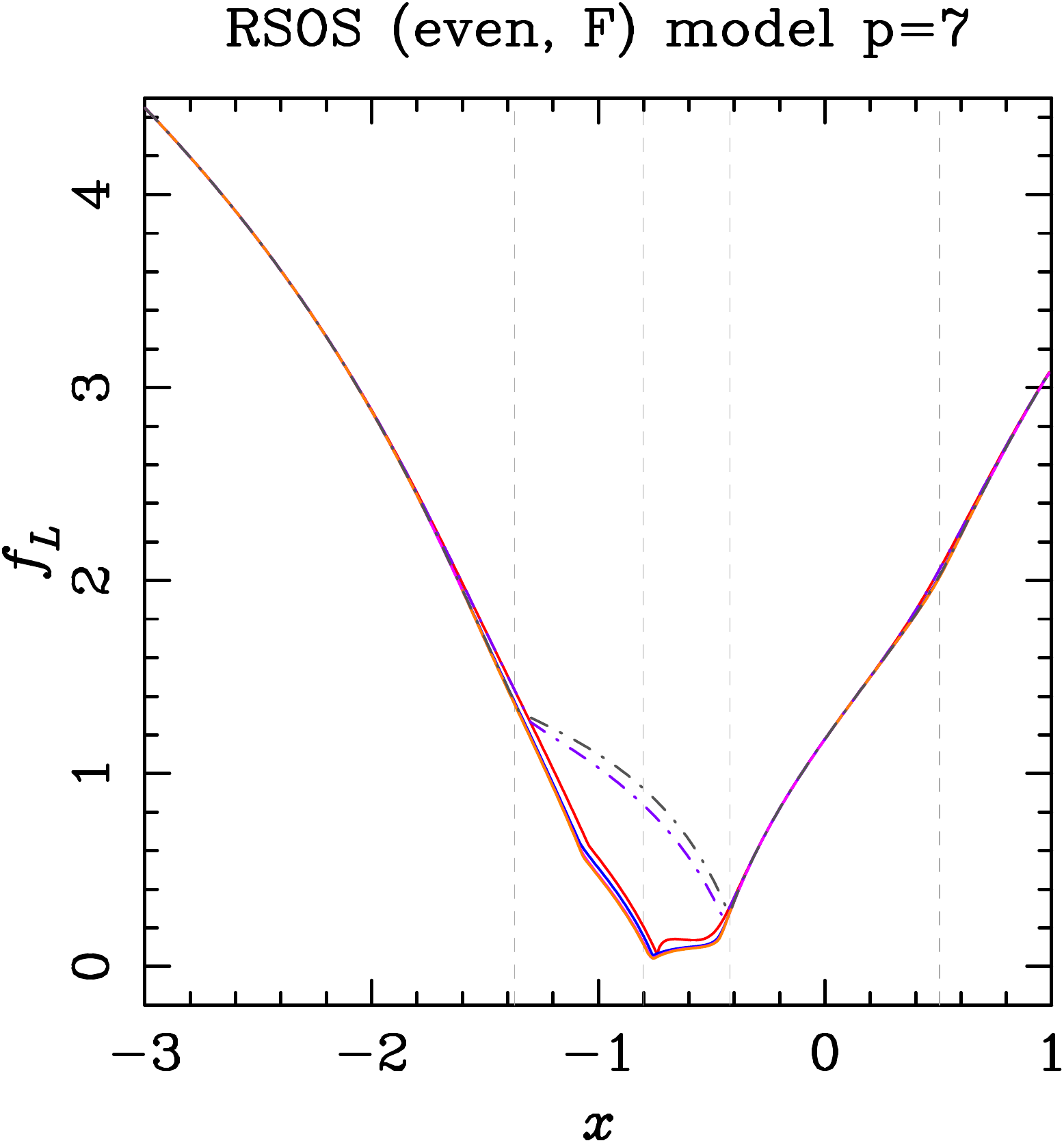} & 
  \includegraphics[width=200pt]{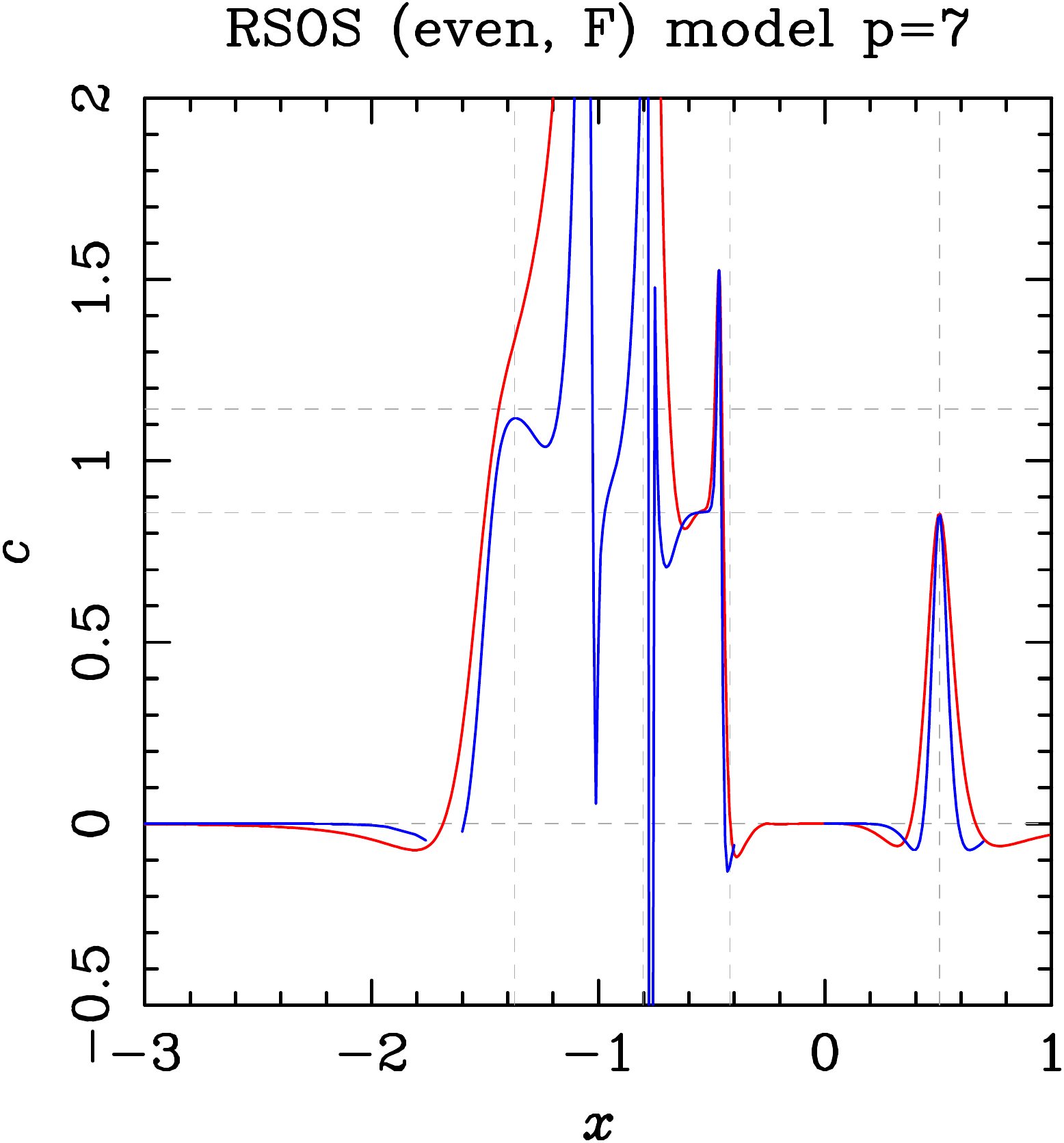} \\ 
  \qquad (a) & \qquad (b) \\
  \end{tabular}
\end{center}
\caption{
  Free energy and central charge for the RSOS model with $p=7$ on a 
  triangular lattice with toroidal boundary conditions and widths 
  $L \equiv 0 \bmod{3}$. 
  (a) The solid curves depict the free energy $f_L^\text{(RSOS,even,F)}$ 
      of the even sector of the RSOS model with all symmetries taken into 
      account. We show the data for widths $L=3$ (red), $L=6$ (blue),  
      $L=9$ (pink), and $L=12$ (orange). The dot-dashed curves show the 
      free energy in the FK representation $f_L^\text{(FK)}$ for 
      widths $L=3$ (violet), and $L=6$ (dark grey). 
  (b) Central charge obtained from the Ansatz \eqref{eq.CFT_Ansatz} and  
      the RSOS data in the (even, F) sector. We display the results for 
      $L_\text{min}=3$ (red) and $L_\text{min}=6$ (blue).  
  The vertical dashed grey lines in both panels show the roots of Baxter's 
  cubic \eqref{cubic_tri}, 
  and the position of the AF critical curve $x_\text{AF}\approx -0.41812$. 
  The horizontal lines in (b) mark the expected results,
  $c_\text{FM}=6/7$ and $c_\text{PF}=8/7$.
}
\label{fig:p=7}
\end{figure}

%
%
\subsubsection{$Q=B_7$} \label{sec:RSOS_p=7}

The situation is very similar to the $p=5$ case. Again we find that
$f^\text{(RSOS,even,F)}_L=f^\text{(RSOS,odd,F)}_L$. If we compare 
any of them with the free energy of the Potts model with $Q=B_7$ in 
the FK representation $f^\text{(FK)}_L$, we see in figure~\ref{fig:p=7}(a)  
that they disagree in the interval $x\in [x_{-},x_\text{AF}]$, where 
$x_{-}=v_{-}/\sqrt{B_7}\approx -1.3686$, and $x_\text{AF}\approx -0.41812$.

\begin{figure}[htb]
\begin{center}
  \begin{tabular}{cc}
  \includegraphics[width=200pt]{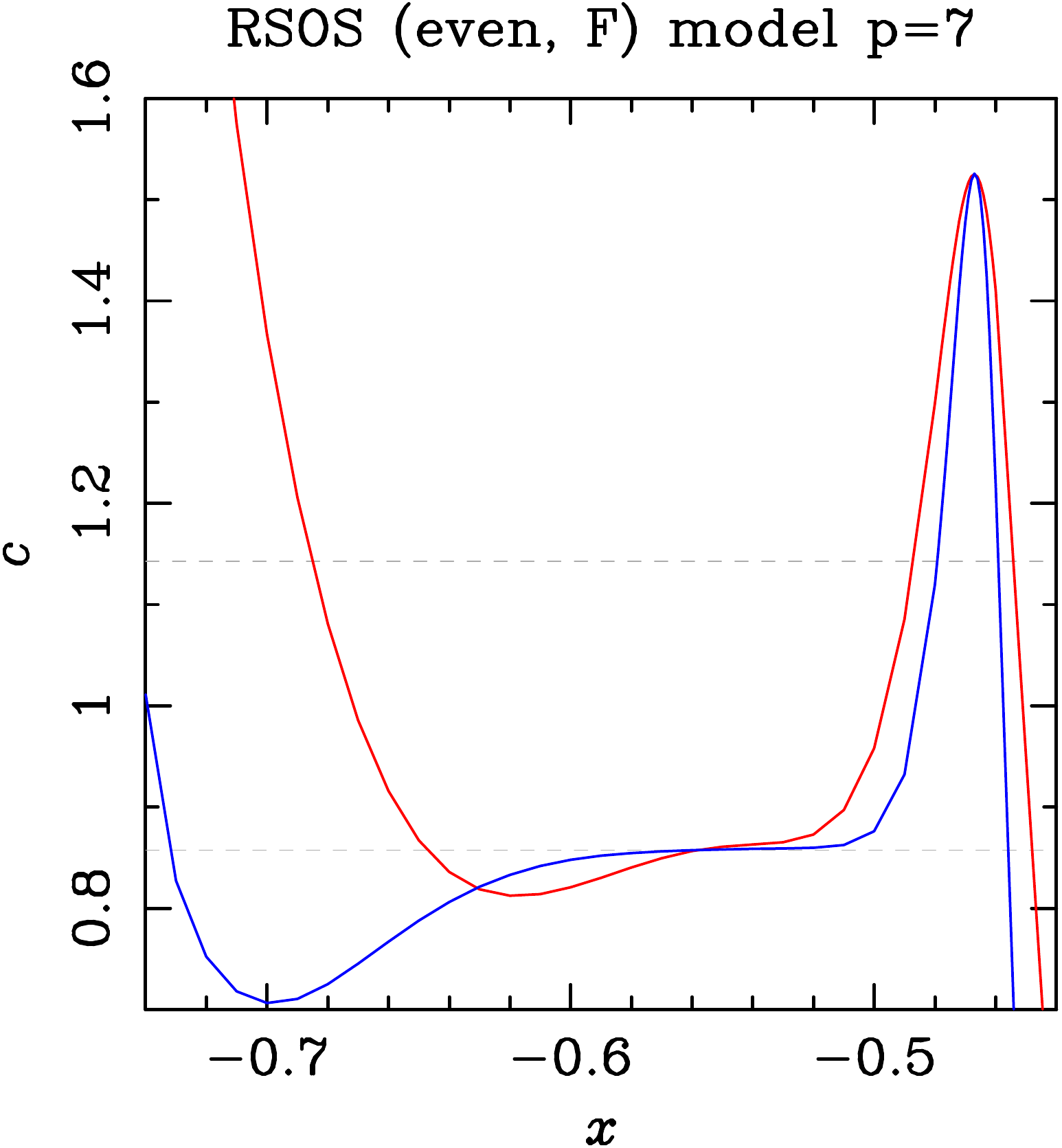} & 
  \includegraphics[width=206pt]{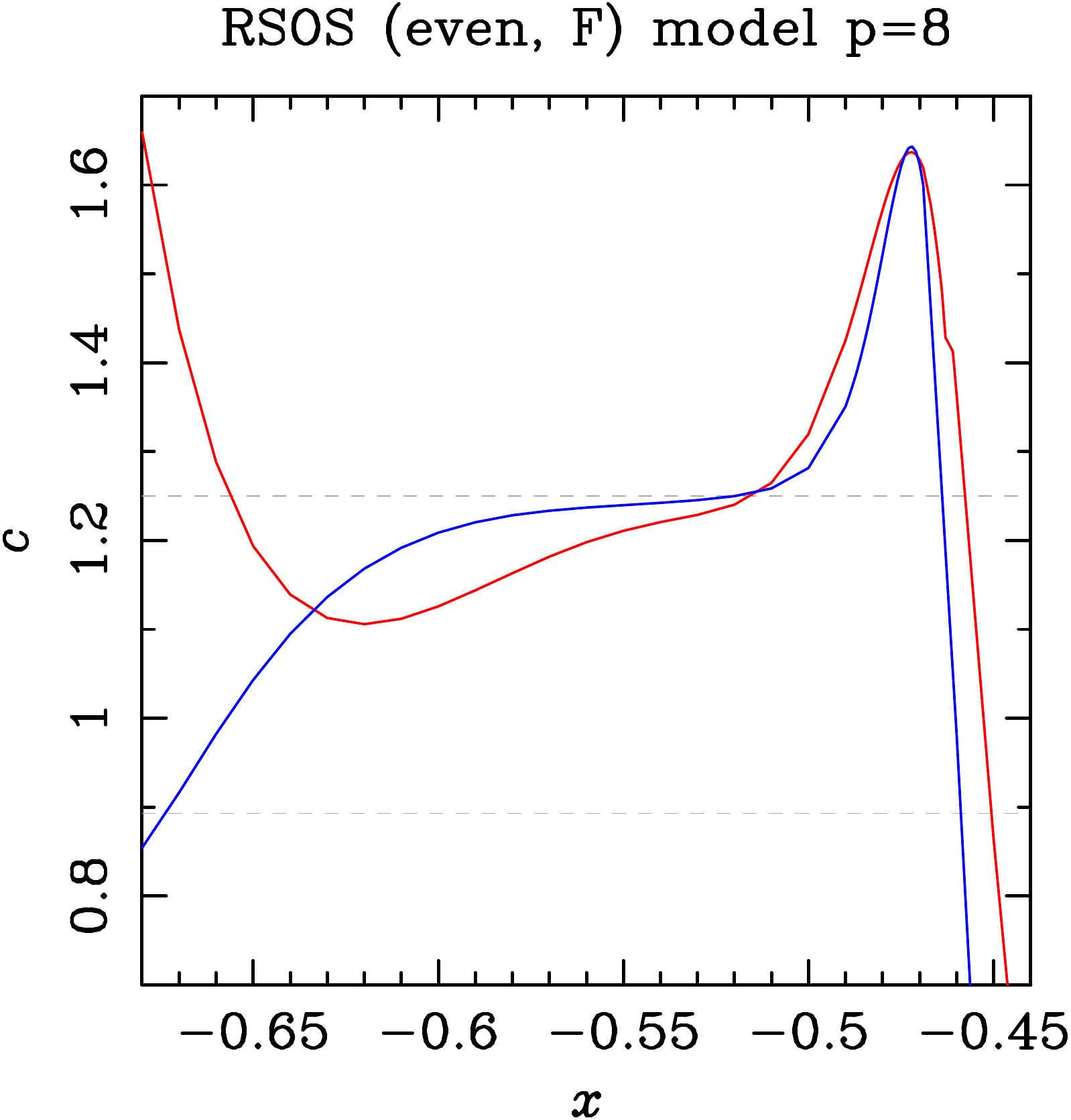} \\ 
  \qquad (a) & \quad\; (b) \\
  \end{tabular}
\end{center}
\caption{
  Zooms of figure~\ref{fig:p=7}(b) in the interval $x \in [-0.74,-0.44]$ (a),
  and of figure~\ref{fig:p=8}(b) in the interval $x \in [-0.68,-0.44]$ (b).
  In each panel, we show the central charge obtained from the Ansatz 
  \eqref{eq.CFT_Ansatz} 
  with $L_\text{min}=3$ (red), $L_\text{min}=6$ (blue), 
  and $L_\text{min}=9$ (pink).    
  Vertical and horizontal dashed grey lines have the same meaning as in 
  figure~\ref{fig:p=5}.
}
\label{fig:p=78_zoom}
\end{figure}

We have fitted the free-energy data $f^\text{(RSOS,even,F)}_L$ to the
CFT Ansatz \eqref{eq.CFT_Ansatz}; the results are displayed in 
figure~\ref{fig:p=7}(b). In addition to the expected peak at the 
ferromagnetic critical point with the usual central charge 
$c_\text{FM}=6/7\approx 0.8571$, we find a similar structure to that of 
the $p=5$ case:
\begin{enumerate}
 \item A broad peak close to $x_{-}= v_{-}/\sqrt{B_7}$ with 
       central charge $c=1.12(1)$. This agrees with the general 
       result $c_\text{PF} = 8/7 \approx 1.1429$.
 \item Just to the right of that peak, it is possible that a plateau is 
       nascent in the $L=12$ data, but it would clearly require larger 
       system sizes to confirm the reality of this feature.
 \item A very narrow plateau in the interval $x\in [-0.58,-0.51]$ with
       an approximately constant central charge $c=0.859(2)$. This value 
       is compatible with $c_\text{FM} = 6/7 \approx 0.8571$.
       [See figure~\ref{fig:p=78_zoom}(a)].
  \item A narrow peak at $x=-0.4669(2)$ with central charge $c \approx 1.53$. 
        Given the scarcity of available sizes, it does not appear reasonable 
        to give any error bar on this result, or even claim convergence 
        to a finite value. In any case, there is 
        no obvious candidate CFT with this value of the central charge.
        Notice that this peak is below our estimate for the AF critical 
        curve $x_\text{AF}\approx -0.41812$. 
        (See figure~\ref{fig:p=7}).
        
\end{enumerate} 

The lower bound of the AF regime $x=-1/\sqrt{B_7}\approx -0.5550$ is 
again rather close to the lower end of the plateau (ii).

%
%

%
%
\begin{figure}[htb]
\begin{center}
  \begin{tabular}{cc}
  \includegraphics[width=200pt]{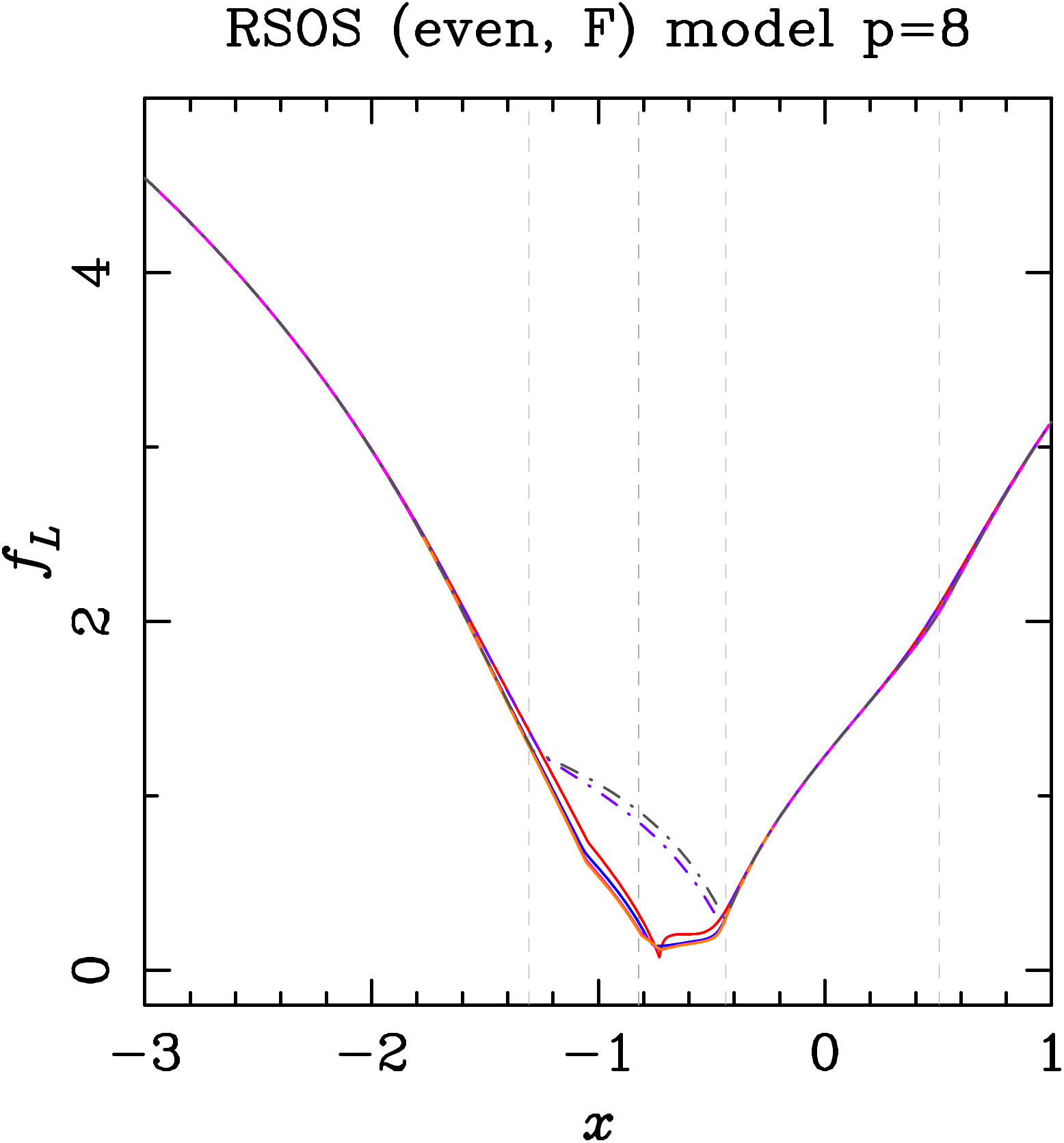} & 
  \includegraphics[width=200pt]{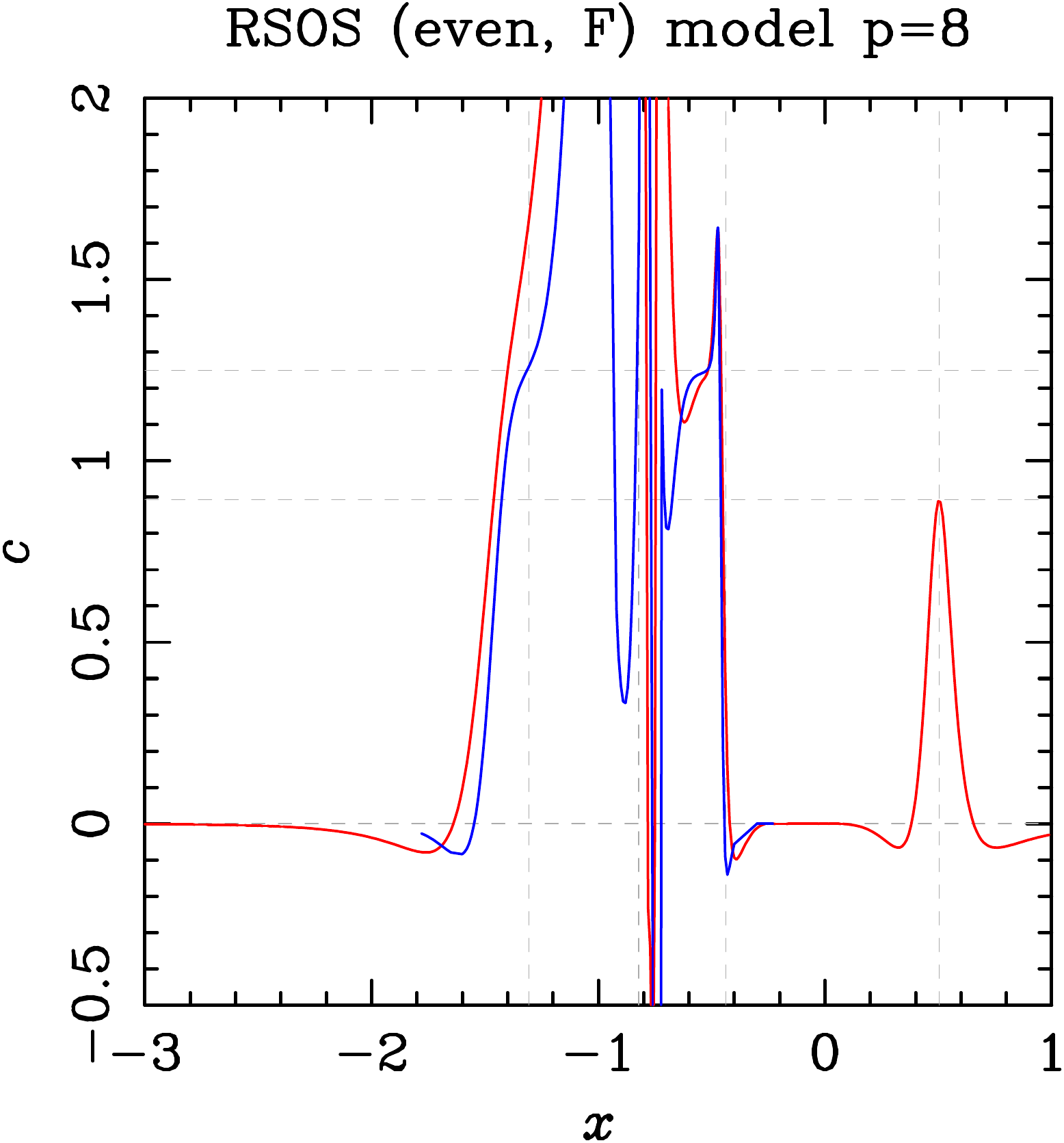} \\ 
  \qquad (a) & \qquad (b) \\
  \end{tabular}
\end{center}
\caption{
  Free energy and central charge for the RSOS model with $p=8$ on a 
  triangular lattice with toroidal boundary conditions and widths 
  $L \equiv 0 \bmod{3}$. 
  (a) The solid curves depict the free energy $f_L^\text{(RSOS,even,F)}$ 
      of the even sector of the RSOS model with all symmetries taken into 
      account. We show the data for widths $L=3$ (red), $L=6$ (blue),  
      $L=9$ (pink), and $L=12$ (orange). The dot-dashed curves show the 
      free energy in the FK representation $f_L^\text{(FK)}$ for 
      widths $L=3$ (violet), and $L=6$ (dark grey). 
  (b) Central charge obtained from the Ansatz \eqref{eq.CFT_Ansatz} and  
      the RSOS data in the (even, F) sector. We display the results for 
      $L_\text{min}=3$ (red) and $L_\text{min}=6$ (blue).  
  The vertical dashed grey lines in both panels show the roots of Baxter's 
  cubic \eqref{cubic_tri}, and the position of the AF critical curve $x_\text{AF}\approx -0.43783$. 
  The horizontal lines in (b) mark the expected results,
  $c_\text{FM}=25/28$ and $c_\text{PF}=5/4$.
}
\label{fig:p=8}
\end{figure}

%
%
\subsubsection{$Q=B_8$} \label{sec:RSOS_p=8}

Again, the situation is very similar to the previous two cases ($p=5$ and 
$p=7$). We find that
$f^\text{(RSOS,even,F)}_L=f^\text{(RSOS,odd,F)}_L$, and if we compare 
any of them with the free energy of the Potts model with $Q=B_8$ in 
the FK representation $f^\text{(FK)}_L$, we see in figure~\ref{fig:p=8}(a)  
that they strongly disagree in the interval $x\in [x_{-},x_\text{AF}]$, where 
$x_{-}=v_{-}/\sqrt{B_8}\approx -1.3066$, and $x_\text{AF}\approx -0.43783$ 
is our estimate for the position of the AF critical curve. 
There is another interval where we
found differences, $x\in [-0,33,-0.05]$, but in this case the differences
decrease as $L$ increases. 

We have fitted the free-energy data $f^\text{(RSOS,even,F)}_L$ to the
CFT Ansatz \eqref{eq.CFT_Ansatz}; the results are displayed in 
figure~\ref{fig:p=8}(b). In this case, the results are less clear than
for $p=5,7$; the FSS corrections seem to be stronger, so we would need
larger sizes to draw more definitive conclusions. 
Indeed, we find a ferromagnetic peak with the expected central charge 
$c=25/28\approx 0.8929$. Although there is no trace of the
peak at the lower branch of the cubic $x_{-} \approx -1.3066$, the 
curve for the largest system sizes convincingly passes through the
right value, $c_\text{PF} = 5/4$.
The plateau starting close to the zero-temperature AF point
$x\approx -0.5412$ is at best emergent, in the form of a shoulder.
The clearest feature that remains is a sharp peak at
$x=-0.4722(1)$, with central charge $c \approx 1.64$. 
Like in the $p=7$ case, larger sizes would be necessary to 
corroborate this result.
As for the other values of $p$, this peak is below the AF critical curve
$x_\text{AF}\approx -0.43783$ (see figure~\ref{fig:p=8}).

%
%
\section{Physical conclusions}
\label{sec:concl}

The first goal of this paper was to compute the AF critical curve for the
triangular-lattice $Q$-state Potts model. This was not, by any means,
a simple computation because of the existence of a T-point. 
We have used two complementary methods: limiting curves obtained from the 
TM approach, and the graph polynomial method. 

The TM method required to build the 
TM for triangular-lattice strip graphs with toroidal boundary conditions
in the FK representation. We chose these boundary conditions as they are
expected to give results closer to the thermodynamic limit 
(as shown in \cite{JacobsenSalas07} for the chromatic polynomial);
but they require to work with larger connectivity spaces and the use of 
many technicalities \cite{JacobsenSalas07}. These two factors
severely limit the maximum strip width one can deal with symbolically, but
nevertheless, one can increase this width by performing the computation 
numerically. 
With this technique we were able to obtain an accurate approximation to
the AF critical curve for $0\le Q \lesssim 3.6$. The T-point $T=(Q_T,v_T)$ 
[cf.~\eqref{def_T}] is rather close to the upper limit of such interval,
so we expect large FSS corrections to take place, as well as parity effects
mod~3 for the interval $Q\in [3.6,4]$. 

\begin{figure}
\begin{center}
\includegraphics[width=200pt]{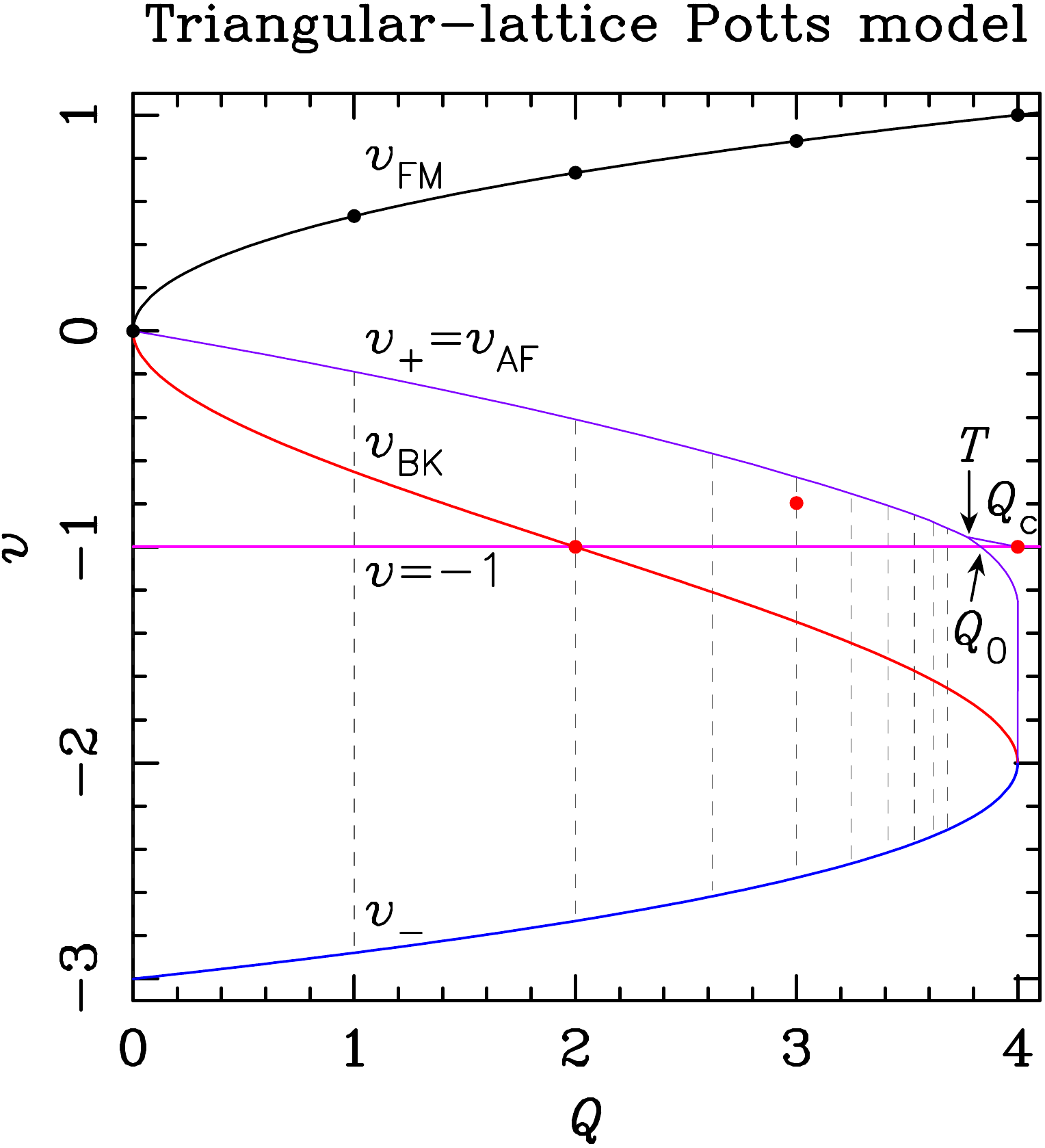}
\end{center}
\caption{Phase diagram of the triangular-lattice $Q$-state Potts 
model in the real $(Q,v)$ plane. (Compare to figure~\ref{fig:sq-tripd}.)
The cubic \eqref{cubic_tri} corresponds
to the three thick lines labeled $v_\text{FM}$, $v_\text{BK}$, and $v_{-}$, 
respectively. The horizontal line corresponds to $v=-1$ 
\eqref{chrom_tri}. The curve between $v_\text{FM}$ and  $v_\text{BK}$ is 
our best numerical estimate of the AF curve $v_+ =v_\text{AF}$
for this lattice. This curve has a 
bifurcation point T, leading to two branches: one goes to
the point $(Q_0,-1)$ [cf. \eqref{eq.Q0}], while the other goes to the 
critical point $(4,-1)$. The region defined by these three points and 
$(4,v_1)$ [cf. \eqref{def_v1}] 
is governed by regime~IV, while the BK phase corresponds to 
regime~I. The vertical dashed lines correspond to the 
Beraha numbers $B_p$ [cf.~\eqref{eq.Bp}], shown here for $3\le k\le 11$. 
The solid dots depict the known critical points for integer values of $Q$. 
}
\label{fig:tripd}
\end{figure}

On the other hand, the graph polynomial method provides also very accurate 
estimates for the AF critical curve, specially for the interval 
$Q\in [0,2]$; but it missed some regions, like the curve in the 
interval $Q\in (2,3)$, and the branch going from the T point down to 
$(4,v_1)$ [cf~\eqref{def_v1}]. By combining the data obtained from these
two methods, we were able to have a rather accurate estimate for the position
of the AF critical curve in the $(Q,v)$-plane. This curve is shown in
figure~\ref{fig:tripd}. The phase diagram of the triangular-lattice $Q$-state
Potts model is now complete at least for $Q\in[0,4]$. 

%
%

It is worth noticing that now we have a consistent renormalisation-group 
(RG) flow at least for $Q\in [0,4]$, which is rather similar to (but richer 
than) the RG flow for the same model on the square lattice. 
For a given value of $Q$ (which is not changed by the RG flow), we find
that the temperature is a relevant operator on the ferromagnetic and AF 
critical curves ($v_\text{FM}$ and $v_\text{AF}$, respectively), and on the
lower branch of the cubic curve $v_{-}$. For $v_\text{FM}$, the RG flow is
towards the trivial zero-temperature $v\to\infty$ and infinite-temperature 
$v=0$ fixed points. For $v_{-}$ the RG flow is towards the BK phase and the 
trivial unphysical $v\to\infty$ fixed points.

Inside the BK phase, there is
a non-trivial fixed point, which can be identified with the curve $v_\text{BK}$.
Note that the integrable chromatic line $v=-1$ is also in the BK universality
class \cite{VJS16} inside Regime 1, i.e., for $0 \le Q \le Q_0$, with $Q_0$ 
given by \eqref{eq.Q0}. However, since only the middle branch $v_\text{BK}$ 
of the cubic \eqref{cubic_tri} exists for all $Q \in [0,4]$, it must be the 
latter that acts as the RG attractor inside the BK phase.

Finally, above the AF critical curve, the RG flow is towards the 
trivial infinite-temperature $v=0$ fixed point. 
Below that curve, for $0 \le Q < Q_0$ it flows to the BK phase fixed point 
$v_\text{BK}$, and for $Q_0 < Q \le 4$ it flows to the chromatic polynomial 
fixed point $v=-1$ governed by the so-called Regime~IV \cite{VJS16}. Let us
recall that in both phases (BK and Regime~IV) the temperature is an 
irrelevant operator.

Although the phase diagram for the triangular-lattice $Q$-state Potts model
is thus well understood for $Q\in[0,4]$, we find that for $Q>4$ there are 
additional
features whose full comprehension would require more work.
In the TM approach, we find `phases' characterised by a pair of 
complex conjugate dominant eigenvalues (instead of the more common situation 
in which there is a single dominant eigenvalue). These phase are denoted 
by $\ell^*$ (with $0\le \ell \le 3$) in figure~\ref{Figure_tri_TM}. These
phases even appear close to the curve $v_{-}$ for $0\le Q\le 1$, although 
their extension seems to shrink as the strip width increases. This means
that partition-function zeros will accumulate not only along the usual 
limiting curves, but also throughout these regions. This unusual feature 
previously appeared when studying the phase diagram in the
$(Q,v)$-plane of the $Q$-state Potts model on the generalised 
Petersen graphs \cite{Salas13-1}. In the critical-polynomial approach,
we also find new structures for $Q > 4$ [see figure~\ref{Figure_tri_CP}(a)].
In both approaches, the phase-diagram features for $Q>4$ look like a band 
extending towards large $Q$ (perhaps even to $Q\to\infty$); but the FSS 
corrections are rather large and prevent us from drawing any definitive
 conclusion
[see figures~\ref{Figure_tri_TM_all}(a) and~\ref{Figure_tri_CP}(a).] 

\begin{figure}
\begin{center}
\includegraphics[width=200pt]{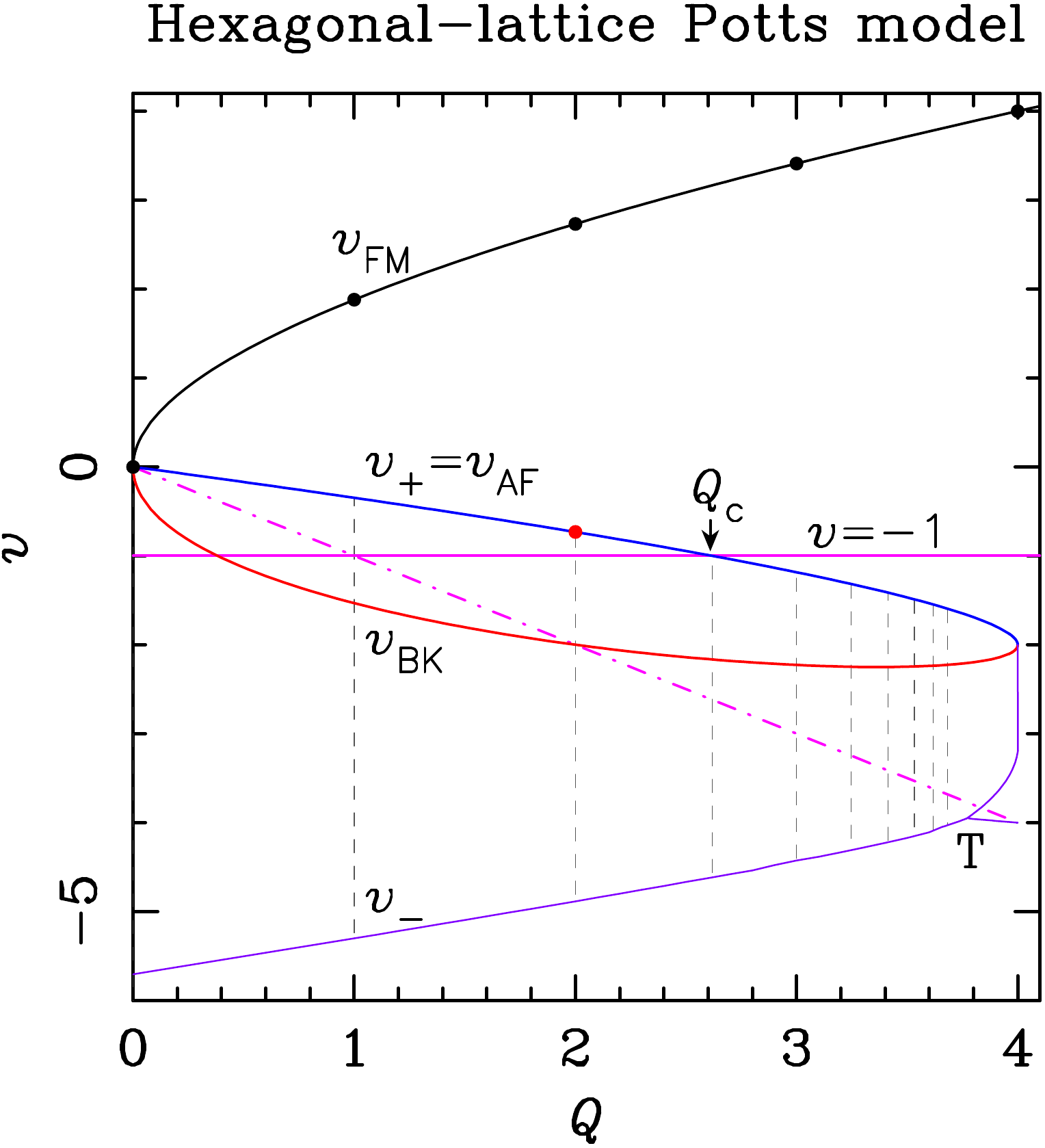}
\end{center}
\caption{Phase diagram of the hexagonal-lattice $Q$-state Potts 
model in the real $(Q,v)$ plane. The lines use the same colour code as
in figure~\ref{fig:tripd} to make the comparison easier. The cubic 
\eqref{cubic_tri} now corresponds the `dual cubic' $v^3 -3 Q v = Q^2$.
Its three branches are $v_\text{FM}$, $v_\text{AF}$, and $v_\text{BK}$. 
The dual of the AF critical curve of the triangular-lattice model now
plays the role of $v_{-}$.  
The horizontal (pink) line corresponds to $v=-1$, and $Q_c(\text{hex})=B_5$.
The inclined dot-dashed (pink) line corresponds to $v=-Q$ for $0\le Q\le 4$,
where the model is integrable. 
The vertical dashed lines correspond to the 
Beraha numbers $B_p$ [cf.~\eqref{eq.Bp}], shown here for $3\le p\le 11$. 
The solid dots depict the known critical points for integer values of $Q$. 
}
\label{fig:hcpd}
\end{figure}

The features of the phase diagram of the triangular-lattice $Q$-state 
Potts model described this far are expected to be
independent of the boundary conditions used in the computations
(once all the FSS effects are removed). Therefore, we can use the above 
conclusions to deduce the phase diagram of the hexagonal-lattice model as well. 
Even though duality for the Potts model only applies to planar graphs
\cite{Wu82}, we can nevertheless apply it to the phase diagram in 
the thermodynamic limit. Therefore, any curve $v(Q)$ in the 
triangular-lattice phase diagram will transform into its dual curve
$v^*(Q) = Q/v(Q)$ in the hexagonal-lattice phase diagram. In this way
we deduce the phase diagram of the hexagonal-lattice $Q$-state Potts model
(depicted in figure~\ref{fig:hcpd}) from figure~\ref{fig:tripd}.  
Indeed, the RG flows are similar to those for the triangular-lattice 
discussed above. 

%
%

Notice that duality relates $v_\text{FM}$ and $v_\text{BK}$ in both models.
More precisely, under duality  
$v_\text{F,tri} \leftrightarrow v_\text{F,hex}$ and 
$v_\text{BK,tri} \leftrightarrow v_\text{BK,hex}$. 
Therefore, we expect the same critical behavior on both
lattices for both the ferromagnetic critical curve $v_\text{FM}$ and the BK
phase. Indeed, there is no surprise in these observations.   

On the other hand, duality interchanges the role played by $v_\text{AF}$ and
$v_{-}$: i.e.,
$v_\text{AF,tri}  \leftrightarrow v_{-\text{,hex}}$ and
$v_{-\text{,tri}} \leftrightarrow v_\text{AF,hex}$. In 
\cite{JacobsenSaleur06} it was numerically shown that 
$v_{-\text{,tri}}(Q)$ belongs to the same universality class as the AF 
critical curve for the square-lattice model 
\cite{IkhlefJacobsenSaleur08,IkhlefJacobsenSaleur12} for the same value of $Q$. 
Notice that for the hexagonal lattice $Q_{\rm c}=B_5$, while for the square lattice,
$Q_{\rm c}=3$. Therefore, we conjecture that
the AF critical curves $v_\text{AF}(Q)$ for the hexagonal- and square-lattice 
$Q$-state Potts models belong to the same universality class at least for 
$Q\in [0,B_5]$.  

The second main goal of this paper is to study the triangular-lattice 
RSOS model on the torus. Indeed, when $Q=B_p$ [cf.~\eqref{eq.Bp}] 
for integer $q\ge 3$, \emph{and} we are in a probabilistic regime 
(i.e., $v\ge 0$ or $x \ge 0$), then the $Q$-state Potts model and the 
corresponding RSOS model of type $A_{p-1}$ give the same free energy
\cite{JacobsenRichard05}. But in the AF or the unphysical regimes, this 
property does not have to hold in general. In particular, this fact opens 
the possibility of finding distinct critical phenomena; as it already happened 
when the RSOS model was studied on the square lattice \cite{JacobsenSaleur06}. 
Moreover, in these latter regimes, universality is not expected to hold, so 
it seems very interesting to consider the triangular-lattice RSOS model on the
torus, and see whether there are new features and/or they are universal or
model-dependent.   

In this paper, we have built the TM for the triangular-lattice RSOS model
on the torus for $4\le p\le 8$, and strip widths $L=3,6,9,12,15$ (to avoid
mod~3 parity effects). Unfortunately, as $p$ increases, the number of states
also increases exponentially fast, so for $p=7,8$ we could only compute data
up to $L=12$. This means in practice that the features of the RSOS model
(e.g., its central charge as a function of $x$) become less 
amenable to numerical investigations as $p$
increases. The behavior of this model (compared to that of the $Q$-state 
Potts model in the FK and/or spin representation) is somewhat different 
depending on whether $Q$ is an integer or not. 

When $Q$ is an integer (i.e., $p=3,4,6$), we have compared the RSOS results
to those coming from TM in both the FK and spin representations. In the
former case, we have explicitly checked for $L=3$ that there are eigenvalue
cancellations, and vanishing amplitudes, as it happens for other 
strip graph families when $Q=B_p$ 
\cite{JacobsenSalas06,JacobsenSalas07,Salas13-1}. But we have used the 
results obtained from the TM in the spin representation (for widths up
to $L=12$) as a comparison for the RSOS results.  
The first non-trivial value corresponds to $p=4$ ($Q=2$). Even though we 
obtained a phase diagram with no surprises at all, it is remarkable that 
we have numerically observed that the full 
spectrum of the RSOS model on the torus is strictly contained in the 
union of the spectra of the corresponding Ising models on the torus
with periodic and anti-periodic boundary conditions, respectively,
in accord with recent results on braid translation in the blob algebra \cite{BGJSV17}.
This feature is also present for $p=6$ ($Q=3$), but in this case, the sum
is over three boundary conditions we can impose for the spin model on the
torus: i.e., periodic, $\mathbb{Z}_2$- and $\mathbb{Z}_3$-twisted 
boundary conditions.

Still for $p=6$, in addition to the expected peaks in the central charge
located at $x_\text{FM}$ and $x_{-}$, we find two interesting features
that merit some comments. First, there is a region approximately in the
interval $x\in [-1.07,-0.77]$ where the absolute value of the central charge 
takes large values that cannot be considered physical. This region is 
contained in the BK phase, and it is an open question to know what is the
real behavior of the RSOS model in that interval. Secondly, we find a sharp
peak at $x=-0.46010(2)$, which coincides within errors with the position 
$x = -0.460106(12)$ of the (weak) first-order phase transition found 
in \cite{Adler95}. The value of the central charge at this peak is 
rather stable $c = 1.265(15)$. The interpretation of this numerical result
is not clear, and we leave it as an open problem for future considerations.   
 
When $Q$ is a non-integer (i.e., $p=5,7,8$), we can only compare our
results with those coming from the TM in the FK representation. In particular,
we find that for $L=3$ in the FK representation, there are no
eigenvalue cancellations and/or vanishing amplitudes when $Q=B_p$.  
This feature can be explained if we take into account that the toroidal
triangular-lattice strip graphs are not planar. In this case, we 
should expect these cancellations to occur only at positive integers,
as conjectured in \cite{Salas13-1}. 
By simple inspection of figures~\ref{fig:p=5}(a), \ref{fig:p=7}(a), 
and~\ref{fig:p=8}(a), it is clear that the free energies for the RSOS and
Potts models differ in the whole BK phase: i.e. in the 
interval $x \in [x_{-},x_\text{AF}]$. Outside this interval (including of 
course the probabilistic regime $x\ge 0$), the ground states of those two 
models have the same free energy. Notice that (as in the $p=6$ case) there 
is a region around $x=-1$ where the central charge becomes unexpectedly
large in absolute value. Again, we need further studies with larger widths 
to disentangle the physics in this region. 

Even though our results are less clear as $p$ is increased, there are
clear indications (at least for $p=5,7$) that there are regions with
\emph{new} physics. Let us review these findings in some detail:
\begin{itemize}
  \item For $p=5$ it is clear that, in addition to the expected central
        charge peak at $x=x_{-}$ with value given by $c_\text{PF}$, there
        is a plateau for $x \in [x_{-},-1.14]$, where the central charge 
        stays approximately constant: i.e., $c=0.70(1)$, which agrees 
        within errors with $c_\text{FM}=7/10$. 
        For $p=7$, the peak at $x_{-}$ can be seen, but we need data with 
        larger values of $L$ to see whether there is a similar plateau to
        the right of the peak. 

  \item For $p=5,7,8$ there is a sharp peak just below the AF critical
        curve $v_\text{AF}$. Its position and the central charge attained
        at its maximum is $x=-0.4833(1)$ and $c=1.198(2)$ for $p=5$; 
        $x=-0.4669(2)$ and $c\approx 1.53$ for $p=7$; and 
        $x=-0.4722(1)$ and $c\approx 1.64$ for $p=8$. As $p$ increases, 
        the position of the maximum becomes more negative, and at the same
        time, it becomes closer to $v_\text{AF}$. In addition, the value of the 
        central charge increases as $p$ increases. It is worth mentioning that 
        the peak found for the first-order phase transition at $p=6$ 
        ($x=-0.46010(2)$ and $c = 1.265(15)$) fits perfectly in this
        general behavior. This seems to be a general feature of the model,
        but additional work is needed in order to investigate the
        nature of these peaks: i.e., whether there is a first-order phase transition,
        or a critical point, or a transition that is first- and second-order
        at the same time (like the AF critical curve for the square-lattice
        $Q$-state Potts model \cite{JacobsenSaleur06}). If (some of) the 
        transitions are second order, then one would need to identify the 
        corresponding CFT. 

  \item For $p=5,7,8$ there are clear indications of the existence of a 
        narrow plateau close and to the left of the previous peak with
        values that seem to alternatively agree with either 
        $c_\text{PF}$ or $c_\text{FM}$.  
        In particular, for $p=5$ the plateau corresponds to the interval  
        $x\in [-0.61,-0.57]$ with central charge $c=0.800(2)$, which agrees 
        within errors with $c_\text{PF}=4/5$. For $p=7$ the plateau is 
        given by the interval $x\in [-0.58,-0.51]$ with central charge 
        $c=0.859(2)$, which is compatible with $c_\text{FM}=6/7$.
        Finally, for $p=8$ we one find some hints of a nascent plateau  
        around $x\approx -1.54$ with $c\approx c_\text{PF} = 5/4$. We need 
        more data to give a more precise account of this feature.  
\end{itemize}

The understanding of these new critical regions, and the RG flow
among them remain as interesting open problems, which are outside the scope of
this paper.   

\bigskip

%
%
\appendix 
\section{Amplitudes for the triangular-lattice Potts model} 
\label{app.coef}

In this appendix we will display which amplitudes $\alpha_k$ appear in the
partition function for the $Q$-state Potts model on a triangular-lattice
strip graph of width $L$ and periodic boundary conditions in the FK 
representation \eqref{def_Z_vs_eigen}. Let us start with the particular
case $v=-1$ \cite{JacobsenSalas07}, 
and then we will move on to the general case. 

\subsection{Chromatic polynomial}

Chang and Shrock \cite{Shrock_06} found a family of basic amplitudes
$\beta^{(\ell)}$ for $\ell\ge 0$ that could be written in terms of the 
amplitudes $\alpha^{(\ell)}$ for strip graphs with cyclic boundary 
conditions \cite{Shrock_06b}:
\begin{equation}
\label{def_betas}
\beta^{(\ell)} \;=\; \begin{cases} 
      \alpha^{(\ell)} & \text{for $\ell=0,1$,} \\ 
      \alpha^{(\ell)} - \alpha^{(\ell-1)} + (-1)^\ell \, \alpha^{(1)} 
                      & \text{for $\ell\ge 2$.}
      \end{cases}
\end{equation}
The amplitudes $\alpha^{(\ell)}$ are expressed in terms of the 
Chebyshev polynomial of second kind $U_n$:
\begin{equation}
\alpha^{(\ell)}(Q) \;=\; U_{2\ell}\left( \frac{\sqrt{Q}}{2} \right) \,.
\end{equation}
In this paper we need the amplitudes $\beta^{(\ell)}$ for $0\le \ell \le 6$
\cite[equation (2.31)]{JacobsenSalas06}.

Moreover, Chang and Shrock noted that for $\ell\ge 2$, these amplitudes 
split in several parts. Richard and Jacobsen proved this
fact, and gave an explicit expression for the relevant amplitudes 
$\beta^{(\ell)}_m$ for $m\mid \ell$ \cite[equation (4.9)]{Richard07}:  
\begin{equation}
\label{def_betasBis}
\beta^{(\ell)}_m \;=\; \sum\limits_{d>0 \colon d\mid \ell} 
  \frac{ \mu\left( \frac{m}{m \wedge d} \right) \, 
         \phi\left(\frac{\ell}{d} \right)}
       {\ell\, \phi\left(\frac{m}{m \wedge d}\right)} \, 
    \widetilde{\beta}^{(\ell)}\,,
\end{equation}  
where $\widetilde{\beta}^{(\ell)} = \beta^{(\ell)}$ for $\ell\ge 2$ 
[cf.~\eqref{def_betas}], and $\widetilde{\beta}^{(1)}=-1$, $m\wedge d$ 
denotes the greatest common divisor of $m,d$, and $\mu$ and $\phi$ are 
the M\"obius and Euler's totient functions, respectively. The amplitudes
for $0\le \ell \le 5$ are listed (with a slightly different notation) 
in \cite[equations (2.32)/(2.33)]{JacobsenSalas07}. 
We will also need the amplitudes for $\ell=6$, which read: 
\begin{subequations}
\label{def_betasTris}
\begin{align}
\beta^{(6)}_1 &\;=\; \frac{1}{6}Q(Q-1)(Q-3)(Q^3-8Q^2+19Q-11)\,, 
   \label{eq.beta6-1} \\
\beta^{(6)}_2 &\;=\; \frac{1}{6}(Q-1)^2(Q-2)(Q^3-8Q^2+17Q-3)\,, 
   \label{eq.beta6-2} \\ 
\beta^{(6)}_3 &\;=\; \frac{1}{6}Q(Q-2)(Q-3)(Q-4)(Q^2-3Q+1) \,,
   \label{eq.beta6-3} \\ 
\beta^{(6)}_6 &\;=\; \frac{1}{6}Q(Q-1)(Q-2)(Q-4)(Q^2-5Q+5) \,,
   \label{eq.beta6-6} 
\end{align}
\end{subequations}
and satisfy $\beta^{(6)}=\beta^{(6)}_1 + \beta^{(6)}_2 + 2\beta^{(6)}_3 + 
2\beta^{(6)}_6$. 

The amplitudes for $\ell=0$ are always $\beta^{(0)}$ for $2\le L\le 6$; and the
other amplitudes are shown in table~\ref{table:ampli.PG}. For those
eigenvalues belonging to a single sector, we always find one of the 
basic amplitudes (i.e., $\beta^{(0)}$, $\beta^{(1)}$, and $\beta^{(\ell)}_m$;  
cf. \eqref{def_betas}/\eqref{def_betasBis}/\eqref{def_betasTris}),
sometimes multiplied by a positive integer multiplicity factor. 
For those eigenvalues belonging to 
two sectors, we always find that the corresponding amplitude is a linear
combination (with positive integer coefficients) of the basic amplitudes
corresponding to those sectors.

%
%
\begin{table}
\centering
\begin{tabular}{rcccc}
\hline\hline
$L$ & $\ell=1$ & $\ell=2$ 
    & $\ell=3$ 
    & $\ell=4$ \\ 
\hline 
2& $\beta^{(1)}$              & $\beta^{(2)}_1$  
 &  
 &  \\
3& $\beta^{(1)}$              & $\beta^{(2)}_1,\beta^{(2)}_2$  
 & $\beta^{(3)}_1$
 & \\
4& $\beta^{(1)}$              & $\beta^{(2)}_1,\beta^{(2)}_2$
 & $\beta^{(3)}_1,\beta^{(3)}_3,2\beta^{(3)}_3$ 
 & $\beta^{(4)}_1$ \\
5& $\beta^{(1)}$              & $\beta^{(2)}_1,\beta^{(2)}_2$
 & $\beta^{(3)}_1,\beta^{(3)}_3,2\beta^{(3)}_3$
 & $\beta^{(4)}_1,\beta^{(4)}_4,2\beta^{(4)}_4,\beta^{(4)}_2$ \\
6& $\beta^{(1)},2\beta^{(1)}$ & $\beta^{(2)}_1,2\beta^{(2)}_1,\beta^{(2)}_2,
                                2\beta^{(2)}_2$  
 & $\beta^{(3)}_1,2\beta^{(3)}_1,\beta^{(3)}_3,2\beta^{(3)}_3$
 & $\beta^{(4)}_1,\beta^{(4)}_4,2\beta^{(4)}_4,\beta^{(4)}_2,3\beta^{(4)}_2$\\
\hline\hline  
$L$ & $\ell=5$ 
    & $\ell=6$ 
    & $\ell=L,L-1$ 
    & $\ell=1,2$ \\
\hline
4& 
 & 
 & $\beta^{(3)}_1+\beta^{(4)}_4$ 
 & \\
5& $\beta^{(5)}_1$
 & 
 & $\beta^{(4)}_1+\beta^{(5)}_5$ 
 & $2(\beta^{(1)} + \beta^{(2)})$ \\
6& $\beta^{(5)}_1,\beta^{(5)}_5,2\beta^{(5)}_5$ 
 & $\beta^{(6)}_1$
 & $\beta^{(5)}_1+\beta^{(6)}_6$
 & \\
\hline\hline 
\end{tabular}
\caption{Amplitudes of the eigenvalues contributing to the chromatic polynomial
according to the sector the eigenvalue belongs to. All the eigenvalues 
belonging to sector $\ell=0$ have amplitude $\beta^{(0)}$. The results for 
$2\le L\le 4$ were obtained by Chang and Shrock 
\cite{Shrock_01a,Shrock_06,Shrock_06b}, while those for $L=5,6$ and the
classification in terms of sectors were obtained in \cite{JacobsenSalas07}. 
}
\label{table:ampli.PG}
\end{table}

\subsection{Partition function}

For the full partition function, we could determine the amplitudes 
only for $2\le L\le 4$. The results are displayed in 
table~\ref{table:ampli.ZG}. By comparing tables~\ref{table:ampli.PG} 
and~\ref{table:ampli.ZG}, we find two differences: 1) for $\ell=L\ge 3$ 
there are more amplitudes in the partition-function case; a fact that is 
quite natural; and 2) each eigenvalue for $L=4$ in the partition-function case
belongs to a single sector (in particular, when we take $v=-1$, 
each eigenvalue corresponding to the amplitude $\beta^{(4)}_2$ becomes 
identical to one of the eigenvalues with amplitude $\beta^{(3)}_1$). 

%
%
\begin{table}
\centering
\begin{tabular}{rcccc}
\hline\hline
$L$ & $\ell=1$ & $\ell=2$ 
    & $\ell=3$ 
    & $\ell=4$ \\ 
\hline 
2& $\beta^{(1)}$              & $\beta^{(2)}_1$  
 &  
 &  \\
3& $\beta^{(1)}$              & $\beta^{(2)}_1,\beta^{(2)}_2$  
 & $\beta^{(3)}_1,\beta^{(3)}_3$
 & \\
4& $\beta^{(1)}$              & $\beta^{(2)}_1,\beta^{(2)}_2$
 & $\beta^{(3)}_1,\beta^{(3)}_3,2\beta^{(3)}_3$ 
 & $\beta^{(4)}_1,\beta^{(4)}_4$ \\
\hline\hline  
\end{tabular}
\caption{Amplitudes of the eigenvalues contributing to the partition function 
for a triangular-lattice strip on a torus of width $2\le L\le 4$. 
They are classified 
according to the sector the eigenvalue belongs to. All the eigenvalues 
belonging to sector $\ell=0$ have amplitude $\beta^{(0)}$. 
}
\label{table:ampli.ZG}
\end{table}

%
%
\ack
We acknowledge useful discussions and correspondence with Yacine Ikhlef, 
Alan Sokal, Eric Vernier, and Paul Zinn-Justin. 
We also thank Alan Sokal for 
the kind permission to use computational resources (provided by Dell
Corporation) at New York University. 

J.S. is grateful for the hospitality from 2009 to 2017
of the Laboratoire de Physique Th\'eorique at the \'Ecole Normale 
Sup\'erieure where part of this work was done. 
 
The research of J.L.J. was supported in part by the Agence Nationale
de la Recherche (grant ANR-10-BLAN-0414: DIME), the Institut
Universitaire de France, and the European Research Council (through the 
advanced grant NuQFT).
The research of J.L.J. and J.S. was supported in part by 
Spanish MINECO grant FIS2014-57387-C3-3-P.
The work of C.R.S was performed under the auspices of the 
U.S.\/ Department of Energy at the Lawrence Livermore National Laboratory 
under Contract No~DE-AC52-07NA27344.

%
%
\section*{References}
\bibliographystyle{iopart-num}
\bibliography{JSS}

\end{document}